\documentclass[11pt,a4paper]{article}
\usepackage{authblk}
\usepackage{lmodern}
\usepackage{jheppub}
\usepackage{amsmath}
\usepackage{mathtools}
\usepackage{tikz}
\usepackage{empheq}
\usepackage{enumitem}
\usepackage{graphics}
\usepackage{wrapfig}
\usepackage{caption}
\usepackage{natbib}
\usepackage{xcolor}
\usepackage{framed}
\usepackage{array}
\definecolor{shadecolor}{gray}{0.925}

\usepackage[cbgreek]{textgreek}

\def\sideremark#1{\ifvmode\leavevmode\fi\vadjust{\vbox to0pt{\vss% the remark
 \hbox to 0pt{\hskip\hsize\hskip1em%                          will appear only
 \vbox{\hsize3cm\tiny\raggedright\pretolerance10000%          on the side
 \noindent #1\hfill}\hss}\vbox to8pt{\vfil}\vss}}}%
\newcommand{\bi}{\begin{itemize}}
\newcommand{\ei}{\end{itemize}}
\newcommand{\bea}{\begin{align}}
\newcommand{\eea}{\end{align}}
\newcommand{\be}{\begin{equation}}
\newcommand{\ee}{\end{equation}}

\makeatletter
\renewcommand*\env@matrix[1][\arraystretch]{%
  \edef\arraystretch{#1}%
  \hskip -\arraycolsep
  \let\@ifnextchar\new@ifnextchar
  \array{*\c@MaxMatrixCols c}}
\makeatother

\author[\ensuremath{a}]{Charlotte SLEIGHT}
\author[\ensuremath{b},\ensuremath{c},\ensuremath{d}]{\quad Massimo TARONNA}

\affiliation[\ensuremath{a}]{Centre for Particle Theory and Department of Mathematical Sciences, \\ Durham University, Durham, DH1 3LE, U.K.}

\affiliation[\ensuremath{b}]{Dipartimento di Fisica ``Ettore Pancini'', Universit\`a degli Studi di Napoli Federico II, \\Monte S. Angelo, Via Cintia, 80126 Napoli, Italy}

\affiliation[\ensuremath{c}]{Scuola Superiore Meridionale, Universit\`a degli Studi di Napoli Federico II,\\ Largo San Marcellino 10, 80138 Napoli, Italy}

\affiliation[\ensuremath{d}]{INFN, Sezione di Napoli, Monte S. Angelo, Via Cintia, 80126 Napoli, Italy}

\emailAdd{charlotte.sleight@durham.ac.uk, massimo.taronna@unina.it,}

\title{\centering \huge From dS to AdS and back}

\abstract{We describe in more detail the general relation uncovered in our previous work between boundary correlators in de Sitter (dS) and in Euclidean anti-de Sitter (EAdS) space, at any order in perturbation theory. Assuming the Bunch-Davies vacuum at early times, any given diagram contributing to a boundary correlator in dS can be expressed as a linear combination of Witten diagrams for the corresponding process in EAdS, where the relative coefficients are fixed by consistent on-shell factorisation in dS. These coefficients are given by certain sinusoidal factors which account for the change in coefficient of the contact sub-diagrams from EAdS to dS, which we argue encode (perturbative) unitary time evolution in dS. dS boundary correlators with Bunch-Davies initial conditions thus perturbatively have the same singularity structure as their Euclidean AdS counterparts and the identities between them allow to directly import the wealth of techniques, results and understanding from AdS to dS. This includes the Conformal Partial Wave expansion and, by going from single-valued Witten diagrams in EAdS to Lorentzian AdS, the Froissart-Gribov inversion formula. We give a few (among the many possible) applications both at tree and loop level. Such identities between boundary correlators in dS and EAdS are made manifest by the Mellin-Barnes representation of boundary correlators, which we point out is a useful tool in its own right as the analogue of the Fourier transform for the dilatation group. The Mellin-Barnes representation in particular makes manifest factorisation and dispersion formulas for bulk-to-bulk propagators in (EA)dS, which imply Cutkosky cutting rules and dispersion formulas for boundary correlators in (EA)dS. Our results are completely general and in particular apply to any interaction of (integer) spinning fields.}

\begin{document}

\begin{flushright}    
\texttt{}
\end{flushright}

\maketitle

\newpage

\section{Introduction}\label{sec::Intro}

de Sitter (dS) space is the maximally symmetric space-time with positive cosmological constant and plays an important role in understanding our Universe. The inflationary paradigm \cite{Guth:1980zm,Linde:1981mu,Albrecht:1982wi,Starobinsky:1982ee} postulates that the very early universe underwent a period of quasi-de Sitter expansion and astronomical observations \cite{SupernovaSearchTeam:1998bnz,SupernovaSearchTeam:1998fmf,SupernovaCosmologyProject:1998vns} indicate that we are currently undergoing a second period of inflationary expansion. Despite this central role, our understanding of Quantum Field Theory in dS space is very primitive compared to its negative curvature cousin, anti-de Sitter (AdS) space. AdS space has a boundary at spatial infinity, where one can identify boundary operators that enjoy an associative and convergent operator product expansion. The AdS isometry group acts as the conformal group in one lower dimension and the AdS/CFT correspondence \cite{Maldacena:1997re,Gubser:1998bc,Witten:1998qj} identifies boundary observables in AdS with correlation functions of operators in a Lorentzian Conformal Field Theory, which are defined completely non-perturbatively by unitarity, conformal symmetry and a consistent operator product expansion. 

In contrast, in dS space there is no notion of spatial infinity and its boundaries are space-like slices that lie in the infinite past and in the infinite future. While the dS isometry group also acts on these boundaries as the conformal group in one lower dimension, in contrast to AdS space the corresponding Conformal Field Theory is Euclidean and is therefore not bound by the standard Wightman and Osterwalder-Schrader axioms \cite{Streater:1989vi,Osterwalder:1973dx,Osterwalder:1974tc,Kravchuk:2021kwe} and, for this reason, the operator product expansion is not necessarily convergent. We thus lack a complete picture of the criteria that should be satisfied by boundary correlation functions of fields in dS space.

In recent years there has been significant effort to bridge the gap in our understanding of boundary correlators in AdS and dS, in many cases drawing on the similarities between them. Boundary correlators in dS, like their AdS counterparts, are strongly constrained by conformal symmetry and the absence of unphysical singularities \cite{Antoniadis:2011ib,Creminelli:2011mw,Maldacena:2011nz,Bzowski:2011ab,Kehagias:2012pd,Mata:2012bx,Kundu:2014gxa,Kundu:2015xta,Ghosh:2014kba,Pajer:2016ieg,Arkani-Hamed:2015bza,Arkani-Hamed:2018kmz,Farrow:2018yni,Baumann:2019oyu,Green:2020ebl}.\footnote{This has also been extended to theories that break the symmetry under special conformal transformations \cite{Green:2020ebl,Goodhew:2020hob,Pajer:2020wxk,Jazayeri:2021fvk,Melville:2021lst,Goodhew:2021oqg,Baumann:2021fxj}.} Furthermore, assuming the Bunch-Davies vacuum at early times, it has been shown \cite{Sleight:2020obc} that diagrams in the perturbative computation of correlators on the (future) boundary of dS can be expressed as a linear combination of Witten diagrams for the same process in Euclidean AdS,\footnote{It should be emphasised that, since unitary irreducible representations of the AdS and dS isometry groups do not coincide, such EAdS Witten diagrams generally are not generated by a theory with a spectrum satisfying the unitarity bound in AdS.} where the relative coefficients account for the change in the coefficients of the contact sub-diagrams when going from EAdS to dS. As shown in \cite{Sleight:2019mgd,Sleight:2019hfp,Sleight:2020obc}, such coefficients are given by certain sinusoidal factors which, as we shall argue in this work, encode (perturbative) unitary time evolution in de Sitter space.\footnote{The implications of perturbative unitarity on dS boundary correlators have also been explored in \cite{Goodhew:2020hob,Cespedes:2020xqq} and see \cite{Hogervorst:2021uvp,DiPietro:2021sjt} for very recent progress at the non-perturbative level.}

dS boundary correlators with Bunch-Davies initial conditions thus have the same singularity structure as their Euclidean AdS counterparts in perturbation theory and the identities between them allow to directly import the wealth of techniques, results and understanding from AdS to dS. Some of these possibilities were already explored in \cite{Sleight:2020obc}. In particular, single-valuedness of AdS boundary correlators in the Euclidean regime implies that they admit an expansion into Conformal Partial Waves/Harmonic Functions \cite{Dobrev:1975ru,Dobrev:1977qv,Mack:2009mi,Costa:2012cb,Caron-Huot:2017vep}. From this one can obtain their expansion into Conformal Blocks which, combined with the requirement of crossing symmetry, lies at the heart of the bootstrap of standard Lorentzian CFTs \cite{Simmons-Duffin:2016gjk,Poland:2018epd}. The fact that diagrams in dS can be expressed as a linear combination of EAdS Witten diagrams implies that dS boundary correlators are also single-valued and hence also admit a Conformal Partial Wave decomposition (at least in perturbation theory). This was applied in \cite{Sleight:2020obc} to obtain the Conformal Partial Wave and Conformal Block decomposition of tree-level exchange diagrams in dS as well as their decomposition under crossing, which are inherited from their EAdS counterparts. Other applications given include the definition of Mellin amplitudes for exchanges in dS, which in AdS have proven to be an instrumental tool owing to the striking similarities with scattering amplitudes in flat space \cite{Mack:2009mi,Mack:2009gy,Penedones:2010ue,Paulos:2011ie,Fitzpatrick:2011ia,Fitzpatrick:2011dm,Fitzpatrick:2011hu,Goncalves:2014rfa}.

In parallel to the above perspective, there has also been considerable progress in developing dS techniques that are inspired by the successes and the strengths of the scattering amplitudes programme in flat space, mostly at the level of the wavefunction coefficients. These include: Cosmological polytopes (positive geometries) \cite{Arkani-Hamed:2017fdk,Arkani-Hamed:2018bjr,Benincasa:2018ssx,Benincasa:2019vqr,Hillman:2019wgh}, Cosmological optical theorem \cite{Goodhew:2020hob} (see also \cite{Meltzer:2020qbr,Cespedes:2020xqq}), cutting rules and dispersion relations for boundary correlators \cite{Sleight:2019hfp,Sleight:2020obc} and wavefunction coefficients \cite{Meltzer:2020qbr,Jazayeri:2021fvk,Melville:2021lst,Goodhew:2021oqg,Meltzer:2021bmb,Baumann:2021fxj,Meltzer:2021zin}, BCFW-like recursion relations \cite{Arkani-Hamed:2017fdk,Jazayeri:2021fvk,Baumann:2021fxj}, cosmological scattering equations \cite{Gomez:2021qfd}, as well as relations between flat space scattering amplitudes/correlators and wavefunction coefficients \cite{Maldacena:2011nz,Raju:2012zr,Arkani-Hamed:2017fdk,Arkani-Hamed:2018bjr,Benincasa:2018ssx,Baumann:2020dch,Baumann:2021fxj,Bonifacio:2021azc}.

In this work we give an extended discussion of \cite{Sleight:2020obc}, providing further technical details behind the results presented, extending them to arbitrary collections of (integer) spinning fields and give some further applications. That any diagram in dS can be expressed as a linear combination of EAdS Witten diagrams follows from the fact that dS Schwinger-Keldysh propagators with Bunch-Davies initial condition can each be expressed as a linear combination of analytically continued bulk-to-bulk propagators in EAdS for the Dirichlet and Neumann boundary conditions. On a practical level, in \cite{Sleight:2020obc} we found it convenient to formulate such relations using a Mellin-Barnes representation for propagators in EAdS and dS, which is defined in the flat slicing of (EA)dS by taking the Mellin transform with respect to the bulk direction.  At the level of the Mellin-Barnes representation, such relations between propagators in EAdS and the Schwinger-Keldysh formalism of dS amount to the multiplication by a simple phase which allows to immediately re-express any given diagram in dS as a linear combination of EAdS Witten diagrams. In other words, as we shall see, the Mellin-Barnes representation makes manifest seemingly complicated functional relationships between diagrams contributing to boundary correlators in EAdS and dS. This allows us to establish a set of rules to immediately obtain the explicit decomposition of any given diagram in dS with Bunch-Davies initial conditions as a linear combination of Witten diagrams for the same process in EAdS, where the relative coefficients of each Witten diagram are fixed by the Bunch-Davies initial condition and consistent on-shell factorisation of the diagram in dS.

We also propose that the Mellin-Barnes representation of boundary correlators is interesting and useful in its own right. We argue that the Mellin transform in the presence of a scale symmetry plays an analogous role to the Fourier transform when we have translation invariance, potentially making the Mellin-Barnes representation a natural habitat for boundary correlators in (EA)dS --- which themselves are constrained by Dilatation Ward identities. It is also a useful tool to solve constraints from the full conformal symmetry group where, as we shall see, the special conformal Ward identity is reduced to a difference relation. In perturbation theory, complicated bulk integrations are trivialised in the Mellin-Barnes representation, much like how position space integrals are replaced by momentum conserving delta functions in flat space scattering amplitudes, and thus place the bulk and the boundary on a similar footing. In this way we can straightforwardly infer properties of boundary correlators which are inherited from the propagators that compute them, including on-shell factorisation and dispersion formulas and, in the case of de Sitter space, unitary time evolution. Using these properties, one can establish cutting rules for diagrams contributing to boundary correlators in (EA)dS, through which the full diagram can be reconstructed using the Mellin-Barnes dispersion formula \cite{Sleight:2019hfp,Sleight:2020obc}.

An outline / summary of results is as follows:
\begin{itemize}
    \item {\bf In section \ref{sec::adsdsdict}} we begin by reviewing the perturbative computation of Witten diagrams in EAdS and harmonic analysis in Euclidean anti-de Sitter space. 
    
    In section \ref{subsec::props(EA)dS} we review the results of \cite{Sleight:2019mgd,Sleight:2019hfp,Sleight:2020obc}, which extend harmonic analysis and propagators in EAdS to the Schwinger-Keldysh formalism of dS via the Wick rotations \eqref{wickinin}. We introduce the Mellin-Barnes representation, where propagators in (EA)dS take a universal form and the Wick rotations \eqref{wickinin} amount to multiplying by a simple phase. The Mellin-Barnes representation makes manifest the on-shell factorisation of bulk-to-bulk propagators and provides a simple dispersion formula \cite{Sleight:2019hfp,Sleight:2020obc} relating the full propagator to its discontinuity.
    
    In section \ref{subsec::feynrules} the Mellin-Barnes representation is motivated as a natural habitat for boundary correlators constrained by Dilatation Ward identities and various parallels are drawn with the properties of the Fourier transform in the presence of a translation symmetry. We give Feynman rules for the perturbative computation of boundary correlators in the Mellin-Barnes representation, which trivialises complicated integrals over the bulk coordinate -- replacing them with Dirac delta functions analogous to momentum conserving delta functions in flat space.
    
    In section \ref{subsec::cuttinganddisp} we derive Cutkosky cutting rules for the perturbative computation of (EA)dS boundary correlators in the Mellin-Barnes representation and a dispersion formula from which a diagram can be reconstructed from its discontinuities. These are motivated by the factorisation properties and dispersion formula for propagators in (EA)dS reviewed in section \ref{subsec::props(EA)dS}. As an example, we use the cutting rules and dispersion formula to compute tree-level exchanges in (EA)dS, with more complicated examples given in later sections.
    
    \item {\bf In section \ref{sec::3pt}} we consider contact diagram contributions to boundary correlators in (EA)dS.
    
    In section \ref{subsec::CWI3pt} we study how constraints from conformal Ward identities in Fourier space are implemented in the Mellin-Barnes representation, focusing on three-point boundary correlators of scalar fields. 
    
    In sections \ref{subsec::3ptsceads} and \ref{subsec::npt} we review the relations \cite{Sleight:2019mgd,Sleight:2019hfp} between contact diagrams of scalar fields in (EA)dS, which differ by a constant sinusoidal factor accounting for the change in contact diagram coefficient from EAdS to dS. These are derived using relations between bulk-to-boundary propagators in (EA)dS reviewed in section \ref{subsec::props(EA)dS}.
    
    In section \ref{subsec::3ptsaddingspin} we extend these results to contact diagrams generated by any cubic coupling of (integer) spinning fields. To this end we employ the weight-shifting operators of \cite{Sleight:2017fpc} developed in the ambient/embedding space formalism, which provide a kinematic map between on-shell cubic couplings in (EA)dS and three-point confomal structures on the boundary and, in particular, reduce any spinning three-point contact diagram to a boundary differential operator acting on a scalar seed. 
    
    In section \ref{subsec::contactunitarity} we discuss how unitary time evolution in dS is encoded perturbatively by the sinusoidal factors relating contact diagram coefficients in EAdS and dS. Such factors also imply the vanishing of contact diagrams in dS for certain collections of fields and certain boundary dimensions $d$, as already observed in \cite{Sleight:2019mgd,Goodhew:2020hob}.
    
    \item {\bf In section \ref{sec::4ptexch}} we consider four-point tree-level exchanges. 
    
    Inspired by the on-shell factorisation of diagrams in (EA)dS implied by the cutting rules given in section \ref{subsec::bootEAdSexch}, we show that on-shell exchanges in (EA)dS can be fixed by a combination of conformal symmetry, factorisation and boundary conditions. The full exchange is then reconstructed using the dispersion formula given in section \ref{subsec::props(EA)dS}, which we argue is a general relation between a function and its discontinuity in the Mellin-Barnes representation.
    
    In section \ref{subsec::dsasadsexch} we use the relations between three-point contact diagrams in EAdS and dS reviewed in section \ref{sec::3pt} to write the corresponding dS exchange as a linear combination of EAdS exchange Witten diagrams. This extends the result given in \cite{Sleight:2020obc} to (EA)dS exchanges generated by any cubic coupling of fields with arbitrary integer spin.
    
    In section \ref{subsec::factunitcbe} we use the identity between exchange diagrams in (EA)dS derived in the previous section to write down the conformal block expansion of dS exchanges, which is inherited from the (known) conformal block expansions of their EAdS counterparts. This allows us to study on-shell factorisation of the full dS exchange, as opposed to the contributions coming from each individual Schwinger-Keldysh propagator, which at the level of the conformal block decomposition is the factorisation of the conformal block coefficients into the three-point coefficients of the contact diagrams mediating the exchange. We also discuss how such coefficients are constrained by unitary time evolution in dS.

\item {\bf In section \ref{sec::generalalg}} we present a set of rules which, given any diagram in dS with Bunch-Davies initial conditions, allow to write down the precise linear combination of EAdS Witten diagrams that compute it. In particular, while the existence of such a decomposition follows as an immediate consequence of the fact that Schwinger-Keldysh propagators with Bunch-Davies initial condition can be expressed as a linear combination of analytically continued propagators in EAdS \cite{Sleight:2019mgd,Sleight:2019hfp,Sleight:2020obc}, one still has to turn the wheel of the Schwinger-Keldysh formalism to obtain the precise coefficient of each EAdS Witten diagram -- which can get cumbersome very quickly beyond the simplest of diagrams. In this section we take a bootstrap/boundary perspective, showing that the coefficients of each EAdS Witten diagram are fixed by the Bunch-Davies initial condition and on-shell factorisation of the diagram in dS.

In sections \ref{subsecc::candy4pt}, \ref{subsec::exbox4pt} and \ref{subsec::2ptbubbbleex} we give some applications to loop diagrams generated by non-derivative interactions of scalar fields in dS, including the one-loop candy and box diagrams at four-points and the one-loop bubble diagram at two-points. In section \ref{subsec::propprods} we derive some useful identities to compute higher-loop diagrams in dS formed by taking powers of bulk-to-bulk propagators, which we do by using the dS to EAdS rules to import analogous tools and identities directly from EAdS.

\item {\bf In section \ref{sec::analyticity}} we discuss the analyticity of dS boundary correlators with Bunch-Davies initial conditions in perturbation theory, which is inherited from the corresponding Witten diagrams in EAdS. In particular, as already pointed out in \cite{Sleight:2020obc}, perturbative dS boundary correlators are single-valued functions and hence admit an expansion into conformal partial waves like their AdS counterparts. As an example we determine the conformal partial wave expansion of tree-level exchanges in dS, extending the example given in \cite{Sleight:2020obc} to dS exchanges generated by any cubic coupling of fields with arbitrary integer spin. We also briefly discuss some loop examples. Given that diagrams in dS can be expressed as a linear combination of Witten diagrams in EAdS, to finish we discuss the possibility of applying Lorentzian AdS techniques to diagrams in dS, including the celebrated Froissart-Gribov inversion formula \cite{Caron-Huot:2017vep}. We compute the double-discontinuity of the dS exchange, finding that it gives back the corresponding conformal partial wave.

\end{itemize}

In appendix \ref{app} we give various technical details on results for propagators in (EA)dS.

\newpage

\section{Propagators and the Mellin-Barnes representation}
\label{sec::adsdsdict}

{\bf Review AdS.} Let us begin with a review of the perturbative computation of Witten diagrams in EAdS$_{d+1}$. We work in the flat slicing of EAdS$_{d+1}$
\begin{equation}\label{adsflatslice}
    \text{d}s^2_{\text{EAdS}} = \frac{L^2_{\text{AdS}}}{z^2}\left(\text{d}z^2+\text{d}{\bf x}^2\right), 
\end{equation}
where, towards the boundary $z\to 0$ a spin-$J$ field $\varphi_J$ of mass $m$ the solution to the wave equation behaves as
\begin{equation}
    \lim_{z\to 0} \varphi_J\left(z,{\bf x}\right) = O^{+}\left({\bf x}\right)z^{\Delta^+-J}+O^{-}\left({\bf x}\right)z^{\Delta^--J},
\end{equation}
with
\begin{equation}\label{massAdS}
    m^2 L^2_{\text{EAdS}} = - \left(\Delta^+ \Delta^-+J\right), \qquad \Delta^+ + \Delta^- = d, \qquad \Delta^+ \geq \Delta^-,
\end{equation}
and we employed the index free notation as in \eqref{genfuncintr}. The field quantized with $O^{-}\left({\bf x}\right) = 0$ corresponds to the Dirichlet (normalizable) boundary condition, while $O^{+}\left({\bf x}\right) = 0$ corresponds to the Neumann (non-normalizable) boundary condition -- see e.g. \cite{Hartman:2006dy}.

Boundary correlators in EAdS$_{d+1}$ can be computed in perturbation theory via a Witten diagram expansion \cite{Witten:2001ua}, which can be regarded as Feynman diagrams in EAdS$_{d+1}$ with the external legs anchored to the boundary. External legs, connecting a bulk point $\left(z,{\bf x}\right)$ to a point ${\bf x}^\prime$ on the boundary, are assigned bulk-to-boundary propagators $K^{\text{AdS}}_{\Delta,J}$. For scalar fields $\left(J=0\right)$ in the flat slicing \eqref{adsflatslice} the bulk-to-boundary propagator reads: 
\begin{subequations}\label{scalabubo}
\begin{align}
    K^{\text{AdS}}_{\Delta,0}\left(z,{\bf x};{\bf x}^\prime\right) &= \frac{C^{\text{AdS}}_{\Delta,0}}{\left(L_{\text{AdS}}\right)^{(d-1)/2}}\left(\frac{z}{z^2+\left({\bf x}-{\bf x}^\prime\right)^2}\right)^\Delta,\\ \label{normbubo}
    C^{\text{AdS}}_{\Delta,0} &=\frac{\Gamma\left(\Delta\right)}{2\pi^{\frac{d}{2}}\Gamma\left(\Delta-\frac{d}{2}+1\right)},
\end{align}
\end{subequations}
which solves the homogeneous wave equation where $\Delta=\Delta^+/\Delta^-$ for the Dirichlet/ Neumann boundary condition respectively. The bulk-to-boundary propagator $K^{\text{AdS}}_{\Delta,J}$ for a spin-$J$ field can be obtained from that \eqref{scalabubo} for a scalar field with the same scaling dimension by acting with a differential operator \cite{Sleight:2016hyl}, which we shall make use of later on in section \ref{subsec::3ptsaddingspin}.

 \begin{figure}[t]
    \centering
    \captionsetup{width=0.95\textwidth}
    \includegraphics[width=0.6\textwidth]{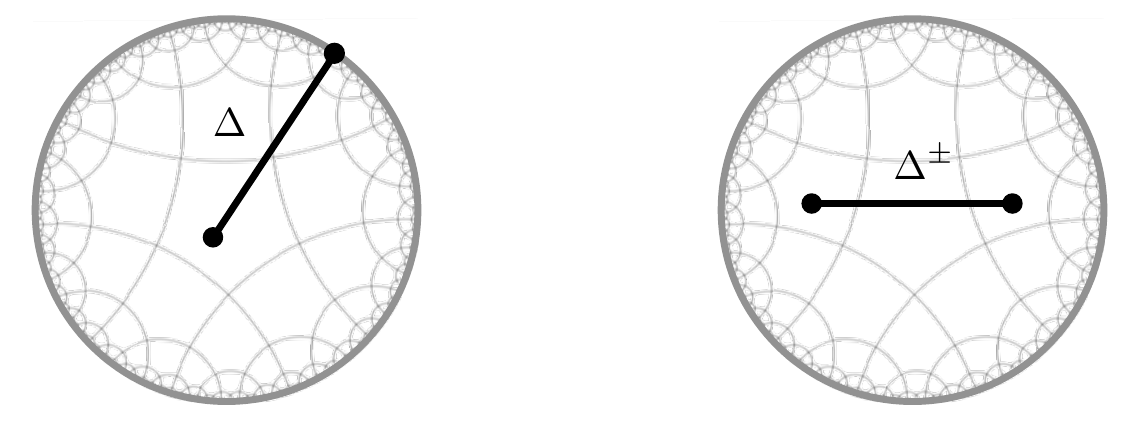}
    \caption{Throughout we represent EAdS pictorially as a Poincar\'e disk with grey circular boundary. \emph{Left:} Bulk-to-boundary propagator in EAdS with boundary condition $\Delta$, where $\Delta=\Delta^\pm$. \emph{Right:} Bulk-to-bulk propagator in EAdS with $\Delta^\pm$ boundary condition.}
    \label{fig::adsprop}
\end{figure}

Internal legs of Witten diagrams, which connect two bulk points, are associated bulk-to-bulk propagators which satisfy the wave equation with a Dirac delta function source:
\begin{equation}\label{bubuadseom}
    \left(\nabla^2-m^2\right)\Pi^{\text{AdS}}_{\Delta,J}\left(x;{\bar x}\right) = -\frac{1}{\sqrt{|g|}}\,\delta^{\left(d+1\right)}\left(x - {\bar x}\right).
\end{equation}
See figure \ref{fig::adsprop} for how we represent bulk-to-boundary and bulk-to-bulk propagators in AdS diagrammatically. Analogous to the exponential plane wave expansion in flat space, it is useful to decompose bulk-to-bulk propagators in a basis of bi-local Harmonic functions $\Omega^{\text{AdS}}_{\nu,J}$. These are trace- and divergence-free, and provide a complete basis of orthogonal solutions to the homogeneous wave equation (see appendix D of \cite{Costa:2014kfa}),
\begin{equation}\label{adsharm}
    \left(\nabla^2+\left(\left(\tfrac{d}{2}+i\nu\right)\left(\tfrac{d}{2}-i\nu\right)+J\right)L^{-2}_{\text{EAdS}}\right)\Omega^{\text{AdS}}_{\nu,J}\left(x;{\bar x}\right)=0.
\end{equation}
See e.g. appendix D of \cite{Costa:2014kfa} for the completeness and orthogonality relations satisfied of AdS harmonic functions and their appendix A for some useful parallels with harmonic functions in flat space. AdS harmonic functions are given explicitly by what is known as the ``split representation" \cite{Leonhardt:2003qu,Penedones:2010ue,Costa:2014kfa},
\begin{equation}\label{splitrep}
    \Omega^{\text{AdS}}_{\nu,J}\left(z,{\bf x};{\bar z},{\bar {\bf x}}\right) = \frac{\nu^2}{\pi} \int d^d{\bf x}^\prime\, K^{\text{AdS}}_{\frac{d}{2}+i\nu,J}\left(z,{\bf x};{\bf x}^\prime\right)K^{\text{AdS}}_{\frac{d}{2}-i\nu,J}\left({\bar z},{\bar {\bf x}};{\bf x}^\prime\right),
\end{equation}
which is a product of a bulk-to-boundary propagators with scaling dimensions $\frac{d}{2}\pm i\nu$ integrated over their common boundary point ${\bf x}^\prime$. 

For the Dirichlet boundary condition, which is normalisable, the decomposition of the bulk-to-bulk propagator for a spin-$J$ field in terms of harmonic functions $\Omega^{\text{AdS}}_{\nu,J}$ is given by the spectral integral \cite{Costa:2014kfa}:
\begin{equation}\label{dricheads}
    \Pi^{\text{AdS}}_{\Delta^+,J}\left(x;{\bar x}\right)=\int^{+\infty}_{-\infty}\frac{d\nu}{\nu^2+\left(\Delta_+-\frac{d}{2}\right)^2}\,\Omega^{\text{AdS}}_{\nu,J}\left(x;{\bar x}\right)+\text{contact},
\end{equation}
where ``+ contact" denotes contact term contributions from harmonic functions of spin $<J$. For the Neumann boundary condition, which is non-normalizable, we have (writing $\Delta^\pm=\frac{d}{2}+i\mu$):
\begin{equation}\label{neueads}
    \Pi^{\text{AdS}}_{\Delta^-,J}\left(x;{\bar x}\right)=\frac{2\pi i}{\mu}
    \Omega^{\text{AdS}}_{\mu,J}\left(x;{\bar x}\right)+\int^{+\infty}_{-\infty}\frac{d\nu}{\nu^2+\left(\Delta_+-\frac{d}{2}\right)^2}\,\Omega^{\text{AdS}}_{\nu,J}\left(x;{\bar x}\right)+\text{contact},
\end{equation}
from which we can infer that a harmonic function $\Omega^{\text{AdS}}_{\nu,J}$ is given by the difference of bulk-to-bulk propagators with $\Delta^+$ and $\Delta^-$ boundary conditions \cite{Costa:2014kfa}:
\begin{equation}\label{idsubharm}
    \Omega^{\text{AdS}}_{\nu,J}\left(x;{\bar x}\right) = \frac{i\nu}{2\pi}\left( \Pi^{\text{AdS}}_{\tfrac{d}{2}+i\nu,J}\left(x;{\bar x}\right)-\Pi^{\text{AdS}}_{\tfrac{d}{2}-i\nu,J}\left(x;{\bar x}\right)\right).
\end{equation}

Harmonic analysis in AdS is at the centre of various techniques to study boundary correlators in EAdS. In recent years it has been shown \cite{Sleight:2019mgd,Sleight:2019hfp,Sleight:2020obc} that the above harmonic analysis on AdS space can be extended to de Sitter space via analytic continuation, which we review in the following section.

\subsection{From EAdS to dS and the Mellin-Barnes representation}
\label{subsec::props(EA)dS}

This section provides further technical details on the intermediate steps taken to obtain the results presented in section 2 of \cite{Sleight:2020obc} and collects together various relevant results of \cite{Sleight:2019mgd,Sleight:2019hfp}.

In de Sitter space we are interested in correlators on the future boundary, which corresponds to $\eta \to 0$ in the flat slicing of dS$_{d+1}$:
\begin{equation}\label{dsflatslice}
    \text{d}s^2_{\text{dS}} = \frac{L^2_{\text{dS}}}{\eta^2}\left(-\text{d}\eta^2+\text{d}{\bf x}^2\right).
\end{equation}
This is related to the EAdS$_{d+1}$ flat slicing \eqref{adsflatslice} by the analytic continuation: 
\begin{equation}\label{analcon}
  z=-i\eta, \qquad  L_{\text{AdS}}=-i L_{\text{dS}}.
\end{equation}
Under the above analytic continuation, boundary correlators in EAdS$_{d+1}$ map to wavefunction coefficients in dS$_{d+1}$ \cite{Maldacena:2002vr,McFadden:2009fg,McFadden:2010vh,Harlow:2011ke,Mata:2012bx,Anninos:2014lwa}. Using that the wavefunction gives a probability amplitude for observing a given set of fluctuations, one can obtain the corresponding boundary correlators in dS$_{d+1}$ from the wavefunction coefficients by  computing the expectation values for the fluctuations -- i.e. by performing an additional path integral \cite{Maldacena:2002vr} -- see e.g. appendix A of \cite{Goodhew:2020hob} for a detailed review. Alternatively one can bypass the wavefunction to compute boundary correlators on the future boundary of de Sitter space directly within the in-in (or Schwinger-Keldysh formalism), which is the approach we take in this work. See \cite{Akhmedov:2013vka,Chen:2017ryl} for reviews. 
In our previous works \cite{Sleight:2019mgd,Sleight:2019hfp,Sleight:2020obc} it was shown that, assuming that all fields started their life in Bunch Davies vacuum at early times,\footnote{In particular, that the vacuum at early times $\eta \to -\infty$ is that of flat Minkowski space, which is sometimes referred to as the Hadamard condition.} the contribution to dS$_{d+1}$ boundary correlators from the $\pm$ branch of the in-in contour can be reached from the flat slicing \eqref{adsflatslice} of EAdS$_{d+1}$ via the following Wick rotation (see figure \ref{fig::ininbranch}):
\begin{equation}\label{wickinin}
   \pm \; \text{branch:} \qquad z = -\eta e^{\pm \frac{i \pi}{2}}.
\end{equation}
In other words, to reach the ``$+$ branch'' of the in-in contour one Wick rotates the bulk radial coordinate $z$ anti-clockwise, while to reach the ``$-$ branch" one Wick rotates clockwise.

\begin{figure}[t]
    \centering
    \captionsetup{width=0.95\textwidth}
    \includegraphics[width=0.6\textwidth]{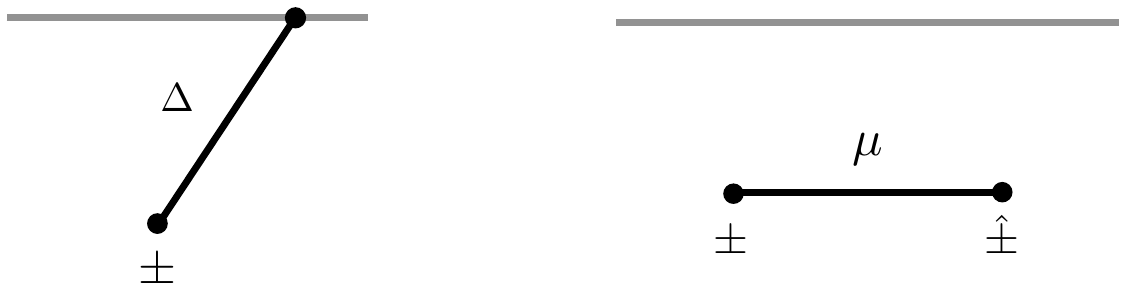}
    \caption{\emph{Left:} Diagrammatic representation of the bulk-to-boundary propagator \eqref{adsodsbubo} in dS where the point in the bulk sits in the $\pm$ branch of the in-in contour. \emph{Right:} Bulk-to-bulk propagator in dS for a field of mass \eqref{dSmass}, where one bulk point sits on the $\pm$ branch of the in-in contour and the other on the ${\hat \pm}$ branch.}
    \label{fig::dsprop}
\end{figure}

According to the above prescription, bulk-to-boundary propagators on the $\pm$-branches of the in-in contour in dS$_{d+1}$ can be obtained from their counterpart $K^{\text{AdS}}_{\Delta,J}$ in Euclidean AdS via the following analytic continuation in the bulk radial coordinate $z$ \cite{Sleight:2019mgd,Sleight:2019hfp}:\footnote{In particular, equation \eqref{adsodsbubo} is equation (2.38) of \cite{Sleight:2019hfp}, where $c^{\text{dS-AdS}}_{\Delta}$ combines the coefficients given in (2.91) and (2.31) of that paper. Strictly speaking, and as in (2.38) of \cite{Sleight:2019hfp}, when considering correlators of bulk fields at late times $\eta_0$ one should also multiply each external leg by a factor $\left(-\eta_0\right)^{\Delta-J}$. For convenience we leave such factors of $\eta_0$ implicit in this work.}
   \begin{align}\label{adsodsbubo}
   K^{\pm}_{\Delta,J}\left(\eta,{\bf x};{\bf x}^\prime\right) &=  c^{\text{dS-AdS}}_{\Delta}\,e^{\mp \Delta\tfrac{i\pi }{2}}K^{\text{AdS}}_{\Delta,J}\left(-\eta e^{\pm \frac{i \pi}{2}},{\bf x};{\bf x}^\prime\right),
\end{align} 
where the factor 
\begin{align}\label{cseadstods}
   c^{\text{dS-AdS}}_{\Delta} =\frac{C^{\text{dS}}_{\Delta,J}}{C^{\text{AdS}}_{\Delta,J}}= \frac{\Gamma\left(\frac{d}{2}-\Delta\right)\Gamma\left(\Delta-\frac{d}{2}+1\right)}{2\pi}=\frac{1}{2}\csc\left(\left(\tfrac{d}{2}-\Delta\right)\pi\right),
\end{align}
accounts for the change in normalisation as we move from AdS to dS. The dS two-point coefficient $C^{\text{dS}}_{\Delta,J}$ was given in section 2.4 of \cite{Sleight:2019hfp}. Note that the ratio \eqref{cseadstods} has the useful property that:
\begin{equation}
    c^{\text{dS-AdS}}_{\Delta^\pm} = \mp \,c^{\text{dS-AdS}}_{\Delta^\mp}.
\end{equation}
The mass of a spin-$J$ field in dS$_{d+1}$ is related to $\Delta^\pm$ via
\begin{equation}\label{dSmass}
    m^2L^2_{\text{dS}}=\Delta^+\Delta^-+J,
\end{equation}
which is obtained from the AdS expression \eqref{massAdS} via the analytic continuation \eqref{analcon}. From this point on wards, unless stated explicitly otherwise we set the curvature radius to 1. 

The relation \eqref{adsodsbubo}, via the split representation \eqref{splitrep}, allows us to define harmonic functions in dS$_{d+1}$ on the $\pm {\hat \pm}$ branches of the in-in contour \cite{Sleight:2020obc}:
\begin{multline}\label{analharmadstods}
 \Omega^{\pm, {\hat \pm}}_{\nu,J}\left(\eta, {\bf x};{\bar \eta},  {\bar {\bf x}}\right) = c^{\text{dS-AdS}}_{\frac{d}{2}+i\nu}c^{\text{dS-AdS}}_{\frac{d}{2}-i\nu}  e^{\mp \tfrac{i\pi }{2}\left(\tfrac{d}{2}+i\nu\right)}e^{{\hat \mp} \tfrac{i\pi }{2}\left(\tfrac{d}{2}-i\nu\right)}\\ \times \Omega^{\text{AdS}}_{\nu,J}\left(-\eta e^{\pm \frac{i \pi}{2}}, {\bf x};-{\bar \eta} e^{{\hat \pm} \frac{i \pi}{2}}, {\bar {\bf x}}\right),
\end{multline}
which, in turn, provides a split representation \cite{Sleight:2020obc}:\footnote{Notice that the dS Harmonic functions defined in this way satisfy orthogonality and completeness relations inherited from their AdS counterparts. This allows to apply Harmonic analysis directly on the in-in contour!}
\begin{equation}
    \Omega^{\pm, {\hat \pm}}_{\nu,J}\left(\eta, {\bf x};{\bar \eta},  {\bar {\bf x}}\right) =  \frac{\nu^2}{\pi} \int d^d{\bf x}^\prime\, K^{\pm}_{\frac{d}{2}+i\nu,J}\left(\eta,{\bf x};{\bf x}^\prime\right)K^{{\hat \pm}}_{\frac{d}{2}-i\nu,J}\left({\bar \eta},{\bar {\bf x}};{\bf x}^\prime\right),
\end{equation}
which extends the split representation \eqref{splitrep} of harmonic functions in EAdS$_{d+1}$ to the in-in formalism of dS space.

\begin{figure}[t]
    \centering
    \captionsetup{width=0.95\textwidth}
    \includegraphics[width=0.85\textwidth]{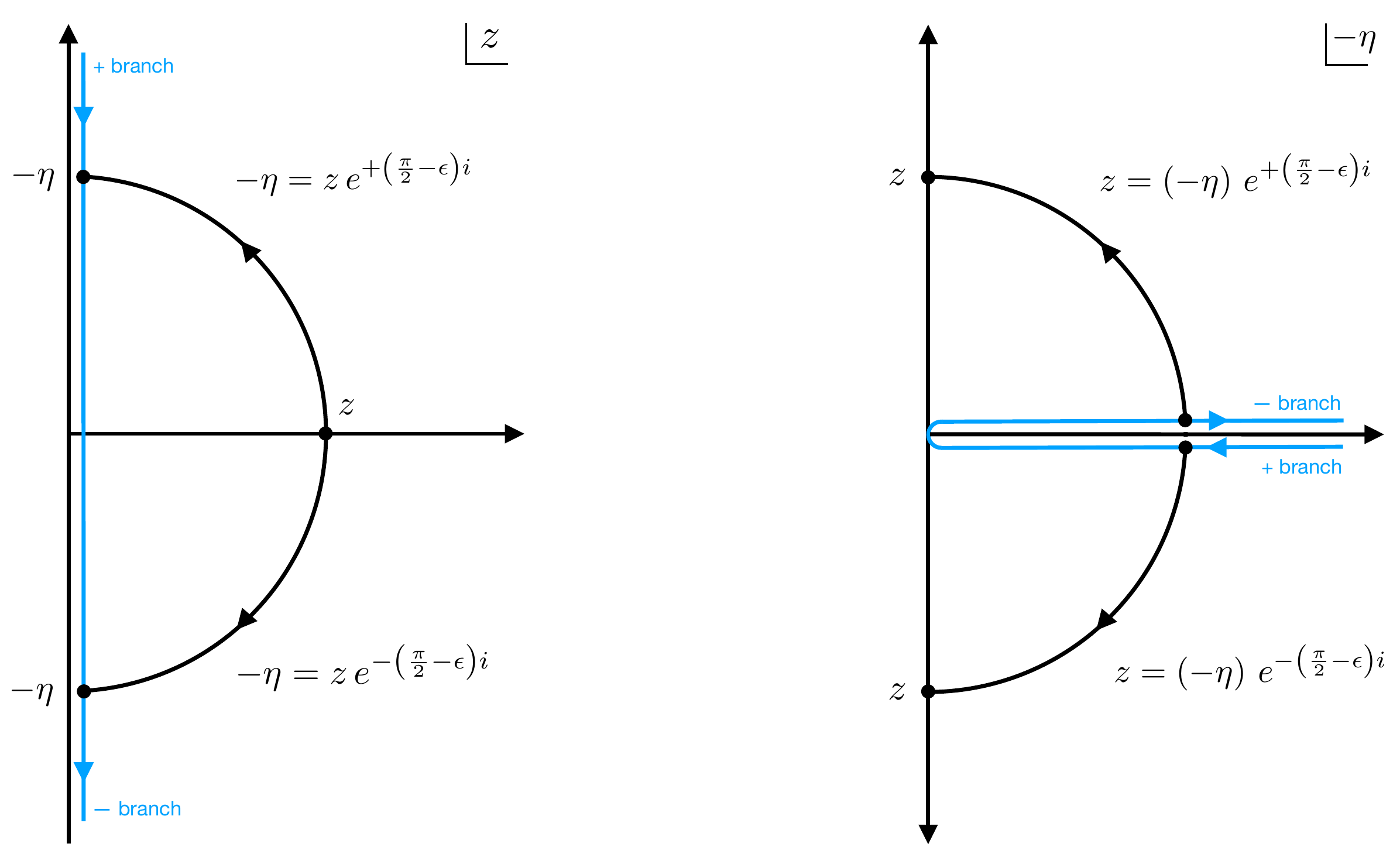}
    \caption{This figure displays the Wick rotations \eqref{wickinin} from the perspective of the complex $z$ plane (left) and the complex $\left(-\eta\right)$ plane (right). \emph{Left:} Starting from the flat slicing \eqref{adsflatslice} of EAdS, where $z \in \left[0,\infty\right)$, by Wick rotating anti-clockwise one lands on the $+$ branch of the in-in contour in the flat slicing \eqref{dsflatslice} of dS and by Wick rotating clockwise one lands on the $-$ branch of the in-in contour. \emph{Right:} From the $-(+)$ branch of the in-in contour one arrives to EAdS by rotating (anti-)clockwise. Similar figures were given in \cite{Sleight:2019mgd,Sleight:2019hfp}, where further details can also be found.}
    \label{fig::ininbranch}
\end{figure}

From a cosmological perspective, fixed time correlators in de Sitter space are usually studied in Fourier space due to translation invariance. In the presence of scale symmetry we propose\footnote{This proposal will be further substantiated in section \ref{subsec::feynrules}.} that it is convenient to furthermore adopt a \emph{Mellin-Barnes representation} 
\cite{Sleight:2019mgd,Sleight:2019hfp,Sleight:2020obc}, which is defined as the Mellin transform with respect to the bulk radial coordinate. For the bulk-to-boundary propagators, this is defined as
\begin{subequations}
\begin{align}
    K^{\text{AdS}}_{\Delta,J}\left(z,{\bf k}\right) &= \int^{+i\infty}_{-i\infty}\frac{ds}{2\pi i}\,K^{\text{AdS}}_{\Delta,J}\left(s,{\bf k}\right)z^{-2s+\tfrac{d}{2}},\\
    K^{\pm}_{\Delta,J}\left(\eta,{\bf k}\right) &= \int^{+i\infty}_{-i\infty}\frac{ds}{2\pi i}\,K^{\pm}_{\Delta,J}\left(s,{\bf k}\right)\left(-\eta\right)^{-2s+\tfrac{d}{2}},
\end{align}
\end{subequations}
where the bulk radial coordinate, both in EAdS and dS, is replaced by a variable $s$, which we refer to as an \emph{external Mellin variable}. For scalar fields ($J=0$) we have:\footnote{One obtains \eqref{MBeadssc} by noting that the bulk-to-boundary propagator for a scalar field of generic mass is given by a modified Bessel function of the second kind (or equivalently the ``Bessel K function") \cite{Gubser:1998bc} and \eqref{MBeadssc} follows from its Mellin-Barnes integral representation.} 
\begin{equation}\label{MBeadssc}
    K^{\text{AdS}}_{\Delta,0}\left(s,{\bf k}\right) = \frac{\Gamma\left(s+\tfrac{1}{2}\left(\tfrac{d}{2}-\Delta\right)\right)\Gamma\left(s-\tfrac{1}{2}\left(\tfrac{d}{2}-\Delta\right)\right)}{2\Gamma\left(\Delta-\frac{d}{2}+1\right)}\left(\frac{k}{2}\right)^{-2s+\Delta-\tfrac{d}{2}},
\end{equation}
where $k=|{\bf k}|$. At the level of the Mellin-Barnes representation the relation \eqref{adsodsbubo} between bulk-to-boundary propagators in EAdS$_{d+1}$ and the in-in formalism of dS$_{d+1}$ translates into a simple phase:
\begin{equation}\label{buborel}
    K^{\pm}_{\Delta,J}\left(s,{\bf k}\right)=c^{\text{dS-AdS}}_{\Delta}\, e^{\mp\left(s+\tfrac{1}{2}\left(\Delta-\tfrac{d}{2}\right)\right)\pi i}\, K^{\text{AdS}}_{\Delta,J}\left(s,{\bf k}\right).
\end{equation}
It follows that each external leg in the Mellin-Barnes representation of boundary correlators in (EA)dS$_{d+1}$ is characterised by two infinite families of poles in the corresponding external Mellin variable:
\begin{equation}\label{polesbuboeads}
    s = \pm \tfrac{1}{2}\left(\Delta-\tfrac{d}{2}\right)-n, \quad n = 0, \,1,\,2,\, \ldots\,,
\end{equation}
whose residues generate the expansion of the bulk-to-boundary propagator for small $kz$.

 \begin{figure}[t]
    \centering
    \captionsetup{width=0.95\textwidth}
    \includegraphics[width=\textwidth]{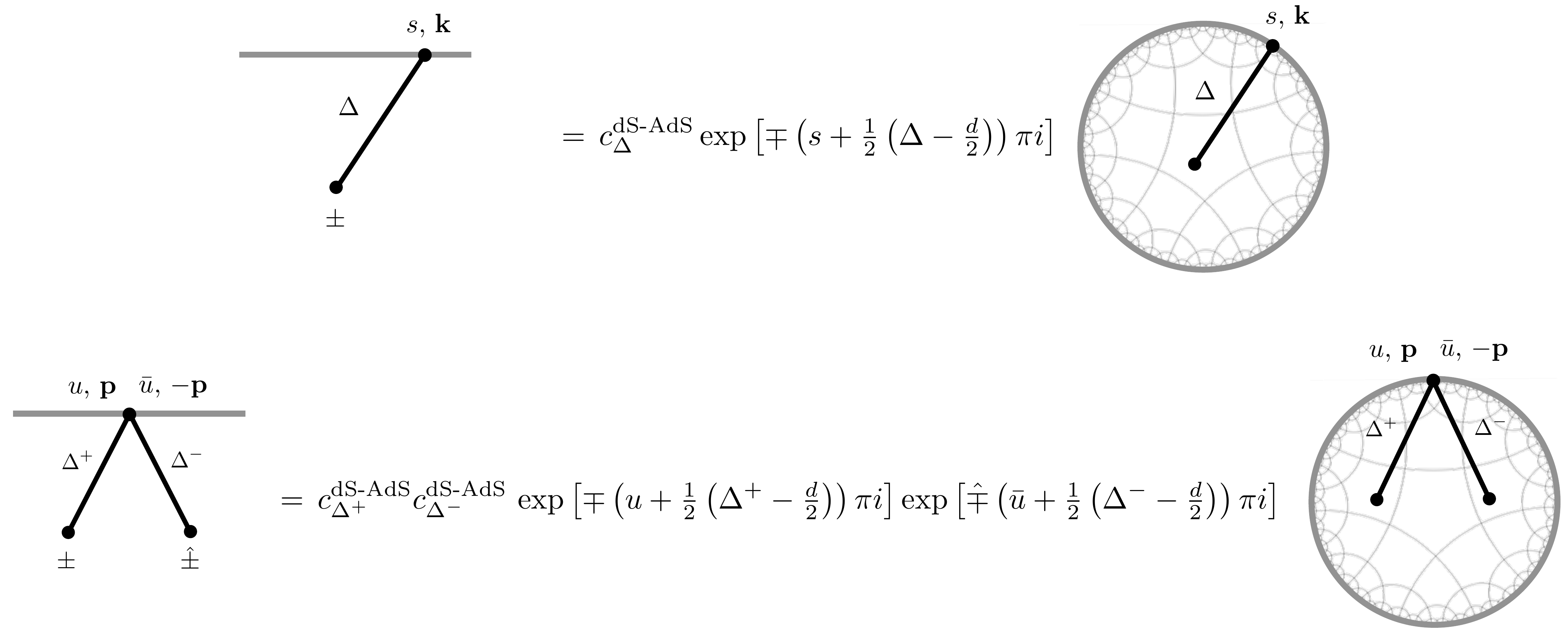}
    \caption{At the level of the Mellin-Barnes representation the relation between bulk-to-boundary propagators and harmonic functions in EAdS and the in-in formalism of dS amounts to a multiplication by a phase, which implements the Wick rotations \eqref{wickinin}, and by a constant $c^{\text{dS-AdS}}_{\Delta^\pm}$ which accounts for the change in two-point function coefficient from EAdS to dS. }
    \label{fig::harmfact}
\end{figure}

Each internal leg is instead described by a pair of \emph{internal Mellin variables} $u,\, {\bar u}$. This can be understood from the split representation \eqref{splitrep} of harmonic functions, which in Fourier space becomes a product of bulk-to-boundary propagators:
\begin{subequations}
\begin{align}
    \Omega^{\text{AdS}}_{\nu,J}(z,{\bf p};{\bar z},-{\bf p}) &=  \,\frac{\nu^2}{\pi}  K^{\text{AdS}}_{\frac{d}{2}+i\nu,J}\left(z,{\bf p}\right)K^{\text{AdS}}_{\frac{d}{2}-i\nu,J}({\bar z},-{\bf p}),\\
    \Omega^{\pm, {\hat \pm}}_{\nu,J}(\eta,{\bf p};{\bar \eta},-{\bf p}) &=  \,\frac{\nu^2}{\pi}  K^{\pm}_{\frac{d}{2}+i\nu,J}\left(\eta,{\bf p}\right)K^{{\hat \pm}}_{\frac{d}{2}-i\nu,J}({\bar \eta},-{\bf p}).
\end{align}
\end{subequations}
Its Mellin-Barnes representation is thus inherited from that of its two constituent bulk-to-boundary propagators:
\begin{subequations}
\begin{align}
    \Omega^{\text{AdS}}_{\nu,J}(z,{\bf p};{\bar z},-{\bf p}) &=  \int^{+i\infty}_{-i\infty} \frac{du}{2\pi i}\frac{d{\bar u}}{2\pi i}\,\Omega^{\text{AdS}}_{\nu,J}\left(u, {\bf p};{\bar u},-{\bf p}\right)z^{-2u+\tfrac{d}{2}}{\bar z}^{-2{\bar u}+\tfrac{d}{2}},\\
    \Omega^{\pm, {\hat \pm}}_{\nu,J}(\eta,{\bf p};{\bar \eta},-{\bf p}) &=  \int^{+i\infty}_{-i\infty} \frac{du}{2\pi i}\frac{d{\bar u}}{2\pi i}\,\Omega^{\pm, {\hat \pm}}_{\nu,J}\left(u, {\bf p};{\bar u},-{\bf p}\right)\left(-\eta\right)^{-2u+\tfrac{d}{2}}\left(-{\bar \eta}\right)^{-2{\bar u}+\tfrac{d}{2}},
\end{align}
\end{subequations}
where
\begin{subequations}\label{spliteadsds}
\begin{align}
    \Omega^{\text{AdS}}_{\nu,J}\left(u, {\bf p};{\bar u},-{\bf p}\right) &=  \frac{\nu^2}{\pi}  K^{\text{AdS}}_{\frac{d}{2}+i\nu,J}\left(u, {\bf p}\right)K^{\text{AdS}}_{\frac{d}{2}-i\nu,J}\left({\bar u}, -{\bf p}\right),\\
    \Omega^{\pm, {\hat \pm}}_{\nu,J}\left(u, {\bf p};{\bar u},-{\bf p}\right) &=  \frac{\nu^2}{\pi}  K^{\pm}_{\frac{d}{2}+i\nu,J}\left(u, {\bf p}\right)K^{{\hat \pm}}_{\frac{d}{2}-i\nu,J}\left({\bar u}, -{\bf p}\right).
\end{align}
\end{subequations}
At the level of the Mellin-Barnes representation the relation \eqref{analharmadstods} between harmonic functions in EAdS$_{d+1}$ and the in-in formalism in dS$_{d+1}$ reads:
\begin{equation}\label{dsharmphases}
    \Omega^{\pm, {\hat \pm}}_{\nu,J}\left(u, {\bf p};{\bar u},-{\bf p}\right)=c^{\text{dS-AdS}}_{\frac{d}{2}+i\nu}c^{\text{dS-AdS}}_{\frac{d}{2}-i\nu} e^{\mp\left(u+\tfrac{i\nu}{2}\right)\pi i}e^{{\hat \mp}\left({\bar u}-\tfrac{i\nu}{2}\right)\pi i}\,\Omega^{\text{AdS}}_{\nu,J}\left(u, {\bf p};{\bar u},-{\bf p}\right).
\end{equation}
The Mellin-Barnes representation of the Harmonic function thus has poles both in $u$ and ${\bar u}$ given by those \eqref{polesbuboeads} of its constituent bulk-to-boundary propagators:
\begin{subequations}\label{polesMBharm}
\begin{align}
    u &= \pm \frac{i\nu}{2}-n, \quad n = 0, \,1,\,2,\, \ldots,\\
    {\bar u} &= \pm \frac{i\nu}{2}-{\bar n}, \quad {\bar n} = 0, \,1,\,2,\, \ldots\,.
\end{align}
\end{subequations}

 \begin{figure}[t]
    \centering
    \captionsetup{width=0.95\textwidth}
    \includegraphics[width=\textwidth]{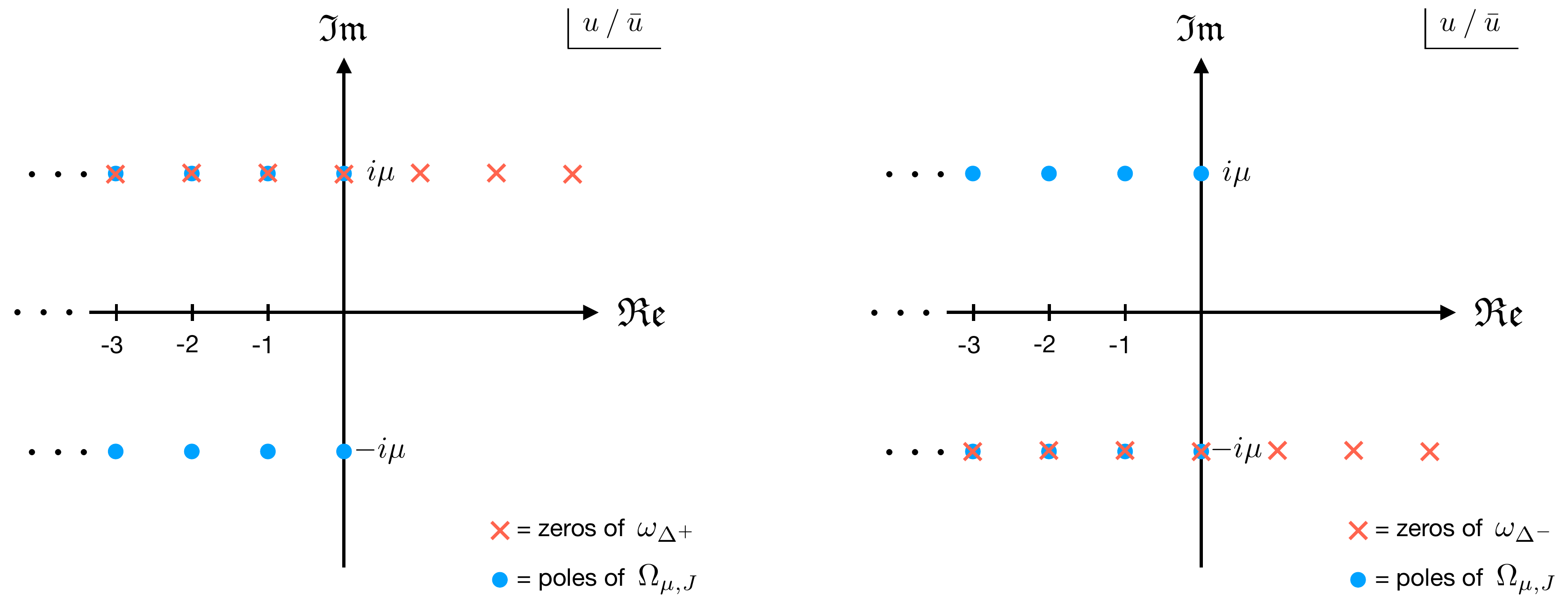}
    \caption{In both plots the blue dots represent the poles \eqref{polesMBharm} in the internal Mellin variables $u$ and ${\bar u}$. \emph{Left}: The crosses mark the zeros of the projector $\omega_{\Delta^+}\left(u,{\bar u}\right)$ which overlap with the poles of $\Omega^{\text{AdS}}_{\mu,J}$ in the upper-half plane that violate the $\Delta^+$ boundary condition. \emph{Right}: The crosses mark the zeros of the projector $\omega_{\Delta^-}\left(u,{\bar u}\right)$ which overlap with the poles of $\Omega^{\text{AdS}}_{\mu,J}$ in the lower-half plane that violate the $\Delta^-$ boundary condition.}
    \label{fig::polesuubar}
\end{figure}

Note that the integral over the spectral parameter $\nu$ in the harmonic function decomposition of the bulk-to-bulk propagators \eqref{dricheads} and \eqref{neueads} in EAdS$_{d+1}$ can be evaluated in closed form at the level of the Mellin-Barnes representation (for details see appendix \ref{app::bubuEAdS}). This gives an expression that places the propagators for the $\Delta^+$ (Dirichlet) and $\Delta^-$ (Neumann) boundary conditions on the same footing: 
\begin{equation}\label{bubuadsmom}
   \Pi^{\text{AdS}}_{\Delta^\pm,J}(z,{\bf p};{\bar z},-{\bf p}) =  \int^{+i\infty}_{-i\infty} \frac{du}{2\pi i}\frac{d{\bar u}}{2\pi i}\,\Pi^{\text{AdS}}_{\Delta^\pm,J}\left(u,{\bf p};{\bar u},-{\bf p}\right)z^{-2u+\tfrac{d}{2}}{\bar z}^{-2{\bar u}+\tfrac{d}{2}},
\end{equation}
where, parameterizing $\Delta^{\pm}=\frac{d}{2}\pm i\mu$, we have
\begin{equation}\label{MBrepbubuads}
   \Pi^{\text{AdS}}_{\Delta^\pm,J}\left(u,{\bf p};{\bar u},-{\bf p}\right) = \csc\left(\pi\left(u+{\bar u}\right)\right) \omega_{\Delta^{\pm}}\left(u,{\bar u}\right) \Gamma\left(i\mu\right)\Gamma\left(-i\mu\right) \Omega^{\text{AdS}}_{\mu,J}\left(u,{\bf p};{\bar u},-{\bf p}\right),
\end{equation}
up to contact terms. The functions $\omega_{\Delta^{\pm}}\left(u,{\bar u}\right)$ project onto the $\Delta^\pm$ boundary conditions and are given explicitly by
\begin{equation}\label{projectors}
    \omega_{\Delta^{\pm}}\left(u,{\bar u}\right) = 2 \sin\left(\pi\left(u \mp \tfrac{i\mu}{2}\right)\right)\sin\left(\pi\left({\bar u} \mp \tfrac{i\mu}{2}\right)\right).
\end{equation}
 In particular, the harmonic functions $\Omega^{\text{AdS}}_{\mu,J}$ solve the homogeneous wave equation \eqref{adsharm} with a specific linear combination of $\Delta^\pm$ boundary conditions, as exhibited by the identity \eqref{idsubharm}. The zeros of each sine function in \eqref{projectors} then overlap with the poles \eqref{polesMBharm} in the Mellin-Barnes representation of the Harmonic function that would violate the Dirichlet or Neumann boundary condition. See figure \ref{fig::polesuubar}. 

 \begin{figure}[t]
    \centering
    \captionsetup{width=0.95\textwidth}
    \includegraphics[width=0.55\textwidth]{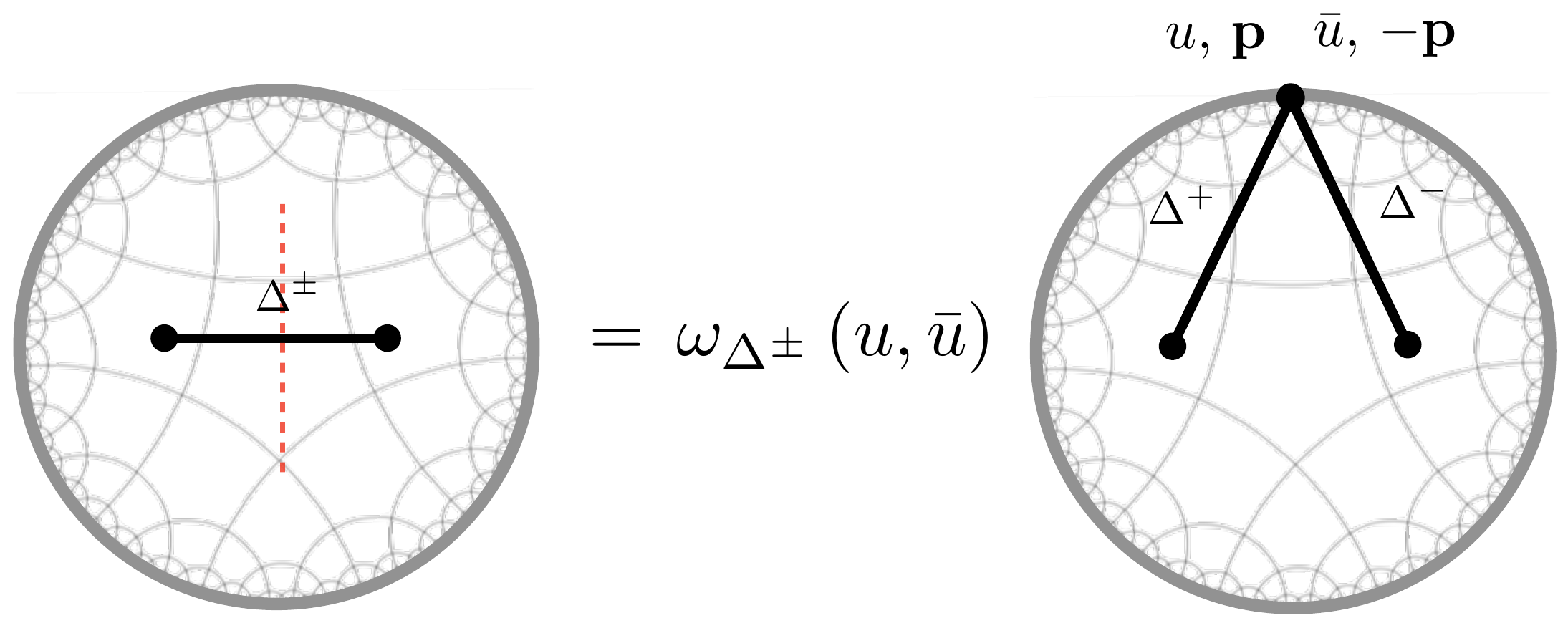}
    \caption{Diagrammatically we represent the discontinuity \eqref{discs} of internal lines by a perpendicular red dashed line. The discontinuity of the $\Delta^\pm$ bulk-to-bulk propagator in EAdS is factorised in $u$ and ${\bar u}$, which follows from the split representation of the harmonic function \eqref{spliteadsds} and the factorised form of the projectors $\omega_{\Delta^\pm}\left(u,{\bar u}\right)$.}
    \label{fig::discadsprop}
\end{figure}
 
At the level of the Mellin-Barnes representation the identity \eqref{idsubharm} reduces to the following trigonometric identity:
\begin{equation}\label{omomp}
    \omega_{\Delta^+}-\omega_{\Delta^-}=-2\sin\left(\pi \mu i \right)\sin\left(\pi\left(u+{\bar u}\right)\right),
\end{equation}
where, in particular, the factor $\sin\left(\pi\left(u+{\bar u}\right)\right)$ cancels the factor $\csc\left(\pi\left(u+{\bar u}\right)\right)$ in the bulk-to-bulk propagators \eqref{MBrepbubuads} to give the harmonic function on the l.h.s. of the identity \eqref{idsubharm}. Since the harmonic functions $\Omega^{\text{AdS}}_{\mu,J}$ satisfy the source-free wave equation \eqref{adsharm}, this indicates that the role of the factor $\csc\left(\pi\left(u+{\bar u}\right)\right)$ in \eqref{MBrepbubuads} is to ensure that the bulk-to-bulk propagators satisfy the wave equation with a source \eqref{bubuadseom} --- they encode the contact terms in the exchange process. This can also be understood from taking the discontinuity with respect to ${\sf s}=p^2$, defined as: 
\begin{equation}\label{discs}
    -2i \text{Disc}_{{\sf s}}\left[f\left({\bf s}\right)\right] = f\left(e^{i\pi}{\bf s}\right)-f\left(e^{-i\pi}{\bf s}\right).
\end{equation}
In particular, using that 
\begin{equation}
    \text{Disc}_{{\sf s}}\left[p^{-2\left(u+{\bar u}\right)}\right] = \sin\left(\pi\left(u+{\bar u}\right)\right)p^{-2\left(u+{\bar u}\right)},
\end{equation}
the discontinuity of the bulk-to-bulk propagators \eqref{bubuadsmom} is given by \cite{Sleight:2020obc}:
\begin{equation}\label{disceads}
\text{Disc}_{{\sf s}}\left[\Pi^{\text{AdS}}_{\Delta^\pm,J}(u,{\bf p};{\bar u},-{\bf p})\right] = \omega_{\Delta^{\pm}}\left(u,{\bar u}\right) \Gamma\left(i\mu\right)\Gamma\left(-i\mu\right) \Omega^{\text{AdS}}_{\mu,J}\left(u,{\bf p};{\bar u},-{\bf p}\right),
\end{equation}
where we see that the factor $\csc\left(\pi\left(u+{\bar u}\right)\right)$ has been cancelled and the propagator is completely factorised.\footnote{This is because both the projectors $\omega_{\Delta^\pm}\left(u,{\bar u}\right)$ and the harmonic function $\Omega^{\text{AdS}}_{\mu,J}\left(u,{\bf p};{\bar u},-{\bf p}\right)$ are factorised in $u$ and ${\bar u}$. That the discontinuity \eqref{disceads} of the $\Delta^\pm$ bulk-to-bulk propagators factorises has also been observed more recently in \cite{Meltzer:2020qbr,Baumann:2021fxj,Goodhew:2021oqg,Meltzer:2021zin}.} The discontinuities \eqref{disceads} satisfy the homogeneous wave equation \eqref{adsharm} with $\Delta^\pm$ boundary conditions and encode the physical exchanged single particle state subject to those boundary conditions. They are the ``on-shell propagators". This observation, originally presented in \cite{Sleight:2019hfp,Sleight:2020obc}, gives a dispersion formula for the bulk-to-bulk propagator through which it is reconstructed from its discontinuity (or ``on-shell part"): 
\begin{subequations}\label{dispprop}
\begin{align}
    \Pi^{\text{AdS}}_{\Delta^\pm,J}(z,{\bf p};{\bar z},-{\bf p}) &=  \int^{+i\infty}_{-i\infty} \frac{du}{2\pi i}\frac{d{\bar u}}{2\pi i}\,\Pi^{\text{AdS}}_{\Delta^\pm,J}\left(u,{\bf p};{\bar u},-{\bf p}\right)z^{-2u+\tfrac{d}{2}}{\bar z}^{-2{\bar u}+\tfrac{d}{2}},\\
    \Pi^{\text{AdS}}_{\Delta^\pm,J}\left(u,{\bf p};{\bar u},-{\bf p}\right)&=\csc\left(\pi\left(u+{\bar u}\right)\right)\text{Disc}_{{\sf s}}\left[\Pi^{\text{AdS}}_{\Delta^\pm,J}(u,{\bf p};{\bar u},-{\bf p})\right].
\end{align}
\end{subequations}
This is the Mellin-Barnes counterpart of the bulk-to-bulk propagator dispersion formulas given in the more recent \cite{Meltzer:2021bmb,Meltzer:2021zin}. In section \ref{subsec::cuttinganddisp} we will relate the prescription \cite{Sleight:2019hfp,Sleight:2020obc} above to that of \cite{Meltzer:2021zin}.

In summary, the Mellin-Barnes representation for a bulk-to-bulk propagator in EAdS$_{d+1}$ subject to a generic linear combination of $\Delta^+$ and $\Delta^-$ boundary conditions reads \cite{Sleight:2020obc}:
\begin{shaded}\label{AdSbubuMB}
\begin{subequations}
\begin{align}
    \Pi^{\text{AdS}}_{\alpha \Delta^++\beta \Delta^-,J}\left(u,{\bf p};{\bar u},-{\bf p}\right) &=  \underbrace{\csc\left(\pi\left(u+{\bar u}\right)\right)}_{\text{contact terms}}\underbrace{\text{Disc}_{{\sf s}}\left[\Pi^{\text{AdS}}_{\alpha \Delta^++\beta \Delta^-,J}(u,{\bf p};{\bar u},-{\bf p})\right]}_{\text{on-shell propagator}},\\ \text{Disc}_{{\sf s}}\left[\Pi^{\text{AdS}}_{\alpha \Delta^++\beta \Delta^-,J}(u,{\bf p};{\bar u},-{\bf p})\right]&=\underbrace{\left(\alpha\, \omega_{\Delta^{+}}\left(u,{\bar u}\right)+\beta\, \omega_{\Delta^{-}}\left(u,{\bar u}\right)\right)}_{\text{boundary conditions}}\, \\ & \hspace*{4cm}\times \Gamma\left(i\mu\right)\Gamma\left(-i\mu\right) \underbrace{\Omega^{\text{AdS}}_{\mu,J}\left(u,{\bf p};{\bar u},-{\bf p}\right)}_{\text{harmonic function}}.\nonumber
\end{align}
\end{subequations}
\end{shaded}

\begin{figure}[t]
    \centering
    \captionsetup{width=0.95\textwidth}
    \includegraphics[width=0.85\textwidth]{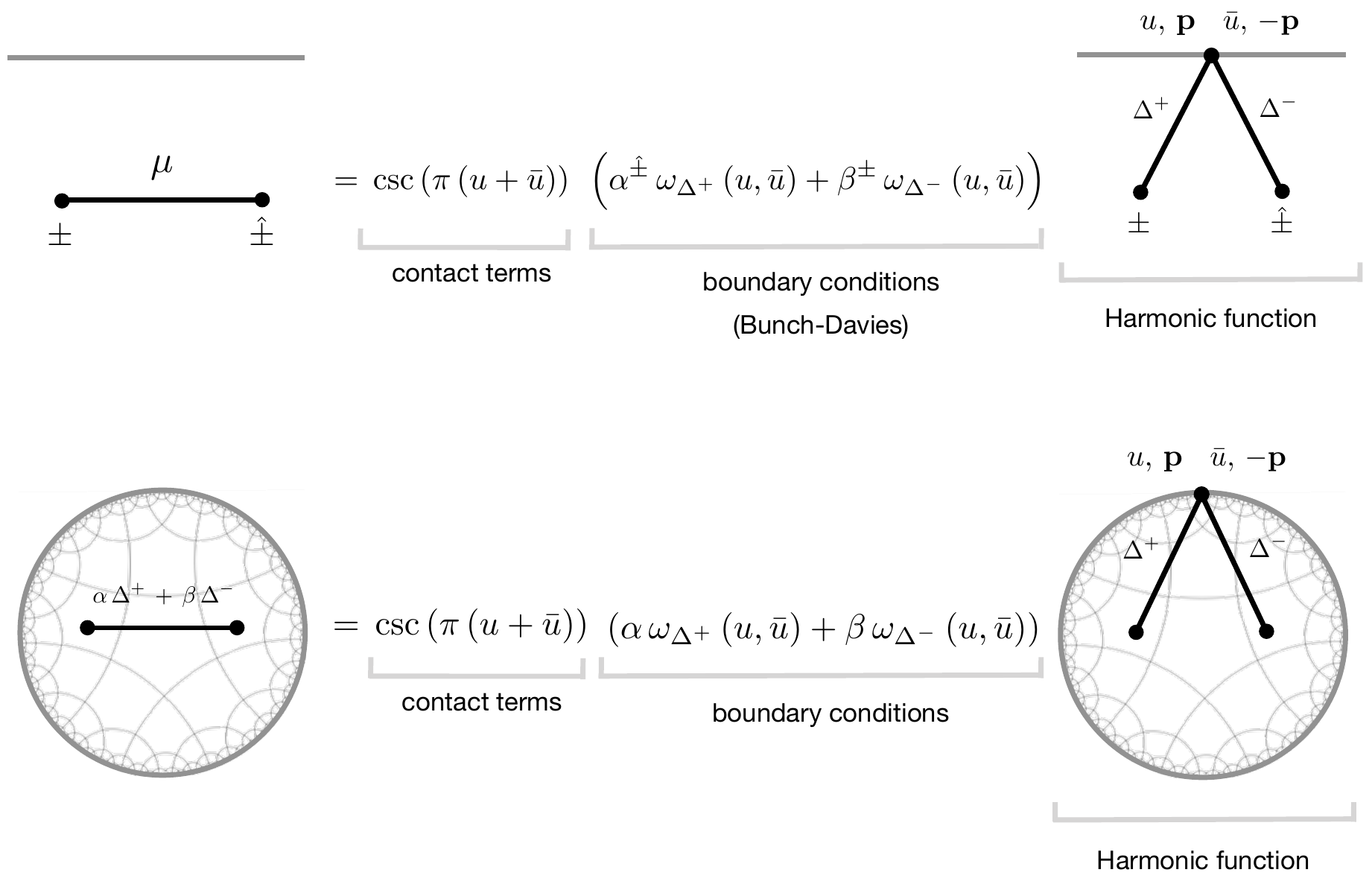}
    \caption{Bulk-to-bulk propagators in (EA)dS take a universal form in the Mellin-Barnes representation, which on-shell is given by the harmonic function multiplied by the appropriate linear combination of projectors $\omega_{\Delta^\pm}\left(u,{\bar u}\right)$ implementing the boundary condition. The full propagator is reconstructed from its on-shell part by multiplying with the factor $\csc\left(\pi\left(u+{\bar u}\right)\right)$.}
    \label{fig::dsadsbubuprops}
\end{figure}

Moving to de Sitter space, as noted in \cite{Sleight:2020obc}, in the Bunch-Davies vacuum bulk-to-bulk propagators can be expressed as a specific linear combination of Wick rotated \eqref{wickinin} Dirichlet/Neumann bulk-to-bulk propagators \eqref{bubuadsmom}. In particular, if we would like to obtain an expression for the dS bulk-to-bulk propagators similar to \eqref{AdSbubuMB} but in terms of the dS harmonic function \eqref{dsharmphases}, for the bulk-to-bulk propagator on the $\pm\,{\hat \pm}$ branches of the in-in contour we an ansatz of the form:
\begin{multline}\label{dslcadsbubu}
    \Pi^{\pm {\hat \pm}}_{\mu,J}\left(\eta,{\bf p};{\bar \eta},-{\bf p}\right)=  c^{\text{dS-AdS}}_{\frac{d}{2}+i\mu}c^{\text{dS-AdS}}_{\frac{d}{2}-i\mu}\,e^{\mp \tfrac{i\pi }{2}\left(\tfrac{d}{2}+i\mu\right)}e^{ {\hat \mp} \tfrac{i\pi }{2}\left(\tfrac{d}{2}-i\mu\right)}\\ \times \left[\alpha^{\pm\,{\hat \pm}}\, \Pi^{\text{AdS}}_{\frac{d}{2}+i\mu,J}\left(-\eta e^{\pm \frac{i \pi}{2}},{\bf p};-{\bar \eta}e^{{\hat \pm} \frac{i \pi}{2}},-{\bf p}\right)+\beta^{\pm\,{\hat \pm}}\, \Pi^{\text{AdS}}_{\frac{d}{2}-i\mu,J}\left(-\eta e^{\pm \frac{i \pi}{2}},{\bf p};-{\bar \eta}e^{{\hat \pm} \frac{i \pi}{2}},-{\bf k}\right)\right],
\end{multline}
which is a linear combination of $\Delta^\pm$ EAdS bulk-to-bulk propagators Wick rotated according to \eqref{wickinin}. To determine the coefficients $\alpha^{\pm\,{\hat \pm}}$ and $\beta^{\pm\,{\hat \pm}}$, we take the Mellin transform:
\begin{multline}\label{dslcadsbubuMB}
    \Pi^{\pm,\, {\hat \pm}}_{\mu,J}\left(u,{\bf p};{\bar u},-{\bf p}\right)= c^{\text{dS-AdS}}_{\frac{d}{2}+i\mu}c^{\text{dS-AdS}}_{\frac{d}{2}-i\mu}\,e^{\mp\left(u+\tfrac{i\mu}{2}\right)\pi i}e^{{\hat \mp}\left({\bar u}-\tfrac{i\mu}{2}\right)\pi i}\,\\ \times \left[\alpha^{\pm\,{\hat \pm}}\, \Pi^{\text{AdS}}_{\frac{d}{2}+i\mu,J}\left(u,{\bf p};{\bar u},-{\bf p}\right)+\beta^{\pm\,{\hat \pm}}\, \Pi^{\text{AdS}}_{\frac{d}{2}-i\mu,J}\left(u,{\bf p};{\bar u},-{\bf p}\right)\right]\,,
\end{multline}
and compare with the Mellin-Barnes representation of the bulk-to-bulk propagators \eqref{AdSbubuMB} in AdS. One finds (for a detailed derivation see appendix \ref{app::bubudS}):
\begin{subequations}
\begin{align}
\alpha^{\pm\, \pm}&=\frac{1}{c^{\text{dS-AdS}}_{\frac{d}{2}-i\mu}}e^{\pm \pi \mu}, &&\beta^{\pm \,\pm}=\frac{1}{c^{\text{dS-AdS}}_{\frac{d}{2}+i\mu}}e^{\mp \pi \mu},
\\
     \alpha^{\pm \mp} &=\frac{1}{c^{\text{dS-AdS}}_{\frac{d}{2}-i\mu}}e^{\mp \pi \mu}, &&\beta^{ \pm \mp}=\frac{1}{c^{\text{dS-AdS}}_{\frac{d}{2}+i\mu}}e^{\mp \pi \mu}.
\end{align}
\end{subequations}
which were originally given in \cite{Sleight:2020obc} (equation (2.17)) but with a different normalisation for the dS harmonic functions \eqref{dsharmphases}. Note that these can be written more compactly as:
\begin{equation}\label{coeffsab}
    \alpha^{{\hat \pm}}:= \alpha^{\pm \,{\hat \pm}}, \qquad \beta^{\pm}:= \beta^{\pm\,{\hat \pm}}.
\end{equation}

We therefore have that the bulk-to-bulk in-in propagators in dS$_{d+1}$ are given by the following linear combination of bulk-to-bulk propagators in EAdS$_{d+1}$ analytically continued according to \eqref{wickinin}:
\begin{multline}\label{analcontbubu}
    \Pi^{\pm {\hat \pm}}_{\mu,J}\left(\eta,{\bf p};{\bar \eta},-{\bf p}\right)=  c^{\text{dS-AdS}}_{\frac{d}{2}+i\mu}e^{\mp \tfrac{i\pi }{2}\left(\tfrac{d}{2}+i\mu\right)}e^{ {\hat \mp} \tfrac{i\pi }{2}\left(\tfrac{d}{2}+i\mu\right)}\, \Pi^{\text{AdS}}_{\frac{d}{2}+i\mu,J}\left(-\eta e^{\pm \frac{i \pi}{2}},{\bf p};-{\bar \eta}e^{{\hat \pm} \frac{i \pi}{2}},-{\bf p}\right)\\+\,\left(\mu \rightarrow -\mu\right)\,\,,
\end{multline}
where, as for the bulk-to-boundary propagators \eqref{adsodsbubo}, the coefficients $c^{\text{dS-AdS}}_{\frac{d}{2}\pm i\mu}$ account for the change in two-point function coefficient as we move from EAdS to dS. Note that the Bunch-Davies vacuum selects in-in propagators that are symmetric under $\mu \to -\mu$ (or equivalently $\Delta^+ \leftrightarrow \Delta^-$). Note also that the precise linear combination of analytically continued $\Delta^\pm$ AdS propagators depends on the branches of the in-in contour. This means that the bulk-to-bulk propagators in the Schwinger-Keldysh formalism are not the analytic continuation of one and the same bulk-to-bulk propagator in anti-de Sitter space.\footnote{For related discussions see \cite{Bousso:2001mw,Spradlin:2001nb,Balasubramanian:2002zh}.} I.e. they are each analytic continuations of different linear combinations of $\Delta^\pm$ AdS propagators. From the form \eqref{dslcadsbubuMB} of our ansatz it follows that bulk-to-bulk in the Bunch-Davies vacuum can also be written in the form \eqref{AdSbubuMB} too \cite{Sleight:2020obc}: 
\begin{shaded}
\begin{subequations}\label{bubodSinin}
\begin{align}
    \Pi^{\pm,\, {\hat \pm}}_{\mu,J}\left(u,{\bf p};{\bar u},-{\bf p}\right) &=  \underbrace{\csc\left(\pi\left(u+{\bar u}\right)\right)}_{\text{contact terms}}\text{Disc}_{{\sf s}}\underbrace{\left[\Pi^{\pm,\, {\hat \pm}}_{\mu,J}\left(u,{\bf p};{\bar u},-{\bf p}\right)\right]}_{\text{on-shell propagator}},\\
    \text{Disc}_{{\sf s}}\left[\Pi^{\pm,\, {\hat \pm}}_{\mu,J}\left(u,{\bf p};{\bar u},-{\bf p}\right)\right]&=
    \underbrace{\left(\alpha^{{\hat \pm}}\, \omega_{\Delta^{+}}\left(u,{\bar u}\right)+\beta^{\pm}\, \omega_{\Delta^{-}}\left(u,{\bar u}\right)\right)}_{\text{boundary conditions}}\\ & \hspace*{4cm}\times \Gamma\left(i\mu\right)\Gamma\left(-i\mu\right) \underbrace{\Omega^{\pm,\, {\hat \pm}}_{\mu,J}\left(u,{\bf p};{\bar u},-{\bf p}\right)}_{\text{harmonic function}}, \nonumber
\end{align}
\end{subequations}
\end{shaded}
\noindent where the coefficients \eqref{coeffsab} ensure that the propagators satisfy the Hadamard condition required by the Bunch Davies vacuum and hence also in the de Sitter case serve to implement boundary conditions.

The above results, first presented in \cite{Sleight:2020obc} (and which build on those of \cite{Sleight:2019mgd,Sleight:2019hfp}), show that in perturbation theory any boundary correlator in dS$_{d+1}$ can be expressed as a linear combination of Witten diagrams for the same process in EAdS$_{d+1}$.\footnote{These results, first obtained in \cite{Sleight:2019mgd,Sleight:2019hfp,Sleight:2020obc}, were very recently presented (in the case of scalar fields) as one of the main results of the work \cite{DiPietro:2021sjt} by V. Gorbenko, S. Komatsu and L. di Pietro. See e.g. their section 3.3, where their main equations (3.16), (3.19), (3.20) and figure 4 are equivalent to (2.15), (2.16) of \cite{Sleight:2020obc}, (2.38) and figure 2 of \cite{Sleight:2019hfp}, which we reviewed in equations \eqref{adsodsbubo}, \eqref{analharmadstods}, \eqref{bubodSinin} and figure \ref{fig::ininbranch}.} For any given diagram in dS, one simply replaces each dS propagator with their expression in terms of EAdS propagators reviewed above. In \cite{Sleight:2020obc} this was applied to write tree-level exchanges in dS as a linear combination of exchange Witten diagrams in EAdS. In practice, this approach can get quite cumbersome since one must sum together various contributions coming from each branch of the in-in contour, which can obscure the properties of the final result. However, given the knowledge that such a decomposition of dS diagrams exists, the precise linear combination of EAdS Witten diagrams can be fixed more directly by the Bunch-Davies initial condition and consistent on-shell factorisation of the dS diagram. This will be explained in section \ref{sec::generalalg}, where we provide a set of rules to immediately write down the decomposition of any given dS diagram as a linear combination of EAdS Witten diagrams.

\subsection{The Mellin-Barnes representation: From the bulk to the boundary}

\label{subsec::feynrules}

In the previous sections we introduced a Mellin-Barnes representation for propagators in (EA)dS$_{d+1}$, which is defined as the Mellin transform with respect to the bulk radial coordinate. In this section we will discuss how in the presence of a scale symmetry such a representation could be regarded as analogous Fourier space when we have a translation symmetry. We shall also provide a set of Feynman rules for the Mellin-Barnes representation, analogous to momentum space Feynman rules in flat space.

If we have a function $f\left(z\right)$, in our conventions its Mellin transform $f\left(s\right)$ is defined as\footnote{And likewise for the dS radial coordinate $\eta$, taking care that it takes negative values $\eta \in \left(-\infty,0\right]$.}${}^{,}$\footnote{Note that the factor of 2 in the definition of the Mellin transform ensures that applying \eqref{MT1} and \eqref{MT2} successively gives the identity in our conventions.} 
\begin{subequations}\label{MT}
\begin{align}\label{MT1}
    f\left(z\right) &= \int^{+i\infty}_{-i\infty} \frac{ds}{2\pi i}\,2\,f\left(s\right)z^{-\left(2s-\frac{d}{2}\right)},\\ \label{MT2}
    f\left(s\right)&=\int^\infty_0 \frac{dz}{z}\,f\left(z\right)z^{2s-\frac{d}{2}}.
\end{align}
\end{subequations}
In particular, by transforming to the Mellin-Barnes representation $f\left(s\right)$ we are replacing the bulk radial coordinate $z$ with a Mellin variable $s$, similar to how we replace the position vector ${\bf x}$ with the momentum vector ${\bf k}$ in going to Fourier space,
\begin{subequations}
\begin{align}
    f\left({\bf x}\right) &= \int \frac{d{\bf k}}{\left(2\pi \right)^d}\,f\left({\bf k}\right)e^{+ i {\bf k} \cdot {\bf x}},\\
    f\left({\bf k}\right)&=\int d {\bf x}\,f\left({\bf x}\right)e^{- i {\bf k} \cdot {\bf x}}.
\end{align}
\end{subequations}
The power law $z^{\mp\left(2s-\frac{d}{2}\right)}$ in \eqref{MT} plays a role analogous to the exponential plane waves $e^{\pm i {\bf k} \cdot {\bf x}}$. In fact, for theories with scale symmetry, i.e. with isometry transformation
\begin{equation}
    z \: \to \: \lambda z, \qquad  {\bf x} \: \to \: \lambda {\bf x},
\end{equation}
the Mellin transform plays an analogous role to Fourier space for theories with translation invariance, which is owing to the scale invariance of the power-law $z^{\pm \left(2s-\frac{d}{2}\right)}$. In particular given a function $f(z): (0,\infty)\to\mathbb{R}$, one can consider the following representation of the dilatation group:
\begin{align}\label{dilrep}
    \mathcal{T}_\lambda [f(z)]=\lambda^{-\tfrac{d}{2}}f(\lambda z)\,,
\end{align}
which preserves the inner product:
\begin{align}
    \left\langle f|g\right\rangle=\int_0^\infty \frac{dz}{z^{d+1}}\,f(z) g^\star(z)\,.
\end{align}
The above inner product allows to define a norm for functions defined on $\mathbb{R}^+$. The space of functions equipped with this norm naturally defines a Hilbert space $L^2(\mathbb{R}^+,\tfrac{dz}{z^{d+1}})$. At this point, much like how the exponential plane waves $e^{\pm i {\bf k} \cdot {\bf x}}$ diagonalise the translation generator, the monomials $z^{\mp \left(2s-\frac{d}{2}\right)}$ diagonalise the generator of dilatations. In particular, for the representation \eqref{dilrep} the dilatation generator is given by the following Hermitean operator:
\begin{align}
    \mathcal{D}_z=i\left(z\partial_z-\tfrac{d}2\right)\,, \qquad  \mathcal{T}_\lambda=e^{\lambda \mathcal{D}_z}.
\end{align}
The Eigenfunctions of the dilatation generator are then easily found to be
\begin{align}
    f_{\alpha}(z)\equiv \langle z|f_\alpha\rangle=z^{-i\alpha+\tfrac{d}{2}}\,,
\end{align}
where we have introduced the position Eigenvectors $|z\rangle$. They are moreover orthonormal:
\begin{align}
    \left\langle f_\alpha|f_\beta\right\rangle=\int_0^\infty\frac{dz}{z} z^{-i(\beta-\alpha)}=\int_{-\infty}^\infty dx\, e^{i(\beta-\alpha)x}=2\pi\delta(\beta-\alpha)\,, 
\end{align}
and satisfy completeness:\footnote{In the second equality we used:
\begin{equation}
    \int_{-\infty}^{+\infty}\frac{d\beta}{2\pi}\,\left(\frac{z_2}{z_1}\right)^{-i\beta} = \int_{-\infty}^{+\infty}\frac{d\beta}{2\pi}\,e^ {i\beta\left( \ln\left(z_1\right)-\ln\left(z_2\right)\right)}=\delta \left( \ln\left(z_1\right)-\ln\left(z_2\right)\right).
\end{equation}}
\begin{align}
    \int_{-\infty}^{+\infty}\frac{d\beta}{2\pi} f_\beta^\star(z_1) f_\beta(z_2)=(z_1 z_2)^{\frac{d}{2}}\delta(\ln(z_1)-\ln(z_2))=z_1^{d+1}\delta(z_1-z_2)\,. \nonumber
\end{align}
The Mellin transform can then be identified with the decomposition of an element of $L^2(\mathbb{R}^+,\frac{dz}{z^{d+1}})$ into the Eigenfunctions $f_{\alpha}$ of the dilatation operator:
\begin{equation}
   \mathcal{M}[g][\alpha]\equiv \left\langle g|f_\alpha\right\rangle\,
   =\int_{0}^\infty\frac{dz}{z^{d+1}}\,g(z) z^{i\alpha+\tfrac{d}{2}}.
\end{equation}
The inverse Mellin transform instead follows from completeness, by multiplying both sides with an Eigenfunction $f_\alpha(\bar{z})$ and integrating in $\alpha$:
\begin{equation}
    \int^{+\infty}_{-\infty}\frac{d\alpha}{2\pi}\,{\bar z}^{-i\alpha+\frac{d}{2}}\mathcal{M}[g][\alpha] = g\left({\bar z}\right).
\end{equation}
One recovers \eqref{MT1} by moving the integration contour in the imaginary direction and redefining $i\alpha=2s$. See e.g. \cite{bertrand:hal-03152634} for further details on this construction of the Mellin transform.

\begin{figure}[t]
    \centering
    \captionsetup{width=0.95\textwidth}
    \includegraphics[width=\textwidth]{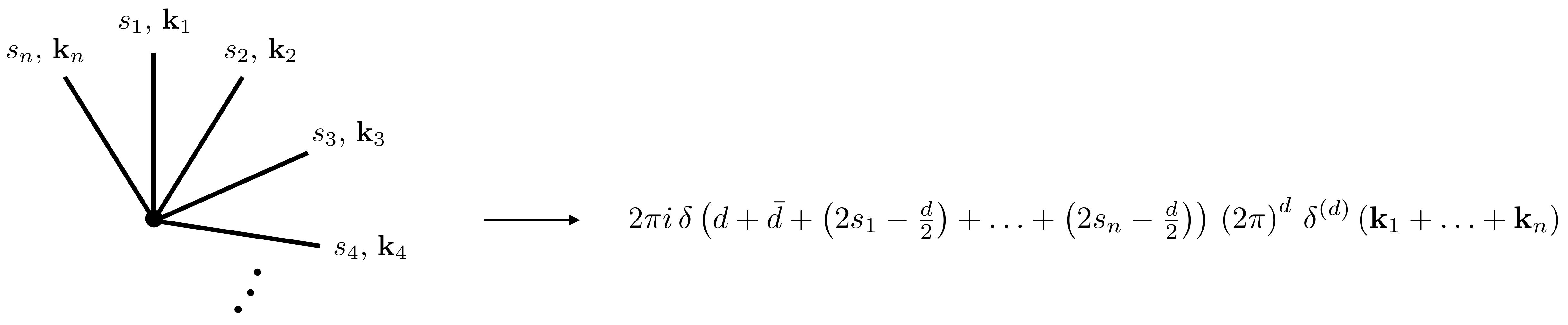}
    \caption{In the Mellin-Barnes representation of diagrams contributing boundary correlators in (EA)dS, in addition to Dirac delta functions enforcing conservation of boundary momenta ${\bf k}_i$ at each vertex, the Dilatation symmetry of (EA)dS requires that there is a Dirac delta function enforcing that the Mellin variables $s_i$ associated to each line add up to a constant. This constant depends on the boundary dimension $d$ and a parameter ${\bar d}$ which is sensitive to spin of the fields meeting at the vertex and the number of derivatives.}
    \label{fig::ddmb}
\end{figure}

By adopting the above Mellin-Barnes representation, the integrals over the bulk radial coordinate trivialise and are given by Dirac delta functions in the Mellin variables, analogous to the momentum conserving delta functions that arise from integrating over position in theories with translation symmetry. To see this it is sufficient to compare the integrals over $z$ and ${\bf x}$ at a bulk vertex with $n$ legs (see figure \ref{fig::ddmb}):
\begin{subequations}
\begin{align}\label{ddmb}
    \int^\infty_0 \frac{dz}{z^{d+{\bar d}+1}}\, z^{-\sum\limits^n_{i=1}\left(2s_i-\frac{d}{2}\right)} &= 2 \pi i\, \delta\left(d+{\bar d }+\sum\limits^n_{i=1}\left(2s_i-\frac{d}{2}\right)\right), \\
   \int d {\bf x}\,f\left({\bf x}\right)e^{- i {\bf x} \cdot \sum\limits^n_{i=1} {\bf k}_i} &= \left(2\pi \right)^d\, \delta^{(d)}\left(\sum\limits^n_{i=1} {\bf k}_i\right),\label{ddfs}
\end{align}
\end{subequations}
where the $i$-th leg is assigned boundary momentum ${\bf k}_i$ and the Mellin variable $s_i$. The symbol ${\bar d}$ parameterises any monomials in $z$ generated by vertices involving tensor fields and/or derivatives, so that for non-derivative interactions for scalar fields we have ${\bar d}=0$.\footnote{For some examples with ${\bar d}\ne0$ see \cite{Sleight:2019hfp,Sleight:2021iix}. In this work we will not need to worry about ${\bar d}$ since we shall take the perspective that diagrams generated by spinning fields and derivative vertices can be obtained by acting with differential ``weight-shifting" operators on a scalar seed diagram generated by non-derivative vertices, where the shifted scaling dimensions of the scalar seed automatically account for ${\bar d}$. See section \ref{subsec::3ptsaddingspin}.} In section \ref{subsec::CWI3pt} we will show explicitly that the Dirac delta functions of the type \eqref{ddmb} in the Mellin variables are required by the Dilatation Ward identities that must be satisfied by boundary correlators, analogous to how translation symmetry implies momentum conservation. Note that, since translation symmetry is broken in the bulk radial direction of EAdS and dS, taking the Fourier transform with respect to the bulk radial coordinate $z$ (or $\eta$ in the case of dS) would not benefit from the convenient properties of Fourier space in the presence of a translation symmetry. For theories with a scale symmetry we are proposing that one can still profit from analogous benefits by instead taking the Mellin transform, which could then be a natural habitat for correlators with a scale symmetry. These and further parallels between the Mellin-Barnes representation in theories with a scale symmetry and Fourier space in theories with a translation symmetry are summarised in the table below.\footnote{Other parallels between momentum space and the Mellin-Barnes representation can be found in \cite{Sleight:2021iix}.} 

\begin{center}
\centering
\begin{tabular}{ | m{18em} | m{18em}|} 
  \hline
 \hfil \emph{Momentum space}  & \hfil \emph{Mellin space}  \\ 
  \hline
 \hfil ${\bf k}$ & \hfil $s$ \\ 
  \hline
\hfil  $e^{\pm i{\bf k}\cdot {\bf x}}$ & \hfil $z^{\mp\left(2s-\tfrac{d}{2}\right)}$  \\ 
\hline
\hfil  $\partial_{{\bf x}}\; \to \; i{\bf k}$ & \hfil $z\partial_z\;\to\;-\left(2s-\frac{d}{2}\right)$\\
\hline
  \hfil  orthonormality & \hfil orthonormality\\ 
  \hfil $\int\,d{\bf x}\, e^{i{\bf x}\cdot {\bf p}}e^{-i{\bf x}\cdot {\bar {\bf p}}}=\left(2\pi\right)^d\delta^{(d)}\left({\bf p}-{\bar {\bf p}}\right)$& \vspace*{0.25cm} \hfil $\int_0^\infty\frac{dz}{z^{d+1}}\,z^{-2s+\tfrac{d}{2}}z^{2{\bar s}+\tfrac{d}{2}}=\pi i\,\delta(s-{\bar s})$\vspace*{0.25cm}\\
  \hline
  \hfil completeness & \hfil completeness \\
    \hfil $\int\,\frac{d{\bf p}}{\left(2\pi\right)^d}\, e^{i{\bf x}\cdot {\bf p}}e^{-i{\bar {\bf x}}\cdot {\bf p} }=\delta^{(d)}\left({\bf x}-{\bar {\bf x}}\right)$& \vspace*{0.25cm} \hfil $2\int_{-i\infty}^{+i\infty}\frac{ds}{2\pi} \,z^{2s+\frac{d}{2}}_1 z^{-2s+\frac{d}{2}}_2=z^{d+1}_1\,\delta(z_1-z_2)$\vspace*{0.25cm}\\
  \hline
  \hfil  translation symmetry & \hfil scale symmetry\\
  \hline
  \hfil  $\left(2\pi \right)^d\, \delta^{(d)}\left(\sum\limits^n_{i=1} {\bf k}_i\right)$ &  \hfil $2 \pi i\, \delta\left(d+{\bar d}+\sum\limits^n_{i=1}\left(2s_i-\frac{d}{2}\right)\right)$\\
  \hline
\end{tabular}
\end{center}

The above discussion naturally suggests a set of Feynman rules for boundary correlators in (EA)dS$_{d+1}$, which we summarise in the following:

\paragraph{Feynman rules for the Mellin-Barnes representation:}
\begin{enumerate}
    \item For a given diagram, for each external leg assign a momentum ${\bf k}_i$ and an external Mellin variable $s_i$. For each internal leg assign a momentum ${\bf p}_j$ and a pair of internal Mellin variables $u_j$, ${\bar u}_j$ distributed according to the split representation \eqref{spliteadsds} of internal legs shown in figure \ref{fig::harmfact}. 
    \item For each vertex include a factor of the Dirac delta function \eqref{ddmb} in the Mellin variables associated to the legs of the vertex and a factor of the momentum conserving Dirac delta function \eqref{ddfs}, as in figure \ref{fig::ddmb}.
    \item For each vertex, multiply by the coupling constant and add the appropriate (polynomial) factors in the momenta (corresponding to spatial derivatives) and the Mellin variables (corresponding to derivatives in the bulk radial coordinate) as described in the table above. Divide by the symmetry factor.
    \item For EAdS diagrams: For each external leg multiply by the corresponding bulk-to-boundary propagator $K_{\delta_i,J_i}\left(s_i,{\bf k}_i\right)$. For each internal leg multiply by the corresponding bulk-to-bulk propagator $\Pi^{\text{AdS}}_{\alpha \Delta^+_j+\beta \Delta^-_j, J_j}\left(u_j,{\bf p}_j;{\bar u}_j,-{\bf p}_j\right)$. For each vertex, multiply by a factor of $\left(-1\right)$ (since the action is Euclidean).
    \item For dS diagrams: For each external leg attached to a vertex on the $\pm$ branch of the in-in contour, multiply by the corresponding bulk-to-boundary propagator $K^\pm_{\delta_i,J_i}\left(s_i,{\bf k}_i\right)$. For each internal leg connecting a vertex on the $\pm$ branch and a vertex on the ${\hat \pm}$ branch, multiply by the corresponding bulk-to-bulk propagator $\Pi^{\pm, {\hat \pm}}_{\mu_j, J_j}\left(u_j,{\bf p}_j;{\bar u}_j,-{\bf p}_j\right)$. For each vertex on the $\pm$ branch of the in-in contour, multiply by a factor $\pm i $. Sum over all branches of the in-in contour.
\end{enumerate}

The above Feynman rules define a Mellin-Barnes representation for any given diagram in (EA)dS with $n$ external legs ${\bf k}_i$ and $m$ internal legs ${\bf p}_j$:\footnote{Note that the Mellin-Barnes representation can be defined as a distribution in the appropriate functional space much like the Fourier transform \cite{bertrand:hal-03152634} usually used in the context of QFT. In fact it is possible to define the Mellin transform for all distributions in $\mathcal{D}_+^\prime$.}
\begin{multline}
   F\left({\bf k}_1,\ldots, {\bf k}_n; {\bf p}_1, \ldots, {\bf p}_m \right) = \int^{+i\infty}_{-i\infty}  \underbrace{\frac{ds_1}{2\pi i}\,\ldots\,\frac{ds_n}{2\pi i}}_{\left[ds_i\right]_n}\, \underbrace{\frac{du_1d{\bar u}_1}{\left(2\pi i\right)^2}\,\ldots\, \frac{du_md{\bar u}_m}{\left(2\pi i\right)^2}}_{\left[du_j\,d{\bar u}_j\right]_m}\,\\
   \times F\left(s_1,{\bf k}_1,\ldots, s_n, {\bf k}_n; u_1, {\bar u}_1,{\bf p}_1, \ldots, u_m, {\bar u}_m, {\bf p}_m \right).
\end{multline}
It is important to appreciate that by adopting the Mellin-Barnes representation the complicated integrals over the bulk radial coordinate are taken care of automatically: They are replaced by Dirac delta functions of the type \eqref{ddmb} in the Mellin variables. This is analogous to the fate of position space integrals when transforming to Fourier space in directions where we have a translation symmetry (like we do have on the boundary), where they get replaced by momentum conserving delta functions. Much like the Fourier space representation of flat space scattering amplitudes, whose properties we can infer from those of Feynman propagators in momentum space (e.g. simple poles in the Mandelstam variables for exchanges at tree level, cutting rules, dispersion relations...), we might then try to use the Mellin-Barnes representation to infer properties of boundary correlators in (EA)dS$_{d+1}$ in a similar fashion -- importing them from the bulk to the boundary. Along these lines, in the following section we will start by using the properties of (EA)dS bulk-to-bulk propagators in the Mellin-Barnes representation to derive cutting rules and dispersion relations to compute boundary correlators in perturbation theory. In section \ref{sec::3pt} we will also see how the Mellin-Barnes representation provides a framework to determine how consistent dS physics is imprinted in the coefficients of boundary contact diagrams, which are the basic building blocks from which other types of diagrams can be constructed.

\subsection{Cutting rules and dispersion}
\label{subsec::cuttinganddisp}

The Mellin-Barnes representation \eqref{AdSbubuMB} and \eqref{bubodSinin} of (EA)dS propagators naturally gives rise to cutting rules and dispersion formulas to compute boundary correlators in (EA)dS. This was explained in \cite{Sleight:2020obc}, where it was applied to compute tree-level exchanges in (EA)dS and below we give a more pedagogical presentation of the general procedure. We also try to make contact with the more recent flurry of activity \cite{Goodhew:2020hob,Jazayeri:2021fvk,Melville:2021lst,Goodhew:2021oqg,Baumann:2021fxj,Meltzer:2021zin} on dS unitarity methods (and related AdS results \cite{Meltzer:2020qbr,Meltzer:2021bmb}), which instead work at the level of the wavefunction. 

\begin{figure}[t]
    \centering
    \captionsetup{width=0.95\textwidth}
    \includegraphics[width=\textwidth]{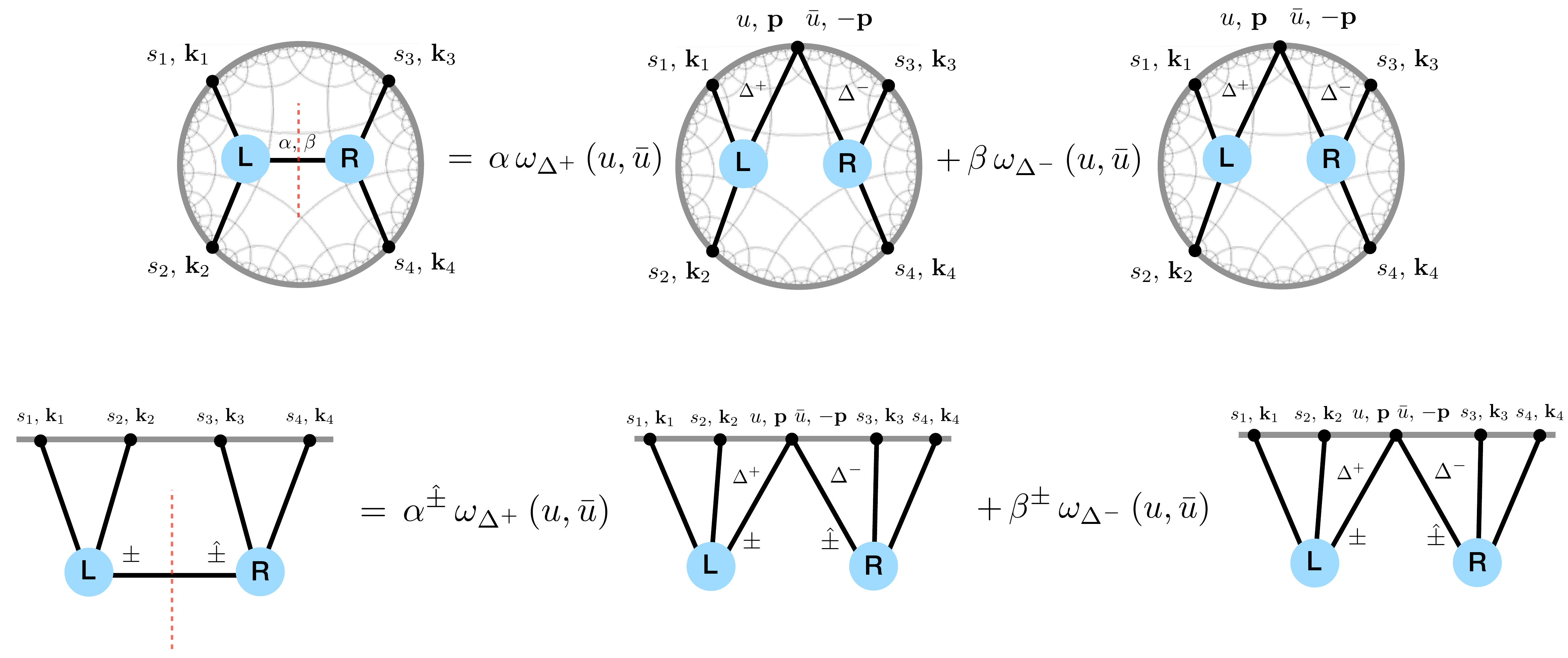}
    \caption{Upon taking the discontinuity \eqref{discs} of an internal line, diagrams contributing to boundary correlators in (EA)dS decompose into a linear combination factorised contributions, one for the exchanged $\Delta^+$ mode and the other for the exchanged $\Delta^-$ mode, which are each given by a product of the two on-shell sub-diagrams to the left and the right of the cut as shown above. In the dS case this factorisation is at the level of each in-in contribution. In the examples below this illustrated for four-point tree-level exchange diagrams. In section \ref{sec::generalalg} we give some examples of loop diagrams.}
    \label{fig::exchdiscfact}
\end{figure}

At the basis of the cutting procedure is the split representation \eqref{spliteadsds} of the harmonic functions in (EA)dS, which factorise into a product of bulk-to-boundary propagators. This in turn implies the factorisation \eqref{disceads} of the bulk-to-bulk propagator \eqref{MBrepbubuads} for the $\Delta^\pm$ mode upon taking its discontinuity or ``cut" (as defined in equation \eqref{discs}). The discontinuity of a bulk-to-bulk propagator in EAdS for a generic linear combination of $\Delta^\pm$ boundary conditions, as well as that of the in-in dS bulk-to-bulk propagators in the Bunch-Davies vacuum, is therefore a linear combination of factorised contributions -- one for the propagating $\Delta^+$ mode and the other for the propagating $\Delta^-$ mode (see figure \ref{fig::exchdiscfact}):
\begin{subequations}\label{factproperty}
\begin{align}
   & \text{Disc}_{{\sf s}}\left[\Pi^{\text{AdS}}_{\alpha \Delta^++\beta \Delta^-,J}(u,{\bf p};{\bar u},-{\bf p})\right] \\& \hspace*{3.5cm}= \alpha\, \text{Disc}_{{\sf s}}\left[\Pi^{\text{AdS}}_{\Delta^+,J}(u,{\bf p};{\bar u},-{\bf p})\right]+\beta\, \text{Disc}_{{\sf s}}\left[\Pi^{\text{AdS}}_{\Delta^-,J}(u,{\bf p};{\bar u},-{\bf p})\right],  \nonumber \\
   & \text{Disc}_{{\sf s}}\left[\Pi^{\pm,\, {\hat \pm}}_{\mu,J}(u,{\bf p};{\bar u},-{\bf p})\right] \\ & \hspace*{3.5cm} = \alpha^{{\hat \pm}}\, \text{Disc}_{{\sf s}}\left[\Pi^{\pm,\, {\hat \pm}}_{\Delta^+,J}(u,{\bf p};{\bar u},-{\bf p})\right]+\beta^{\pm}\, \text{Disc}_{{\sf s}}\left[\Pi^{\pm,\, {\hat \pm}}_{\Delta^-,J}(u,{\bf p};{\bar u},-{\bf p})\right], \nonumber
\end{align}
\end{subequations}
where we defined
\begin{subequations}
\begin{align}
    \text{Disc}_{{\sf s}}\left[\Pi^{\pm,\, {\hat \pm}}_{\Delta^+,J}(u,{\bf p};{\bar u},-{\bf p})\right] &= 
     \omega_{\Delta^{+}}\left(u,{\bar u}\right) \Gamma\left(i\mu\right)\Gamma\left(-i\mu\right) \Omega^{\pm,\, {\hat \pm}}_{\mu,J}\left(u,{\bf p};{\bar u},-{\bf p}\right),\\
     \text{Disc}_{{\sf s}}\left[\Pi^{\pm,\, {\hat \pm}}_{\Delta^-,J}(u,{\bf p};{\bar u},-{\bf p})\right] &= 
     \omega_{\Delta^{-}}\left(u,{\bar u}\right) \Gamma\left(i\mu\right)\Gamma\left(-i\mu\right) \Omega^{\pm,\, {\hat \pm}}_{\mu,J}\left(u,{\bf p};{\bar u},-{\bf p}\right).
\end{align}
\end{subequations}
These identities, in turn, imply that diagrams for boundary correlators in (EA)dS reduce to a linear combination of factorised contributions upon putting an internal leg on-shell, where the precise linear combination is dictated by the boundary condition on the exchanged field. See figure \ref{fig::exchdiscfact}. 

To illustrate, let us consider the four-point tree level exchange, restricting for ease of presentation to scalar fields $\phi_i$, $\phi$, $i=1,2,3,4$, interacting though non-derivative cubic vertices $g_{12\phi}\, \phi_1 \phi_2 \phi$ and $g_{34\phi}\, \phi_3 \phi_4 \phi$. In EAdS$_{d+1}$, applying the Feynman rules of section \ref{subsec::feynrules} we have
\begin{multline}\label{adsexch}
     {\cal A}^{\text{EAdS}}_{\alpha \Delta^++\beta \Delta^-,J}\left(s_1,{\bf k}_1,s_2,{\bf k}_2;s_3,{\bf k}_3,s_4,{\bf k}_4\right)\\ = g_{12\phi}\, 2\pi i\, \delta\left(-\tfrac{d}{2}+2s_1+2s_2+2u\right)\, \left(2\pi\right)^{d}\delta^{(d)}\left({\bf k}_1+{\bf k}_2+{\bf p}\right)\\ \times \,g_{34\phi}\,2\pi i\, \delta\left(-\tfrac{d}{2}+2{\bar u}+2s_3+2s_4\right)\left(2\pi\right)^{d}\delta^{(d)}\left({\bf k}_3+{\bf k}_4-{\bf p}\right)\\ \hspace*{-1cm}\times K^{\text{AdS}}_{\Delta_1,0}\left(s_1,{\bf k}_1\right)\,K^{\text{AdS}}_{\Delta_2,0}\left(s_2,{\bf k}_2\right)\Pi^{\text{AdS}}_{\alpha \Delta^++\beta \Delta^-,0}(u,{\bf p};{\bar u},-{\bf p})K^{\text{AdS}}_{\Delta_3,0}\left(s_3,{\bf k}_3\right)\,K^{\text{AdS}}_{\Delta_4,0}\left(s_4,{\bf k}_4\right).
\end{multline}
For the corresponding exchange in the Bunch-Davies vacuum of dS$_{d+1}$, the contribution from the $\pm\, {\hat \pm}$ branch of the in-in contour reads
\begin{multline}\label{pmdsexch}
     {\cal A}^{\pm,\, {\hat \pm}}_{\mu,J}\left(s_1,{\bf k}_1,s_2,{\bf k}_2;s_3,{\bf k}_3,s_4,{\bf k}_4\right)\\ = g_{12\phi}\, 2\pi i\, \delta\left(-\tfrac{d}{2}+2s_1+2s_2+2u\right)\, \left(2\pi\right)^{d}\delta^{(d)}\left({\bf k}_1+{\bf k}_2+{\bf p}\right)\\ \times \,g_{34\phi}\,2\pi i\, \delta\left(-\tfrac{d}{2}+2{\bar u}+2s_3+2s_4\right)\left(2\pi\right)^{d}\delta^{(d)}\left({\bf k}_3+{\bf k}_4-{\bf p}\right)\\ \times K^{\pm}_{\Delta_1,0}\left(s_1,{\bf k}_1\right)\,K^{\pm}_{\Delta_2,0}\left(s_2,{\bf k}_2\right)\Pi^{\pm,\, {\hat \pm}}_{\mu,0}(u,{\bf p};{\bar u},-{\bf p})K^{{\hat \pm}}_{\Delta_3,0}\left(s_3,{\bf k}_3\right)\,K^{{\hat \pm}}_{\Delta_4,0}\left(s_4,{\bf k}_4\right).
\end{multline}
From the factorisation property \eqref{factproperty} of bulk-to-bulk propagators in (EA)dS, we see that the discontinuity of the EAdS exchange \eqref{adsexch} and the $\pm\, {\hat \pm}$ branch contribution \eqref{pmdsexch} to the dS exchange is a linear combination of factorised contributions, one for the propagation of the $\Delta^+$ mode and the other for the propagation of the $\Delta^-$ mode, given by the product of three-point boundary correlators generated by the cubic vertices that mediate the exchange:
\begin{multline}\label{discadsexch0}
    \hspace*{-1cm}  \text{Disc}_{{\sf s}}\left[{\cal A}^{\text{EAdS}}_{\alpha \Delta^++\beta \Delta^-,0}\left(s_1,{\bf k}_1,s_2,{\bf k}_2;s_3,{\bf k}_3,s_4,{\bf k}_4\right)\right] = \frac{\Gamma\left(1+i\mu\right)\Gamma\left(1-i\mu\right)}{\pi}
   \left(\alpha\, \omega_{\Delta^+}\left(u,{\bar u}\right)+\beta\, \omega_{\Delta^-}\left(u,{\bar u}\right)\right)\\ \times F^{\text{EAdS}}_{\Delta_1,\Delta_2,\Delta^+}\left(s_1,{\bf k}_1,s_2,{\bf k}_2,u,{\bf p}\right)F^{\text{EAdS}}_{\Delta^-,\Delta_3,\Delta_4}\left({\bar u},-{\bf p},s_3,{\bf k}_3,s_4,{\bf k}_4\right),
\end{multline}
and
\begin{multline}\label{discdsexch0}
     \hspace*{-1.5cm}  \text{Disc}_{{\sf s}}\left[{\cal A}^{\pm,\, {\hat \pm}}_{\mu,0}\left(s_1,{\bf k}_1,s_2,{\bf k}_2;s_3,{\bf k}_3,s_4,{\bf k}_4\right)\right] = \left(\pm i\right)\left({\hat \pm} i\right) \frac{\Gamma\left(1+i\mu\right)\Gamma\left(1-i\mu\right)}{\pi}\left(\alpha^{{\hat \pm}}\, \omega_{\Delta^+}\left(u,{\bar u}\right)+\beta^{\pm}\, \omega_{\Delta^-}\left(u,{\bar u}\right)\right)\\ \times F^{\pm}_{\Delta_1,\Delta_2,\Delta^+}\left(s_1,{\bf k}_1,s_2,{\bf k}_2,u,{\bf p}\right)F^{{\hat \pm}}_{\Delta^-,\Delta_3,\Delta_4}\left({\bar u},-{\bf p},s_3,{\bf k}_3,s_4,{\bf k}_4\right),
\end{multline}
where
\begin{multline}
    F^{\bullet}_{\Delta_1,\Delta_2,\Delta}\left(s_1,{\bf k}_1,s_2,{\bf k}_2,u,{\bf p}\right) = - g_{12 \phi}\, 2\pi i\, \delta\left(-\tfrac{d}{2}+2s_1+2s_2+2u\right)\, \left(2\pi\right)^{d}\delta^{(d)}\left({\bf k}_1+{\bf k}_2+{\bf p}\right) \\ \times K^{\bullet}_{\Delta_1,0}\left(s_1,{\bf k}_1\right) K^{\bullet}_{\Delta_2,0}\left(s_2,{\bf k}_2\right)K^{\bullet}_{\Delta,0}\left(u,{\bf p}\right),
\end{multline}
is the Mellin-Barnes representation of the constituent three-point boundary correlators, where for the EAdS diagram we have $\bullet = \text{AdS}$ and for the $\pm\,{\hat \pm}$ contribution to the dS diagram we have $\bullet = \pm\,{\hat \pm}$. In the dS case one should not not forget to sum over all branches of the in-in contour to obtain the full on-shell exchange:
\begin{multline}
   \text{Disc}_{{\sf s}}\left[{\cal A}^{\text{dS}}_{\mu,0}\left(s_1,{\bf k}_1,s_2,{\bf k}_2;s_3,{\bf k}_3,s_4,{\bf k}_4\right)\right]\\=\sum\limits_{\pm\,{\hat \pm}}\left(\pm i\right)\left({\hat \pm} i\right)\text{Disc}_{{\sf s}}\left[{\cal A}^{\pm,\, {\hat \pm}}_{\mu,0}\left(s_1,{\bf k}_1,s_2,{\bf k}_2;s_3,{\bf k}_3,s_4,{\bf k}_4\right)\right].
\end{multline}
Let us note that the discontinuity of the wavefunction coefficient as considered in \cite{Melville:2021lst,Goodhew:2021oqg,Baumann:2021fxj,Meltzer:2021zin}, in contrast to the in-in correlators above, would yield a single factorised contribution given by a product of three-point wavefunction coefficients -- as opposed to the linear combination of factorised contributions that we observe above for the corresponding in-in correlator. This is because bulk-to-bulk propagators for wavefunction coefficients, as opposed to in-in bulk-to-bulk propagators in the Bunch-Davies vacuum, only propagate a single $\Delta^\pm$ mode. 

This cutting procedure naturally extends to the internal lines of any type of diagram contributing to boundary correlators in (EA)dS, which factorise according to the split representation \eqref{spliteadsds} of the harmonic function associated to that internal line and the boundary condition on the exchanged particle -- which is implemented by the appropriate linear combination of projectors $\omega_{\Delta^\pm}\left(u,{\bar u}\right)$. What we learn is that the cut of any internal line in a given diagram is fixed by factorisation (in the sense \eqref{spliteadsds} of harmonic functions) and boundary conditions. While we have derived these properties of perturbative boundary correlators in (EA)dS from the properties of the corresponding (EA)dS bulk-to-bulk propagators, in section \ref{sec::4ptexch} we demonstrate how they can be obtained from a boundary perspective as a consequence of \emph{factorisation}, \emph{conformal symmetry} and \emph{boundary conditions}.

 \begin{figure}[t]
    \centering
    \captionsetup{width=0.95\textwidth}
    \includegraphics[width=.75\textwidth]{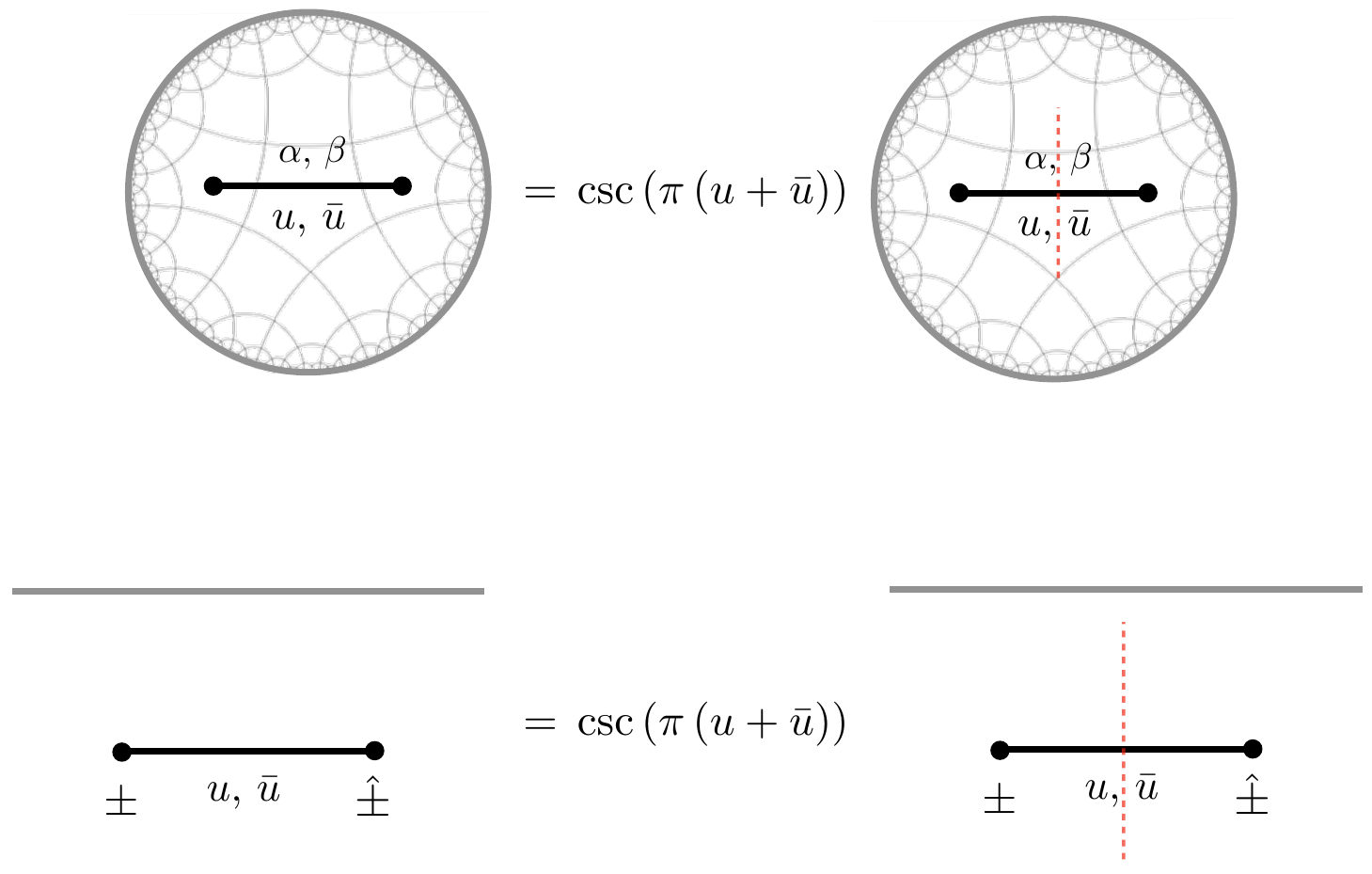}
    \caption{At the level of the Mellin-Barnes representation the full bulk-to-bulk propagator in (EA)dS can be reconstructed from its discontinuity \eqref{discs} by multiplying with the factor $\csc\left(\pi\left(u+{\bar u}\right)\right)$.}
    \label{fig::cutprop}
\end{figure}

The full boundary correlator can then be reconstructed using the dispersion formula \eqref{dispprop} for the bulk-to-bulk propagators. For a given diagram, at the level of the Mellin-Barnes representation for each internal leg that has been placed on shell this amounts to simply multiplying by the cosecant factor $\csc\left(\pi\left(u_i+{\bar u}_i\right)\right)$  where $\left(u_i,{\bar u_i}\right)$ are the pair of internal Mellin variables which describe the on-shell leg under consideration. For example, for exchange diagrams we have
\begin{subequations}\label{dispersionadsdsexch}
\begin{align}
&{\cal A}^{\text{EAdS}}_{\alpha \Delta^++\beta \Delta^-,J}\left(s_1,{\bf k}_1,s_2,{\bf k}_2;s_3,{\bf k}_3,s_4,{\bf k}_4\right)\\ &\hspace*{4cm}= \csc\left(\pi\left(u+{\bar u}\right)\right)\text{Disc}_{{\sf s}}\left[{\cal A}^{\text{EAdS}}_{\alpha \Delta^++\beta \Delta^-,J}\left(s_1,{\bf k}_1,s_2,{\bf k}_2;s_3,{\bf k}_3,s_4,{\bf k}_4\right)\right],\nonumber\\
 & {\cal A}^{\text{dS}}_{\mu,J}\left(s_1,{\bf k}_1,s_2,{\bf k}_2;s_3,{\bf k}_3,s_4,{\bf k}_4\right)\\ &\hspace*{4cm}=\csc\left(\pi\left(u+{\bar u}\right)\right)\text{Disc}_{{\sf s}}\left[{\cal A}^{\text{dS}}_{\mu,J}\left(s_1,{\bf k}_1,s_2,{\bf k}_2;s_3,{\bf k}_3,s_4,{\bf k}_4\right)\right]. \nonumber
\end{align}
\end{subequations}
This extends naturally to any diagram contributing to boundary correlators in (EA)dS, also beyond tree level. Some examples are discussed in section \ref{sec::generalalg}, see e.g. equation \eqref{fig::candydisp} for the four-point one-loop candy diagram.

It is instructive to note that the dispersion formula of \cite{Sleight:2019hfp,Sleight:2020obc} reviewed above and in section \ref{subsec::props(EA)dS} is related to dispersion formulas that were presented recently e.g. in \cite{Meltzer:2021zin}. In particular, in Fourier space the (EA)dS bulk-to-bulk propagator satisfies the following dispersion formula (see e.g. \cite{Meltzer:2021zin} equation (2.28)):
\begin{equation}
  \Pi(z,{\bf p};{\bar z},-{\bf p}) = \frac{1}{2\pi i} \int^\infty_0\,\frac{dk^2}{k^2+p^2+i\epsilon}\text{Disc}_{k^2}\left[\Pi(z,{\bf k};{\bar z},-{\bf k})\right].
\end{equation}
To relate this to the dispersion formula of \cite{Sleight:2019hfp,Sleight:2020obc} given in equation \eqref{dispprop} one simply transforms to the Mellin-Barnes representation:
\begin{equation}
  \Pi(u,{\bf p};{\bar u},-{\bf p}) = \frac{1}{2\pi i} \int^\infty_0\,\frac{dk^2}{k^2+p^2+i\epsilon}\text{Disc}_{k^2}\left[\Pi(u,{\bf k};{\bar u},-{\bf k})\right],
\end{equation}
where, if the strip of analyticity is $\mathfrak{Re}(u+{\bar u}
)\in(0,1)$, one can perform the $k^2$ integral explicitly in terms of a Hypergeometric function which then reduces to:
\begin{align}
    \int_0^\infty dk^2\frac{k^{-2(u+{\bar u})}}{k^2+p^2+i\epsilon}=\left(p^2\right)^{-(u+{\bar u})}\,\pi\csc(\pi(u+{\bar u}))\,,
\end{align}
up to contact terms, which in the Mellin-Barnes representation are encoded in the choice of integration contour (see e.g. appendix C.3 of \cite{Sleight:2019hfp}). This recovers the Mellin-Barnes dispersion formula \eqref{dispprop}.\footnote{Note that our formula is consistent with the definition \eqref{discs} which places the discontinuity on the negative real axis. If the discontinuity is on the positive real axis one can simply send $p^2\to-p^2$.}

\section{Contact diagrams}\label{sec::3pt}

In this section we consider contact diagram contributions to boundary correlators in (EA)dS$_{d+1}$, which are the basic building blocks from which other types of diagrams may be constructed via the cutting rules outlined in section \ref{subsec::cuttinganddisp}. In section \ref{subsec::CWI3pt} we study how the constraints from conformal Ward identities are implemented in the Mellin-Barnes representation focusing on three-point boundary correlators of scalar fields, where it is well known that conformal symmetry constrains their functional form up to a constant \cite{Polyakov:1970xd,Osborn:1993cr}. In sections \ref{subsec::3ptsceads} and \ref{subsec::npt} we review the relations \cite{Sleight:2019mgd,Sleight:2019hfp} between $n$-point contact diagrams of scalar fields in EAdS and dS, derived using the Mellin-Barnes representation. In section \ref{subsec::3ptsaddingspin} we extend these results to contact diagrams generated by any cubic coupling of (integer) spinning fields. In section \ref{subsec::contactunitarity} we discuss the constraints imposed on the coefficients of contact diagrams by unitarity time evolution in de Sitter.

\subsection{Solving three-point Conformal Ward Identities \`a la Mellin-Barnes}
\label{subsec::CWI3pt}

Correlators of quantum fields in (EA)dS must be invariant under the corresponding isometries of (EA)dS. The corresponding generators act as the conformal group on the boundary of (EA)dS which, in addition to the usual translation and rotations, include the generators of dilatations and special conformal transformations which in momentum space read:
\begin{subequations}
\begin{align}
    {\cal D}&= -\left(\Delta-d\right)+k \partial_k, \\
    {\cal K}^i &= 2\left(\Delta-d\right) \partial_{k_i}-2k^j\partial_{k^j} \partial_{k_i}+k^i\partial_{k^j}\partial_{k_j}.\label{SCgen}
\end{align}
\end{subequations}
It is well known that conformal symmetry constrains three-point functions of scalar operators up to coefficient \cite{Polyakov:1970xd,Osborn:1993cr}. In particular, in momentum space, three-point functions of scalar operators with generic scaling dimensions $\Delta_i$ are constrained by the Conformal Ward Identities to be given by Appell's function $F_4$ \cite{Antoniadis:2011ib,Coriano:2013jba,Bzowski:2013sza}. In the following we will re-derive this result by solving the Conformal Ward Identities directly at the level of the Mellin-Barnes representation.

In the usual way translation invariance implies that the three-point function is proportional to a momentum-conserving delta function 
\begin{equation}\label{3ptscalar}
    F_{\Delta_1\,\Delta_2\,\Delta_3}\left({\bf k}_1,{\bf k}_2,{\bf k}_3\right) = \left(2\pi\right)^d \delta^{\left(d\right)}\left({\bf k}_1+{\bf k}_2+{\bf k}_3\right)  F^\prime_{\Delta_1\,\Delta_2\,\Delta_3}\left({\bf k}_1,{\bf k}_2,{\bf k}_3\right).
\end{equation}
For scalar correlators, rotational invariance requires that $F^\prime$ is a function of the magnitudes $k_j=|{\bf k}_j|$,
\begin{equation}
    F^\prime_{\Delta_1\,\Delta_2\,\Delta_3}\left({\bf k}_1,{\bf k}_2,{\bf k}_3\right) = F^\prime_{\Delta_1\,\Delta_2\,\Delta_3}\left(k_1,k_2,k_3\right).
\end{equation}
It is instructive to study the constraints from the Dilatation and Special Conformal Ward identities employing a Mellin-Barnes representation, defined as 
\begin{equation}\label{invMB3pt}
    F^\prime_{\Delta_1\,\Delta_2\,\Delta_3}\left(k_1,k_2,k_3\right) = \int^{+i\infty}_{-i\infty}\,\left[ds_j\right]_3\,F_{\Delta_1\,\Delta_2\,\Delta_3}\left(s_1,k_1,s_2,k_2,s_3,k_3\right),
\end{equation}
where we can write
\begin{equation}
F_{\Delta_1\,\Delta_2\,\Delta_3}\left(s_1,k_1,s_2,k_2,s_3,k_3\right)=F_{\Delta_1\,\Delta_2\,\Delta_3}\left(s_1,s_2,s_3\right)\prod^3_{j=1}\left(\frac{k_j}{2}\right)^{-2s_j+\Delta_j-\tfrac{d}{2}},
\end{equation}
with Mellin variables $s_1$, $s_2$, $s_3$. The function $F_{\Delta_1\,\Delta_2\,\Delta_3}\left(s_1,s_2,s_3\right)$ is the Mellin transform of $F^\prime_{\Delta_1\,\Delta_2\,\Delta_3}\left(k_1,k_2,k_3\right)$ with respect to the $k_j$. The Dilatation Ward identity imposes 
\begin{equation}
    0 = \left(-d+\sum^3\limits_{j=1}{\cal D}_j\right)F^\prime_{\Delta_1\,\Delta_2\,\Delta_3}\left(k_1,k_2,k_3\right),
\end{equation}
which at the level of the Mellin-Barnes representation translates into
\begin{equation}
    0 = \int^{+i\infty}_{-i\infty}\,\left[ds_j\right]_3\, \left(\tfrac{d}{2}-2\left(s_1+s_2+s_3\right)\right)F_{\Delta_1\,\Delta_2\,\Delta_3}\left(s_1,s_2,s_3\right)\,\prod^3_{j=1}\left(\frac{k_j}{2}\right)^{-2s_j+\Delta_j-\tfrac{d}{2}}.
\end{equation}
This implies the following linear constraint on the Mellin variables $s_j$:
\begin{equation}\label{constr}
    s_1+s_2+s_3=\frac{d}{4}.
\end{equation}
This is analogous to momentum conservation imposed by translation invariance. Scale invariance therefore requires
\begin{equation}
  \label{MB3pt} F_{\Delta_1\,\Delta_2\,\Delta_3}\left(s_1,s_2,s_3\right) = 2\pi i\, \delta\left(\tfrac{d}{4}-s_1-s_2-s_3\right) F^\prime_{\Delta_1\,\Delta_2\,\Delta_3}\left(s_1,s_2,s_3\right),
\end{equation}
for some function $F^\prime_{\Delta_1\,\Delta_2\,\Delta_3}\left(s_1,s_2,s_3\right)$ and is the analogue of equation \eqref{3ptscalar} in the case of translation symmetry. 

The poles of the function  $F^\prime_{\Delta_1\,\Delta_2\,\Delta_3}\left(s_1,s_2,s_3\right)$ are fixed by the Ward identity associated to Special Conformal Transformations, which in momentum space reads
\begin{equation}
    0=\left(\sum\limits^3_{a=1}{\cal K}^i_a\right)\, F^\prime_{\Delta_1\,\Delta_2\,\Delta_3}\left(k_1,k_2,k_3\right).
\end{equation}
Following \cite{Bzowski:2013sza}, by taking ${\bf k}_1$ and ${\bf k}_2$ to be independent momenta this reduces to two independent scalar equations
\begin{subequations}
\begin{align}
    0&=\left[\left(\frac{\partial^2}{\partial k^2_1}+\frac{d+1-2\Delta_1}{k_1}\frac{\partial}{\partial k_1}\right)-\left(\frac{\partial^2}{\partial k^2_3}+\frac{d+1-2\Delta_3}{k_3}\frac{\partial}{\partial k_3}\right)\right]F^\prime_{\Delta_1\,\Delta_2\,\Delta_3}\left(k_1,k_2,k_3\right),\\
    0&=\left[\left(\frac{\partial^2}{\partial k^2_2}+\frac{d+1-2\Delta_2}{k_2}\frac{\partial}{\partial k_2}\right)-\left(\frac{\partial^2}{\partial k^2_3}+\frac{d+1-2\Delta_3}{k_3}\frac{\partial}{\partial k_3}\right)\right]F^\prime_{\Delta_1\,\Delta_2\,\Delta_3}\left(k_1,k_2,k_3\right).
\end{align}
\end{subequations}
At the level of the Mellin-Barnes representation these become the following difference relations for $F^\prime_{\Delta_1\,\Delta_2\,\Delta_3}\left(s_1,s_2,s_3\right)$,
\begin{subequations}
\begin{multline}\label{diffr1}
    \left(s_1-1+\tfrac{1}{2}\left(\tfrac{d}{2}-\Delta_1\right)\right)\left(s_1-1-\tfrac{1}{2}\left(\tfrac{d}{2}-\Delta_1\right)\right)F^\prime_{\Delta_1\,\Delta_2\,\Delta_3}\left(s_1-1,s_2,s_3\right) \\ = \left(s_3-1+\tfrac{1}{2}\left(\tfrac{d}{2}-\Delta_3\right)\right)\left(s_3-1-\tfrac{1}{2}\left(\tfrac{d}{2}-\Delta_3\right)\right)F^\prime_{\Delta_1\,\Delta_2\,\Delta_3}\left(s_1,s_2,s_3-1\right),
\end{multline}
\begin{multline}
    \left(s_2-1+\tfrac{1}{2}\left(\tfrac{d}{2}-\Delta_2\right)\right)\left(s_2-1-\tfrac{1}{2}\left(\tfrac{d}{2}-\Delta_2\right)\right)F^\prime_{\Delta_1\,\Delta_2\,\Delta_3}\left(s_1,s_2-1,s_3\right) \\ = \left(s_3-1+\tfrac{1}{2}\left(\tfrac{d}{2}-\Delta_3\right)\right)\left(s_3-1-\tfrac{1}{2}\left(\tfrac{d}{2}-\Delta_3\right)\right)F^\prime_{\Delta_1\,\Delta_2\,\Delta_3}\left(s_1,s_2,s_3-1\right).\label{diffr2}
\end{multline}
\end{subequations}
The difference relation \eqref{diffr1} is solved by
\begin{multline}
    F^\prime_{\Delta_1\,\Delta_2\,\Delta_3}\left(s_1,s_2,s_3\right) =  c\left(s_2\right)\Gamma\left(s_1+\tfrac{1}{2}\left(\tfrac{d}{2}-\Delta_1\right)\right)\Gamma\left(s_1-\tfrac{1}{2}\left(\tfrac{d}{2}-\Delta_1\right)\right)\\ \times \Gamma\left(s_3+\tfrac{1}{2}\left(\tfrac{d}{2}-\Delta_3\right)\right)\Gamma\left(s_3-\tfrac{1}{2}\left(\tfrac{d}{2}-\Delta_3\right)\right)\,p\left(s_1,s_3\right),
\end{multline}
for some function $c\left(s_2\right)$ of $s_2$ and periodic function $p\left(s_1,s_3\right)$ of unit period in $s_1$ and $s_3$. The final difference relation \eqref{diffr2} gives 
\begin{equation}\label{fprime3pt}
    F^\prime_{\Delta_1\,\Delta_2\,\Delta_3}\left(s_1,s_2,s_3\right) = p\left(s_1,s_2,s_3\right)\prod^3_{j=1}\Gamma\left(s_j+\tfrac{1}{2}\left(\tfrac{d}{2}-\Delta_j\right)\right)\Gamma\left(s_j-\tfrac{1}{2}\left(\tfrac{d}{2}-\Delta_j\right)\right),
\end{equation}
for periodic function $p\left(s_1,s_2,s_3\right)$ of unit period in $s_1$, $s_2$ and $s_3$. This should be chosen such that the Mellin integrals converge, with different choices yielding different solutions. See \cite{MellinBook} chapter 4.4. There are four in total since we seek solutions to two second order equations. One of these is:
\begin{equation}\label{constlambda}
    p\left(s_1,s_2,s_3\right) = \lambda_{\Delta_1\,\Delta_2\,\Delta_3}, \qquad \lambda_{\Delta_1\,\Delta_2\,\Delta_3} = \text{const}.
\end{equation}
This is the unique solution with no singularities for collinear momentum configurations e.g. $k_1+k_2=k_3$. The other three are obtained by the analytic continuations
\begin{subequations}
\begin{align}
&k_1 \to -k_1, \quad k_2 \to k_2,\\
&k_1 \to k_1,  \quad \hspace*{0.3cm}k_2 \to -k_2,\\
&k_1 \to -k_1,  \quad k_2 \to -k_2.
\end{align}
\end{subequations}
In total we therefore have,  
\begin{equation}\label{gensol3pt}
    p\left(s_1,s_2,s_3\right)= \lambda_{\Delta_1\,\Delta_2\,\Delta_3}\, e^{-2\alpha \pi s_1}e^{-2\beta \pi s_2}, \qquad \alpha, \beta = 0, 1, \qquad \lambda_{\Delta_1\,\Delta_2\,\Delta_3} = \text{const}.
\end{equation}

Boundary correlators in AdS and in the Bunch Davies vacuum of dS do not have singularities in collapsed triangle configurations \cite{Chen:2006nt,Holman:2007na,LopezNacir:2011kk,Flauger:2013hra,Aravind:2013lra}. In this work we shall therefore always take $\alpha=\beta=0$. In this case, from the Mellin-Barnes representation of the Bessel $K$ function \eqref{MBeadssc} it is straightforward to show that the three-point function \eqref{fprime3pt} recovers the familiar triple-$K$ integral representation \cite{Bzowski:2013sza} for conformal three-point functions of generic scalar operators in momentum space.

\subsection{Three-point contact diagrams of scalar fields in (EA)dS}
\label{subsec::3ptsceads}

It is straightforward to see that the three-point function \eqref{3ptscalar} with $\alpha=\beta =0$ can be interpreted as a three-point Witten diagram generated by the non-derivative cubic vertex of scalar fields $\phi_i$ in EAdS$_{d+1}$:
\begin{equation}\label{V3}
{\cal V} = g\, \phi_1 \phi_2 \phi_3.
\end{equation}
Using the Feynman rules given in section \ref{subsec::feynrules} we have:
\begin{multline}\label{FAdS}
F^{\text{AdS}}_{\Delta_1\,\Delta_2\,\Delta_3}\left(s_1,{\bf k}_1,s_2,{\bf k}_2,s_3,{\bf k}_3\right)= -g\,i \pi\,  \delta\left(\tfrac{d}{4}-s_1-s_2-s_3\right)\left(2\pi\right)^{d}\delta^{(3)}\left({\bf k}_1+{\bf k}_2+{\bf k}_3\right)\\ \times \prod^3_{j=1}\,K^{\text{AdS}}_{\Delta_j,0}\left(s_j,{\bf k}_j\right),
\end{multline}
where the coupling constant $g$ is related to the 3pt coefficient $\lambda^{\text{AdS}}_{\Delta_1\,\Delta_2\,\Delta_3}$ defined in \eqref{constlambda} via:
\begin{equation}\label{AdScccoeff}
\lambda^{\text{AdS}}_{\Delta_1\,\Delta_2\,\Delta_3}  = - g\, \prod^3_{j=1}\frac{1}{2\Gamma\left(\Delta_j-\frac{d}{2}+1\right)},
\end{equation}
which we identified by using the Mellin-Barnes representation \eqref{MBeadssc} of the bulk-to-boundary propagators. Note that the vertex \eqref{V3} is the unique on-shell cubic vertex of scalar fields $\phi_i$.

\begin{figure}[t]
    \centering
    \captionsetup{width=0.95\textwidth}
    \includegraphics[width=\textwidth]{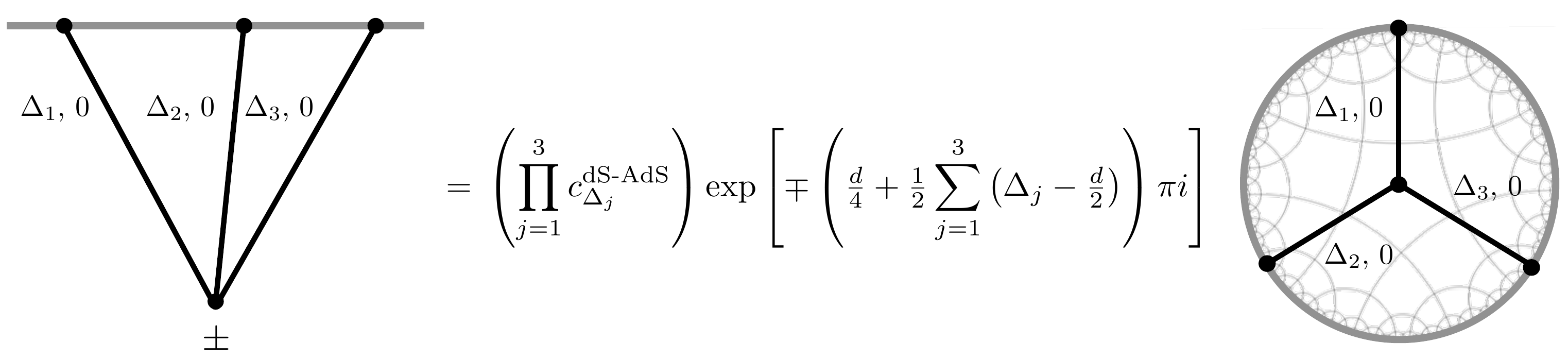}
    \caption{The contribution from the $\pm$ branch of the in-in contour to a three-point contact diagram of scalar fields in dS$_{d+1}$ is given by the corresponding diagram in EAdS$_{d+1}$ multiplied by a constant phase which implements the Wick rotation \eqref{wickinin} and the factors $c^{\text{dS-AdS}}_{\Delta_i}$ which account for the change in two-point coefficient.}
    \label{fig::3ptpm}
\end{figure}

The three-point boundary correlator generated by the same vertex \eqref{V3} in the Bunch Davies vacuum dS$_{d+1}$ can be obtained from its EAdS$_{d+1}$ counterpart \eqref{FAdS} using the Wick rotations \eqref{wickinin} \cite{Sleight:2019hfp,Sleight:2020obc}. In the Mellin-Barnes representation, to obtain the contribution from the $\pm$-branch of the in-in contour we multiply each leg of the Mellin transform \eqref{FAdS} of the EAdS$_{d+1}$ Witten diagram by the by phase and normalization factor $c^{\text{dS-AdS}}_{\Delta_j}$ which converts each of them to a bulk-to-boundary propagator on the $\pm$-branch of the in-in contour:
\begin{multline}
    F^{\pm}_{\Delta_1\,\Delta_2\,\Delta_3}\left(s_1,{\bf k}_1,s_2,{\bf k}_2,s_3,{\bf k}_3\right)= \left(\prod\limits^3_{j=1} c^{\text{dS-AdS}}_{\Delta_j} e^{\mp \left(s_j+\tfrac{1}{2}\left(\Delta_j-\tfrac{d}{2}\right)\right)\pi i}\right)\\ \times F^{\text{AdS}}_{\Delta_1\,\Delta_2\,\Delta_3}\left(s_1,{\bf k}_1,s_2,{\bf k}_2,s_3,{\bf k}_3\right).
\end{multline}
The constraint \eqref{constr} coming from the Dilatation Ward identity enforces that the proportionality factor between $F^{\pm}_{\Delta_1\,\Delta_2\,\Delta_3}\left(s_i,k_i\right)$ and $F^{\text{AdS}}_{\Delta_1\,\Delta_2\,\Delta_3}\left(s_i,k_i\right)$ is a constant \cite{Sleight:2019hfp,Sleight:2020obc},
\begin{multline}\label{dsads3ptphase}
    F^{\pm}_{\Delta_1\,\Delta_2\,\Delta_3}\left(s_1,{\bf k}_1,s_2,{\bf k}_2,s_3,{\bf k}_3\right)= \left(\prod\limits^3_{j=1} c^{\text{dS-AdS}}_{\Delta_j} \right)e^{\mp \left(\tfrac{d}{4}+\tfrac{1}{2}\sum\limits^3_{j=1}\left(\Delta_j-\tfrac{d}{2}\right)\right)\pi i}\\ \times F^{\text{AdS}}_{\Delta_1\,\Delta_2\,\Delta_3}\left(s_1,{\bf k}_1,s_2,{\bf k}_2,s_3,{\bf k}_3\right),
\end{multline}
so that the three-point function \eqref{3ptscalar} with $\alpha=\beta =0$ can also be interpreted as as a three-point boundary correlator generated by the non-derivative cubic vertex of fields $\phi_i$ in dS$_{d+1}$. The full dS correlator is the sum of the $\pm$ branch contributions \cite{Sleight:2019hfp,Sleight:2020obc}
\begin{align}
  \hspace*{-0.5cm} F^{\text{dS}}_{\Delta_1\,\Delta_2\,\Delta_3}\left(s_1,{\bf k}_1,s_2,{\bf k}_2,s_3,{\bf k}_3\right) &= i\,F^{+}_{\Delta_1\,\Delta_2\,\Delta_3}\left(s_1,{\bf k}_1,s_2,{\bf k}_2,s_3,{\bf k}_3\right)-i\,F^{-}_{\Delta_1\,\Delta_2\,\Delta_3}\left(s_1,{\bf k}_1,s_2,{\bf k}_2,s_3,{\bf k}_3\right) \nonumber \\
  &= 2\left(\prod\limits^3_{j=1} c^{\text{dS-AdS}}_{\Delta_j} \right) \sin \left(\tfrac{d}{4}+\tfrac{1}{2}\sum\limits^3_{j=1}\left(\Delta_j-\tfrac{d}{2}\right)\right)\pi \label{adstodsscalar}\\
  & \hspace*{4.5cm}\times F^{\text{AdS}}_{\Delta_1\,\Delta_2\,\Delta_3}\left(s_1,{\bf k}_1,s_2,{\bf k}_2,s_3,{\bf k}_3\right). \nonumber
\end{align}

\begin{figure}[t]
    \centering
    \captionsetup{width=0.95\textwidth}
    \includegraphics[width=\textwidth]{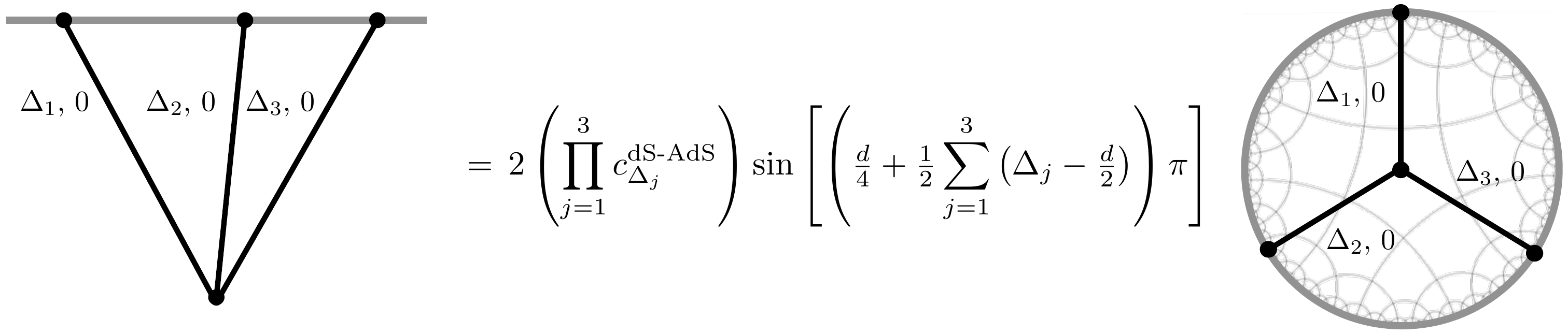}
    \caption{Upon summing the contributions from the $\pm$ branches of the in-in contour to obtain the full dS contact diagram, the constant phases from the individual $\pm$ branch contributions -- which have opposite sign -- combine to give a constant sinusoidal factor multiplying the corresponding contact diagram in EAdS.}
    \label{fig::3ptdsadsfull}
\end{figure}

From this it follows that the 3pt boundary correlator generated by a cubic vertex of scalar fields $\phi_i$ in dS$_{d+1}$ can be obtained from its EAdS$_{d+1}$ counterpart by the following replacement of the 3pt coefficient: 
\begin{equation}\label{3ptadsdsratio}
\lambda^{\text{AdS}}_{\Delta_1\,\Delta_2\,\Delta_3} \quad \to \quad \lambda^{\text{dS}}_{\Delta_1\,\Delta_2\,\Delta_3} =  \lambda^{\text{AdS}}_{\Delta_1\,\Delta_2\,\Delta_3}  \times  2\left(\prod\limits^3_{j=1} c^{\text{dS-AdS}}_{\Delta_j} \right) \sin \left(\tfrac{d}{4}+\tfrac{1}{2}\sum\limits^3_{j=1}\left(\Delta_j-\tfrac{d}{2}\right)\right)\pi\,.
\end{equation}

In the following sections this result is extended to contact diagrams with any number of legs and spinning fields. 

\subsection{Adding legs}
\label{subsec::npt}

The relation \eqref{adstodsscalar} between three-point contact diagrams in EAdS and dS naturally extends to contact diagrams involving any number $n$ of legs --- see e.g. \cite{Sleight:2019mgd} section 3.2, which we review here for completeness.

\begin{figure}[t]
    \centering
    \captionsetup{width=0.95\textwidth}
    \includegraphics[width=\textwidth]{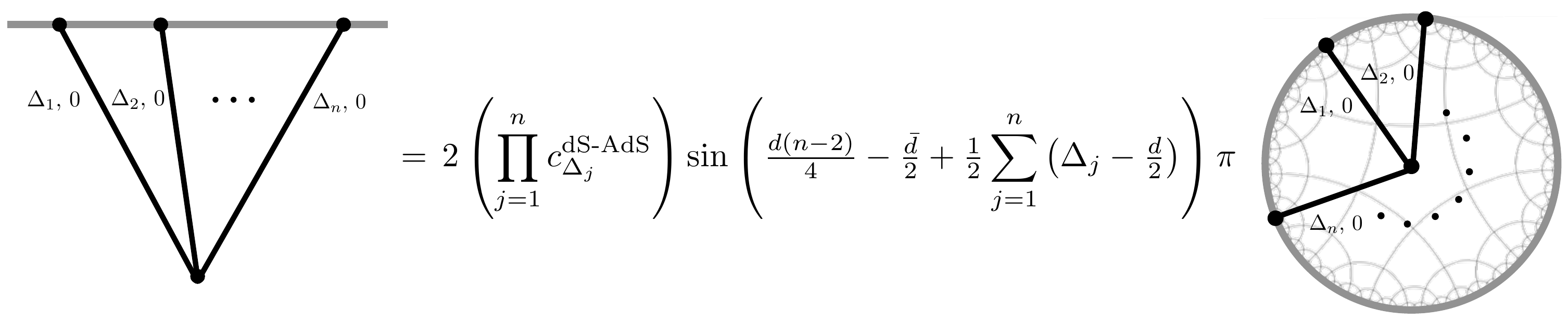}
    \caption{At higher-points, the sinusoidal factor relating the diagrams in dS and EAdS depends on ${\bar d}$, which is sensitive to the number of derivatives in the corresponding vertex. As we will see in the next section, it also depends on the spin of the fields.}
    \label{fig::nptdsads}
\end{figure}

Following the prescription outlined in section \ref{subsec::feynrules}, each leg of a contact diagram is assigned an external Mellin variable $s_i$ and a boundary momentum ${\bf k}_i$ where $i=1, \ldots , n$ for a contact diagram with $n$ legs. As for the $n=3$ case in section \ref{subsec::3ptsceads}, given the Mellin-Barnes representation $F^{\text{AdS}}_{\Delta_1\,\ldots\,\Delta_n}\left(s_1,{\bf k}_1,\ldots,s_n,{\bf k}_n\right)$ for an $n$-point contact diagram involving scalar fields $\phi_i$ in EAdS$_{d+1}$, to obtain the contribution from the $\pm$ branch of the in-in contour to the diagram generated by the same vertex in dS$_{d+1}$ we multiply, for each leg, by the phase \eqref{buborel} which converts each of them to a bulk-to-boundary propagator on the $\pm$ branch of the in-in contour:
\begin{multline}
   F^{\pm}_{\Delta_1\,\ldots\,\Delta_n}\left(s_1,{\bf k}_1,\ldots,s_n,{\bf k}_n\right)= \left(\prod\limits^n_{j=1} c^{\text{dS-AdS}}_{\Delta_j} e^{\mp \left(s_j+\tfrac{1}{2}\left(\Delta_j-\tfrac{d}{2}\right)\right)\pi i}\right)\\ \times F^{\text{AdS}}_{\Delta_1\,\ldots\,\Delta_n}\left(s_1,{\bf k}_1,\ldots,s_n,{\bf k}_n\right).
\end{multline}
The only difference with respect to the $n=3$ case considered in the previous section is that for $n>3$ contact diagrams of scalar fields are not unique due to the possibility of derivative interactions which are non-vanishing on-shell. In general these can have ${\bar d}\ne 0$ in the constraint \eqref{ddmb} imposed on the Mellin variables $s_i$ by scale symmetry, so that 
\begin{multline}
    F^{\bullet}_{\Delta_1\,\ldots\,\Delta_n}\left(s_1,{\bf k}_1,\ldots,s_n,{\bf k}_n\right) = 2\pi i \, \delta\left(d+{\bar d}+\sum\limits^n_{i=1}\left(2s_i-\frac{d}{2}\right)\right)\,\\ \times \left(F^{\bullet}\right)^\prime_{\Delta_1\,\ldots\,\Delta_n}\left(s_1,{\bf k}_1,\ldots,s_n,{\bf k}_n\right),
\end{multline}
where $\bullet=\text{AdS}$ or $\bullet=\pm$. The constant phase that relates $F^{\pm}_{\Delta_1\,\ldots\,\Delta_n}$ and $F^{\bullet}_{\Delta_1\,\ldots\,\Delta_n}$ is therefore
\begin{multline}
    F^{\pm}_{\Delta_1\,\ldots\,\Delta_n}\left(s_1,{\bf k}_1,\ldots,s_n,{\bf k}_n\right)= \left(\prod\limits^n_{j=1} c^{\text{dS-AdS}}_{\Delta_j} \right)e^{\mp \left(\tfrac{d(n-2)}{4}-\tfrac{{\bar d}}{2}+\tfrac{1}{2}\sum\limits^n_{j=1}\left(\Delta_j-\tfrac{d}{2}\right)\right)\pi i}\\ \times F^{\text{AdS}}_{\Delta_1\,\ldots\,\Delta_n}\left(s_1,{\bf k}_1,\ldots,s_n,{\bf k}_n\right),
\end{multline}
so that the full dS contact diagram is related to its EAdS counterpart by
\begin{align}
    F^{\text{dS}}_{\Delta_1\,\ldots\,\Delta_n}\left(s_1,{\bf k}_1,\ldots,s_n,{\bf k}_n\right) &= i F^{+}_{\Delta_1\,\ldots\,\Delta_n}\left(s_1,{\bf k}_1,\ldots,s_n,{\bf k}_n\right)-i F^{-}_{\Delta_1\,\ldots\,\Delta_n}\left(s_1,{\bf k}_1,\ldots,s_n,{\bf k}_n\right),\nonumber \\
    &=\frac{\lambda^{\text{dS}}_{\Delta_1\, \ldots \, \Delta_n}}{\lambda^{\text{AdS}}_{\Delta_1\, \ldots \, \Delta_n}}\,F^{\text{AdS}}_{\Delta_1\,\ldots\,\Delta_n}\left(s_1,{\bf k}_1,\ldots,s_n,{\bf k}_n\right),
\end{align}
where 
\begin{equation}\label{npointsine}
\lambda^{\text{dS}}_{\Delta_1\,\ldots\,\Delta_n} =  \lambda^{\text{AdS}}_{\Delta_1\,\ldots\,\Delta_n}  \times  2\left(\prod\limits^n_{j=1} c^{\text{dS-AdS}}_{\Delta_j} \right) \sin \left(\tfrac{d(n-2)}{4}-\tfrac{{\bar d}}{2}+\tfrac{1}{2}\sum\limits^n_{j=1}\left(\Delta_j-\tfrac{d}{2}\right)\right)\pi\,,
\end{equation}
which extends \eqref{3ptadsdsratio} to $n$-point contact diagrams of scalar fields $\phi_i$. See figure \ref{fig::nptdsads}. Note that contact diagrams generated by the non-derivative $n$-point interaction $\phi_1 \ldots \phi_n$ have ${\bar d}=0$. 

It would be interesting to extend the approach of section \ref{subsec::CWI3pt} to solve Conformal Ward Identities for higher-point contact diagrams using the Mellin-Barnes representation. We shall not explore this direction further here since the relation \eqref{npointsine} between contact diagrams in (EA)dS is sufficient for the purposes of this paper. Solutions to the Conformal Ward Identities for four-point boundary contact diagrams in Fourier space were classified in \cite{Arkani-Hamed:2018kmz} for conformally coupled scalars. 

One can also obtain the Mellin-Barnes representation of any contact diagram by simply applying the Feynman rules for the Mellin-Barnes representation given in section \ref{subsec::feynrules}. Applying these rules, the Mellin-Barnes representation for the contact diagram generated by the non-derivative vertex $g\,\phi_1 \ldots \phi_n$ in EAdS is simply
\begin{multline}\label{nptcontMB}
   {}^{\phi_1 \ldots \phi_n}F^{\text{AdS}}_{\Delta_1\,\ldots\,\Delta_n}\left(s_1,{\bf k}_1,\ldots,s_n,{\bf k}_n\right)\\ = -g\, 2\pi i\,\delta\left(d+\left(2s_1-\tfrac{d}{2}\right)+\ldots+\left(2s_n-\tfrac{d}{2}\right)\right)\left(2\pi\right)^d\delta^{(d)}\left({\bf k}_1+\ldots+{\bf k}_n\right)\\\times K^{\text{AdS}}_{\Delta_1,0}\left(s_1,{\bf k}_1\right) \ldots K^{\text{AdS}}_{\Delta_n,0}\left(s_1,{\bf k}_n\right).
\end{multline}
Contact diagrams generated by derivative interactions differ from the above via polynomials in the external Mellin variables $s_i$ and the parameter ${\bar d}$. In section 3.2 of \cite{Sleight:2021iix} one can find examples of the Mellin-Barnes representation for contact diagrams generated by derivative interactions in the $n=4$ case. 

\subsection{Adding spin}
\label{subsec::3ptsaddingspin}

In section \ref{subsec::CWI3pt} we reviewed how conformal symmetry constrains the boundary three-point function of scalar fields in (EA)dS up to a coefficient \cite{Polyakov:1970xd,Osborn:1993cr}. Similarly, conformal symmetry constrains correlators of spinning fields to a linear combination of tensorial structures with unfixed relative coefficients \cite{Polyakov:1970xd,Mack:1976pa,Sotkov:1976xe,Osborn:1993cr,Erdmenger:1996yc,Maldacena:2011nz,Costa:2011mg}. 

Such solutions are most straightforwardly obtained by uplifting to an ambient $\left(d+2\right)$-dimensional flat Minkowski space \cite{Dirac:1936fq}, where the $SO\left(d+1,1\right)$ isometry acts linearly and the symmetry constraints become as trivial as those of Lorentz symmetry \cite{Costa:2011mg}. In \cite{Sleight:2017fpc} it was shown that ambient space makes manifest the kinematic map between consistent on-shell cubic vertices of spinning fields in (EA)dS and spinning three-point conformal structures on the boundary, which immediately identifies the boundary three-point functions generated by a given cubic vertex in (EA)dS -- and vice versa. The ambient formalism also naturally identifies boundary differential operators that would generate such spinning three-point functions when they act on a scalar seed.\footnote{This is in the same spirit as the various types of ``weight-shifting" operators that by now are widely used in the CFT and (A)dS literature, both in position \cite{Costa:2011dw,Sleight:2017krf,Castro:2017hpx,Sleight:2017fpc,Chen:2017yia,Karateev:2017jgd,Costa:2018mcg} and momentum \cite{Isono:2018rrb,Arkani-Hamed:2018kmz,Isono:2019ihz,Sleight:2019hfp,Baumann:2019oyu,Baumann:2020dch,Sleight:2021iix} space.} As we shall see in this section, we can recycle these results to extend those for scalar three-point functions in sections \ref{subsec::CWI3pt} and \ref{subsec::3ptsceads} to the three-point boundary correlators generated by any cubic coupling of an arbitrary triplet of massive spinning fields. This is explained in the following, where we begin by reviewing known results on generating functions for on-shell cubic couplings of spinning fields in flat space -- which are then uplifted to (EA)dS using the ambient space formalism. 

This section is mostly a technical review of existing results in the literature and one can safely skip to the final result \eqref{spinning3pt} where, for a given cubic coupling of spinning fields in (EA)dS, they are applied to relate the corresponding 3pt boundary correlator in dS to its EAdS counterpart.

\paragraph{Cubic vertices for spinning fields: Flat space.} In the following we shall closely follow the presentation of section 3 in \cite{Joung:2012fv}. When dealing with spinning fields it is employ an index-free notation, where a spin-$J$ field $\varphi_{\mu_1 \dots \mu_J}\left(x\right)$ is represented by a generating function:
\begin{equation}\label{genfuncintr}
    \varphi_{\mu_1 \ldots \mu_J } \; \longrightarrow \; \varphi_J\left(x;u\right) = \frac{1}{J!}  \varphi_{\mu_1 \ldots \mu_J }\left(x\right)\, u^{\mu_1} \ldots u^{\mu_J},
\end{equation}
where we introduced a constant auxiliary vector $u^\mu$. In terms of generating functions, the cubic coupling a massive fields of spins $J_1$-$J_2$-$J_3$ in flat space can be expressed in the form (using point splitting)
\begin{equation}\label{flatcubic}
    {}^{(\text{flat})}{\cal V}_{J_1,J_2,J_3}\left(x\right) = C_{J_1,J_2,J_3}\left(\partial_{x_i};\partial_{u_i}\right)\, \varphi_{J_1}\left(x_1;u_1\right)\varphi_{J_2}\left(x_2;u_2\right)\varphi_{J_3}\left(x_3;u_3\right)\Big|_{{}^{x_i=x}_{u_i = 0}},
\end{equation}
where, restricting to parity invariant interactions, $C_{J_1,J_2,J_3}$ is a function of the following 12 (6+9+6) Lorentz scalars:
\begin{equation}
    \partial_{x_i} \cdot \partial_{x_j}, \qquad \partial_{u_i} \cdot \partial_{x_j}, \qquad \partial_{u_i} \cdot \partial_{u_j}.
\end{equation}
Such functions $C_{J_1,J_2,J_3}$ however are not all physically distinguishable. There are two sources of ambiguity. One is owing to the triviality of total derivatives (or equivalently integration by parts), which correspond to terms in $C_{J_1,J_2,J_3}$ involving the combination:
\begin{equation}
    \partial_{x_1}+\partial_{x_2}+\partial_{x_3}. 
\end{equation}
This ambiguity can be fixed by, say, replacing $\partial_{u_i} \cdot \partial_{x_{i-1}}$ with its representation in terms of the other Lorentz scalars:
\begin{equation}
    \partial_{u_i} \cdot \partial_{x_{i-1}} = -\partial_{u_i} \cdot \partial_{x_{i+1}}-\partial_{u_i} \cdot \partial_{x_{i}}+\underbrace{\partial_{u_i} \cdot \left(\partial_{x_1}+\partial_{x_2}+\partial_{x_3}\right)}_{\text{total derivative}}.
\end{equation}
The second ambiguity is the possibility to perform non-linear field re-definitions, which generate fictitious interaction terms that vanish on the free equations of motion. This ambiguity can therefore be fixed by neglecting terms that vanish on-shell, which in the function $C_{J_1,J_2,J_3}$ corresponds to dropping terms involving the operator $\partial_{x_i} \cdot \partial_{x_j}$ since we can write:
\begin{equation}
    \partial_{x_i} \cdot \partial_{x_{i-1}} = \frac{1}{2}\left(\partial^2_{x_{i+1}}-\partial^2_{x_{i}}-\partial^2_{x_{i-1}}\right)+\underbrace{\frac{1}{2}\left(\partial_{x_1}+\partial_{x_2}+\partial_{x_3}\right) \cdot \left(\partial_{x_i}+\partial_{x_{i-1}}-\partial_{x_{i+1}}\right)}_{\text{total derivative}},
\end{equation}
where, using the linear equations of motion, the terms $\partial^2_{x_i}$ can be replaced by $\left(\partial_{u_i} \cdot \partial_{x_i}\right)$ (divergence) and $\partial^2_{u_i}$ (trace).

Upon fixing the above ambiguities the function $C_{J_1,J_2,J_3}$ can only depend on 12 (3+3+6) Lorentz scalars:
\begin{equation}
    \partial_{u_i} \cdot \partial_{x_{i+1}}, \qquad \partial_{u_i} \cdot \partial_{x_{i}}, \qquad \partial_{u_i} \cdot \partial_{u_{j}}.
\end{equation}
On shell, the fields $\varphi_{J_i}$ satisfy the traceless and divergenceless constraints:
\begin{subequations}
\begin{align}
   \left(\partial_x \cdot \partial_u\right) \varphi_{J}\left(x,u\right)&=0,\\
   \left(\partial_u \cdot \partial_u\right) \varphi_{J}\left(x,u\right)&=0.
\end{align}
\end{subequations}
We refer to the part of $C_{J_1,J_2,J_3}$ that does not involve such divergence and trace terms as the \emph{traceless and transverse part}, denoted by $C^{TT}_{J_1,J_2,J_3}$, which indeed is what survives upon eliminating unphysical degrees of freedom. This is a function of 6 (3+3) Lorentz scalars:
\begin{subequations}
\begin{align}
    \partial_{u_1}\cdot  \partial_{x_2}, \qquad \partial_{u_2}\cdot  \partial_{x_3}, \qquad \partial_{u_3}\cdot  \partial_{x_1},\\
    \partial_{u_1}\cdot  \partial_{u_2}, \qquad \partial_{u_2}\cdot  \partial_{u_3}, \qquad \partial_{u_3}\cdot  \partial_{u_1}.
\end{align}
\end{subequations}
To determine the most general form for the function $C^{TT}_{J_1,J_2,J_3}$ it is sufficient to note that if there is a term which is a monomial of degree $n_{j+1}$ in the Lorentz scalar $\partial_{u_i}\cdot  \partial_{u_j}$ then, since the generating functions $\varphi_{J_i}\left(x_i;u_i\right)$ are degree $J_i$ in the auxiliary vectors $u_i$, the same term must also be a monomial of degree $J_i-n_{j-1}-n_{j+1}$ in the remaining Lorentz scalars $\partial_{u_i}\cdot  \partial_{x_{i+1}}$. We can therefore conclude that the general form for the function $C^{TT}_{J_1,J_2,J_3}$ is: 
\begin{equation}\label{CTT}
    C^{TT}_{J_1,J_2,J_3}\left(\partial_{x_i};\partial_{u_i}\right) = \sum\limits_{n_i} g^{n_1,n_2,n_3}_{J_1,J_2,J_3}\,C^{n_1,n_2,n_3}_{J_1,J_2,J_3}\left(\partial_{x_i};\partial_{u_i}\right), 
\end{equation}
where
\begin{multline}\label{CTTn1n2n3}
  C^{n_1,n_2,n_3}_{J_1,J_2,J_3}\left(\partial_{x_i};\partial_{u_i}\right) = \left(\partial_{u_1} \cdot \partial_{x_2}\right)^{{\tilde J}_1}\left(\partial_{u_2} \cdot \partial_{x_3}\right)^{{\tilde J}_2}\left(\partial_{u_3} \cdot \partial_{x_1}\right)^{{\tilde J}_3}\\ \times  \left(\partial_{u_2} \cdot \partial_{u_3}\right)^{n_1}\left(\partial_{u_3} \cdot \partial_{u_1}\right)^{n_2}\left(\partial_{u_1} \cdot \partial_{u_2}\right)^{n_3},
\end{multline}
and for convenience we defined ${\tilde J}_i=J_i - n_{i-1}-n_{i+1}$. The $g^{n_1,n_2,n_3}_{J_1,J_2,J_3}$ are independent coupling constants. The sum over the $n_i$ is given explicitly by:
\begin{equation}
    \sum\limits_{n_i} = \sum\limits^{\text{min}\{J_1,J_2\}}_{n_3=0}\,\sum\limits^{\text{min}\{J_1-n_3,J_3\}}_{n_2=0}\,\sum\limits^{\text{min}\{J_2-n_3,J_3-n_2\}}_{n_1=0}.
\end{equation}
Note that at cubic order there are no non-localities since the Lorentz scalars $\partial_{x_i} \cdot \partial_{x_j}$ do not appear and the others only come with positive power. 

\paragraph{Cubic vertices for spinning fields: (A)dS space.} If we regard AdS$_{d+1}$ and dS$_{d+1}$ as hypersurfaces embedded in a $(d+2)$-dimensional (ambient) flat space,
\begin{equation}\label{HSAdS}
    X^2=L^2_{\text{dS}}, \qquad X^2=-L^2_{\text{AdS}},
\end{equation}
with coordinates $X^A$, the traceless and transverse flat space cubic vertices \eqref{CTT} 
immediately give on-shell cubic vertices of spinning fields in (A)dS$_{d+1}$. The metric of the ambient space is taken to be $\eta_{AB} =\left(-,+,\ldots,+\right)$, with $A, B = 0, \ldots, d+2$, so that we have Euclidean AdS$_{d+1}$ and Lorentzian dS$_{d+1}$.

There is an isomorphism between between symmetric fields $\varphi_{\mu_1 \ldots \mu_J}\left(x\right)$ in (A)dS$_{d+1}$ of mass \eqref{mass(A)dS} and those $\varphi_{A_1 \ldots A_J}\left(X\right)$ in the flat ambient space which satisfy the following tangentiality and homogeneity constraints \cite{Fronsdal:1978vb}:\footnote{In the homogeneity constraint \eqref{homofrons} we could have chosen $\Delta^-$ in place of $\Delta^+$.}
\begin{subequations}\label{THconstr}
\begin{align}
  &\text{Tangentiality:} \hspace*{2cm} \left(X \cdot \partial_U\right)\varphi_{J}\left(X,U\right)=0, \\
  &\text{Homogeneity:} \hspace*{1.2cm} \left(X \cdot \partial_X+\Delta^+\right)\varphi_{J}\left(X,U\right)=0, \label{homofrons}
\end{align}
\end{subequations}
where we packaged the field $\varphi_{A_1 \ldots A_J}\left(X\right)$ in the ambient space counterpart of the generating function \eqref{genfuncintr}:
\begin{equation}
    \varphi_{J}\left(X,U\right) = \frac{1}{J!} \varphi_{A_1 \ldots A_J}\left(X\right) U^{A_{1}} \ldots U^{A_{J}}.
\end{equation}
A traceless and divergence-free ambient field $\varphi_{A_1 \ldots A_J}\left(X\right)$ satisfies the massless Fierz-Pauli system,\footnote{Note that despite the absence of a mass term in \eqref{ambfp} there is no gauge redundancy for generic mass \eqref{mass(A)dS}. This emerges for the values of $\Delta^\pm$ corresponding to (partially-)massless fields. See e.g. section 2.2 of \cite{Joung:2012rv} for details on the description of (partially-)massless fields in the ambient space formalism.}
\begin{subequations}\label{AMBFPJ}
\begin{align}\label{ambfp}
    \partial^2_{X}\, \varphi_{J}\left(X,U\right)=&0,\\
    \left(\partial_X \cdot \partial_U\right)\varphi_{J}\left(X,U\right)=&0,\\
    \left(\partial_U \cdot \partial_U\right)\varphi_{J}\left(X,U\right)=&0.
\end{align}
\end{subequations}
Using that the (commuting) ambient partial derivative is related to the (non-commuting) AdS covariant derivative $\nabla$ via
\begin{equation}\label{ambpartialcov}
    \partial^A_X=\nabla^A+\frac{1}{X^2}\left(X^A X \cdot \partial_X+U^A X \cdot \partial_U-U \cdot X \partial_U\right),
\end{equation}
it is straightforward to show that, upon pull-back to the (EA)dS manifold, an ambient field $\varphi_J\left(X;U\right)$ satisfying the massless Fierz-Pauli system \eqref{AMBFPJ} together with the tangentiality and homogeneity constraints \eqref{THconstr} is identified with a traceless and divergence-free field $\varphi_{\mu_1 \ldots \mu_J}\left(x\right)$ in (A)dS$_{d+1}$ satisfying the massive Fierz-Pauli system:
\begin{subequations}\label{fp(A)dS}
\begin{align}
    \left(\nabla^2-m^2\right)\, \varphi_{J}\left(x,u\right)=&0,\\
    \left(\nabla \cdot \partial_u\right)\varphi_{J}\left(x,u\right)=&0,\\
    \left(\partial_u \cdot \partial_u\right)\varphi_{J}\left(x,u\right)=&0,
\end{align}
\end{subequations}
with mass 
\begin{equation}\label{mass(A)dS}
    m^2 L^2_{\text{(A)dS}} = (-)(\Delta^+\Delta^-+J).
\end{equation}
With the above isomorphism between fields $\varphi_{\mu_1 \ldots \mu_J}\left(x\right)$ on (A)dS$_{d+1}$ and fields $\varphi_{A_1 \ldots A_J}\left(X\right)$ in $\left(d+2\right)$-dimensional flat space subject to the tangentiality and homogeneity constraints \eqref{THconstr}, we can immediately write down cubic vertices of traceless and transverse spinning fields in (A)dS from their flat space counterparts \eqref{CTT} through the replacements:
\begin{subequations}
\begin{align}\label{urepl}
u \quad &\to \quad U,\\ \label{xrepl}
x \quad &\to \quad X,\\
\varphi_J\left(x,u\right) \quad &\to \quad\varphi_J\left(X,U\right) \quad \text{subject to \eqref{THconstr}},
\end{align}
\end{subequations}
giving 
\begin{equation}\label{(a)dscubic}
  \hspace*{-0.25cm}  {}^{\text{(A)dS}}{\cal V}^{TT}_{J_1,J_2,J_3}\left(X\right) = C^{TT}_{J_1,J_2,J_3}\left(\partial_{X_i};\partial_{U_i}\right)\, \varphi_{J_1}\left(X_1;U_1\right)\varphi_{J_2}\left(X_2;U_2\right)\varphi_{J_3}\left(X_3;U_3\right)\Big|_{{}^{X_i=X}_{U_i = 0}},
\end{equation}
where $C^{TT}_{J_1,J_2,J_3}\left(\partial_{X_i};\partial_{U_i}\right)$ is simply \eqref{CTT} with the replacements \eqref{urepl} and \eqref{xrepl}:
\begin{equation}
    C^{TT}_{J_1,J_2,J_3}\left(\partial_{X_i};\partial_{U_i}\right) = \sum\limits_{n_i} g^{n_1,n_2,n_3}_{J_1,J_2,J_3}\,C^{n_1,n_2,n_3}_{J_1,J_2,J_3}\left(\partial_{X_i};\partial_{U_i}\right), 
\end{equation}
with
\begin{multline}\label{(a)dscubicn1n2n3}
  C^{n_1,n_2,n_3}_{J_1,J_2,J_3}\left(\partial_{X_i};\partial_{U_i}\right) = \left(\partial_{U_1} \cdot \partial_{X_2}\right)^{{\tilde J}_1}\left(\partial_{U_2} \cdot \partial_{X_3}\right)^{{\tilde J}_2}\left(\partial_{U_3} \cdot \partial_{X_1}\right)^{{\tilde J}_3}\\ \times  \left(\partial_{U_2} \cdot \partial_{U_3}\right)^{n_1}\left(\partial_{U_3} \cdot \partial_{U_1}\right)^{n_2}\left(\partial_{U_1} \cdot \partial_{U_2}\right)^{n_3}.
\end{multline}
Note that the traceless and transverse part of cubic vertices is sufficient to compute (tree-level) three-point boundary correlators in (A)dS$_{d+1}$ since the external legs are on-shell.

What we see from the above is that the ambient space formalism re-expresses quantities intrinsic to (A)dS in terms of simpler flat-space ones. In particular, in (A)dS, covariant derivatives  $\nabla$ do not commute -- meaning that cubic vertices in (A)dS$_{d+1}$ generally differ from their counterparts \eqref{CTTn1n2n3} in $\left(d+1\right)$-dimensional flat space by a cumbersome tail of lower derivative terms proportional to the cosmological constant. In the ambient space formalism, these are re-packaged\footnote{See appendix B of \cite{Sleight:2016dba} on obtaining the intrinsic expressions for the vertices \eqref{(a)dscubicn1n2n3} in terms of (A)dS covariant derivatives.} into the homogeneous expressions \eqref{(a)dscubicn1n2n3} in terms of the \emph{commuting} partial derivatives of the $(d+2)$-dimensional flat ambient space \eqref{ambpartialcov}.

\begin{figure}[t]
    \centering
    \captionsetup{width=0.95\textwidth}
    \includegraphics[width=0.85\textwidth]{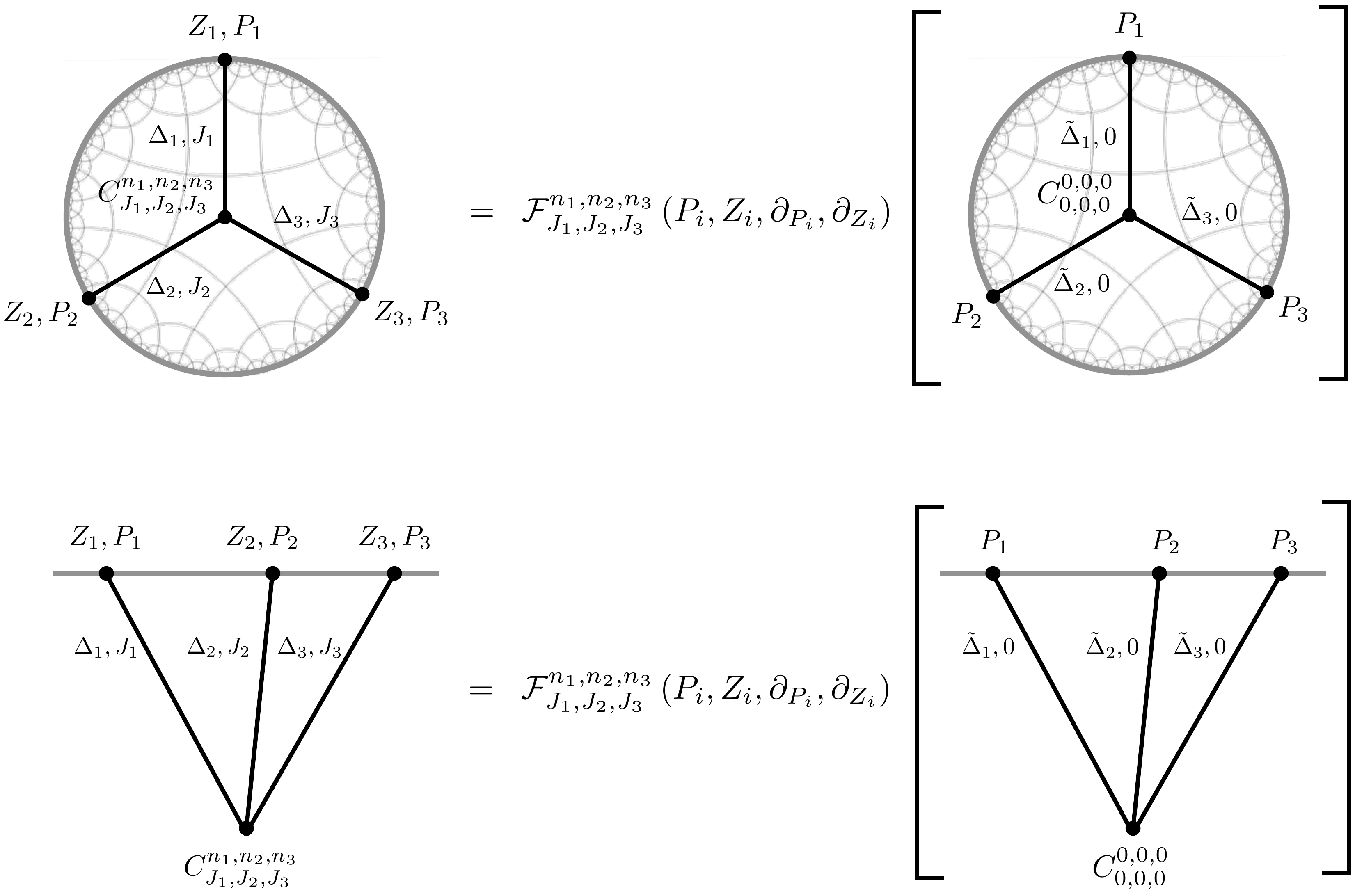}
    \caption{Spinning three-point contact diagrams in (EA)dS generated by the cubic vertex \eqref{(a)dscubicn1n2n3} of fields with spins $J_1$-$J_2$-$J_2$ can be generated by acting with the boundary differential operator \eqref{delta-nonnormspect} on a scalar ``seed", which is a three-point contact diagram generated by the non-derivative cubic interaction $\phi_1\phi_2\phi_3$ of scalar fields with shifted scaling dimensions ${\tilde \Delta}_i$ related to the original scaling dimensions $\Delta_i$ via \eqref{delatilde}.}
    \label{fig::WS}
\end{figure}

\paragraph{Spinning 3pt functions from a scalar seed.} A convenient feature of the ambient space formalism is that, for a given cubic coupling \eqref{(a)dscubicn1n2n3}, it is straightforward to express the corresponding boundary 3pt contact diagram as a \emph{boundary} differential operator acting on a scalar seed (see figure \ref{fig::WS}). In the work \cite{Sleight:2017fpc}, this was used to identify a basis of 3pt conformal structures on the boundary which makes manifest the cubic vertices in (EA)dS$_{d+1}$ that generate them; it provides the \emph{kinematic map} between cubic vertices of spinning fields in (EA)dS$_{d+1}$ and spinning 3pt structures in CFT$_{d}$. This is reviewed in the following, which we then use to extend the results of the previous sections to fields with spin.

In ambient space, the boundary of (A)dS$_{d+1}$ is identified with light rays \cite{Dirac:1936fq}:
\begin{equation}
    P^2=0, \qquad P \, \sim \, \lambda P, \qquad \lambda \ne 0.
\end{equation}
Boundary tensors $F_{A_1 \ldots A_J}\left(P\right)$ are encoded in generating functions
\begin{equation}
    F\left(P;\Xi\right) = \frac{1}{J!}F_{A_1 \ldots A_J}\left(P\right)\Xi^{A_1} \ldots \Xi^{A_J},
\end{equation}
with null, transverse auxiliary vector $\Xi^2=\Xi \cdot P=0$. Like their bulk counterparts \eqref{THconstr}, they must also satisfy the tangentiality and homogeneity constraints:
\begin{subequations}
\begin{align}
    \left(P \cdot \partial_\Xi\right)F\left(P;\Xi\right) &= 0,\\
    \left(P \cdot \partial_P+\Delta\right)F\left(P;\Xi\right) &= 0.
\end{align}
\end{subequations}
For more details see \cite{Costa:2011mg}.

One notes that the bulk-to-boundary propagator for a spin-$J$ field in EAdS$_{d+1}$ can be represented via a differential operator ${\cal D}$ acting on its scalar counterpart, which in ambient space takes the following simple form \cite{Sleight:2016hyl}:
\begin{equation}\label{Dpsrbubo}
    {\cal D}_P=\left(\Xi \cdot U\right)\left(\Xi \cdot \frac{\partial}{\partial \Xi}-P \cdot \frac{\partial}{\partial P}\right)+\left(P \cdot U\right)\left(\Xi \cdot \frac{\partial}{\partial P}\right).
\end{equation}
The ambient expression for the scalar bulk-to-boundary propagator \eqref{scalabubo} in EAdS$_{d+1}$ is simply \cite{Penedones:2010ue}:\footnote{One recovers the expression \eqref{scalabubo} by using that in Poincar\'e coordinates the bulk and boundary points are parameterised by: 
\begin{equation}
    X\left(z,{\bf x}\right) = \frac{L_{\text{AdS}}}{z}\left(\frac{L^2_{\text{AdS}}+z^2+{\bf x}\cdot {\bf x}}{2L_{\text{AdS}}}, {\bf x}, \frac{L^2_{\text{AdS}}-z^2-{\bf x}\cdot {\bf x}}{2L_{\text{AdS}}}\right), \quad P\left({\bf x}\right) =\frac{1}{2}\left(1+{\bf x}^\prime\cdot {\bf x}^\prime, {\bf x}^\prime, 1-{\bf x}^\prime\cdot {\bf x}^\prime\right).
\end{equation}}
\begin{equation}\label{scalarbubo}
    K_{\Delta,0}\left(X;P\right) = \frac{C^{\text{AdS}}_{\Delta,0}}{\left(-2 P \cdot X\right)^\Delta}.
\end{equation}
The $n$-th ambient partial derivative of a scalar bulk-to-boundary propagator can be re-expressed as a degree $n$ monomial in the boundary vector $P$ accompanied by the shift ${\Delta \to \Delta+n}$: 
\begin{equation}\label{nidscabubo}
    \partial^{A_1}_{X} \ldots \partial^{A_n}_{X}\,K_{\Delta,0}\left(X;P\right) = 2^n \left(\Delta\right)_n\,P^{A_1} \ldots  P^{A_n}K_{\Delta+n,0}\left(X;P\right).
\end{equation}
These relations also hold for bulk-to-boundary propagators in dS$_{d+1}$, which as we saw in section \ref{sec::adsdsdict} are related to those in EAdS$_{d+1}$ by Wick rotation. Combining \eqref{Dpsrbubo} and \eqref{nidscabubo}, we see that the 3pt boundary correlator generated by the cubic coupling \eqref{(a)dscubicn1n2n3} of spinning fields in (EA)dS$_{d+1}$ can be obtained from the 3pt correlator generated by the non-derivative cubic interaction of scalar fields with the following shifts with respect to their spinning counterparts:
\begin{equation}
    \Delta_i \: \to \: {\tilde \Delta}_i = \Delta_i+J_{i+2}-n_i - n_{i+1}. \label{delatilde}
\end{equation}

The corresponding boundary differential operator can be read off as (\cite{Sleight:2017fpc} equation (3.18)): 
\begin{multline}\label{fn1n2n3}
    {\cal F}^{n_1,n_2,n_3}_{J_1,J_2,J_3} =  \frac{2^{{\tilde J}_1+{\tilde J}_2+{\tilde J}_3} J_1! J_2! J_3!\left(\Delta_1\right)_{{\tilde J}_1}\left(\Delta_2\right)_{{\tilde J}_2}\left(\Delta_3\right)_{{\tilde J}_3}}{\left(\Delta_1-1\right)_{J_1}\left(\Delta_2-1\right)_{J_2}\left(\Delta_3-1\right)_{J_3}} \\ \times \left({\cal D}_{P_2}\cdot {\cal D}_{P_3}\right)^{n_1}\left({\cal D}_{P_1}\cdot {\cal D}_{P_3}\right)^{n_2}\left({\cal D}_{P_1}\cdot {\cal D}_{P_2}\right)^{n_3}\left( P_1 \cdot {\cal D}_{P_3}\right)^{{\tilde J}_1}\left( P_2 \cdot {\cal D}_{P_1} \right)^{{\tilde J}_2}\left( P_3 \cdot {\cal D}_{P_2} \right)^{{\tilde J}_3},
\end{multline}
where the operator
\begin{equation}
   P_j \cdot {\cal D}_{P_i} = \left(\Xi_i \cdot P_j\right) \left(\Xi_i \cdot \frac{\partial}{\partial \Xi_i}\right)-\left(\Xi_i \cdot P_j\right) \left(P_i \cdot \frac{\partial}{\partial P_i}\right)+\left(P_i\cdot P_j\right)\left(\Xi_i \cdot \frac{\partial}{\partial P_i}\right),
\end{equation}
coincides with the operator introduced in \cite{Costa:2011dw}, which raises the spin of the operator at $P_i$ by one unit and lowers the scaling dimension of the operator at $P_j$ by one unit. The operator ${\cal D}_{P_i} \cdot {\cal D}_{P_{j}}$ instead raises the spin at both points $P_i$ and $P_j$ by one unit.

Note that \eqref{fn1n2n3}, in the same spirit as the work \cite{Costa:2011dw}, provides a differential basis for conformal three-point functions of spinning operators. It however goes beyond \cite{Costa:2011dw} in that it also provides a \emph{kinematic} one-to-one map between the set of spin-$J_1-$spin-$J_2-$spin-$J_3$ (parity-even) cubic vertices \eqref{(a)dscubicn1n2n3} permitted by the (A)dS$_{d+1}$ isometry and the possible spin-$J_1-$spin-$J_2-$spin-$J_3$ (parity-even) three-point structures permitted by conformal symmetry on the $d$-dimensional boundary, where:
\begin{subequations}
\begin{align}
 \partial_{U_{i+1}} \cdot \partial_{X_i}  \qquad &\leftrightarrow \qquad  P_{i+1} \cdot {\cal D}_{P_i}, \\
 \partial_{U_{i}} \cdot \partial_{U_j}  \qquad &\leftrightarrow \qquad  {\cal D}_{P_i} \cdot {\cal D}_{P_{j}}.
\end{align}
\end{subequations}
In other words, expanding conformal three point functions in the differential basis \eqref{fn1n2n3} immediately gives back the cubic vertex in (EA)dS$_{d+1}$ that generates it.

Using the differential operator \eqref{fn1n2n3} one can write down an expression for the Mellin-Barnes representation of the three point boundary correlator in (EA)dS$_{d+1}$ generated by the cubic vertex \eqref{(a)dscubic}:
\begin{multline}\label{spin3ptfromscse}
F^{\text{(A)dS}}_{\Delta_1,\Delta_2,\Delta_3; J_1, J_2, J_3; n_1, n_2, n_3}\left(s_1,{\bf k}_1,s_2,{\bf k}_2,s_3,{\bf k}_3\right) \\ =  {\cal F}^{n_1,n_2,n_3}_{J_1,J_2,J_3}\,F^{\text{(A)dS}}_{{\tilde \Delta}_1,\,{\tilde \Delta}_2,\,{\tilde \Delta}_3}\left(s_1,k_1,s_2,k_2,s_3,k_3\right),
\end{multline}
where to use the operator ${\cal F}^{n_1,n_2,n_3}$ in this equation one should  take the Fourier transform of the expression \eqref{fn1n2n3}, which then acts on the Mellin-Barnes representation of the scalar seed. As in \eqref{AdScccoeff}, the cubic coupling $g^{n_1,n_2,n_3}_{J_1,J_2,J_3}$ of the vertex \eqref{(a)dscubic} is related to the three-point coefficient of the AdS correlator via
\begin{equation}\label{spinningads3pt}
\lambda^{\text{AdS}}_{\Delta_1\,\Delta_2\,\Delta_3} = - g^{n_1,n_2,n_3}_{J_1,J_2,J_3}\, \prod^3_{j=1}\frac{1}{2\Gamma\left({\tilde \Delta}_j-\frac{d}{2}+1\right)}.
\end{equation}
By combining \eqref{spin3ptfromscse} and the relation \eqref{adstodsscalar} between EAdS and dS three-point coefficients of boundary correlators involving only scalar fields, the three-point coefficient $\lambda^{\text{dS}}$ of the boundary correlator in dS generated by the same vertex \eqref{(a)dscubic} is obtained from its AdS counterpart $\lambda^{\text{AdS}}$ via the replacement \cite{Sleight:2019hfp,Sleight:2020obc}
\begin{multline}\label{spinning3pt0}
\lambda^{\text{AdS}}_{\Delta_1\,\Delta_2\,\Delta_3}  \quad \\ \to \quad \lambda^{\text{dS}}_{\Delta_1\,\Delta_2\,\Delta_3}  =  \lambda^{\text{AdS}}_{\Delta_1\,\Delta_2\,\Delta_3}  \times  2\left(\prod\limits^3_{j=1} c^{\text{dS-AdS}}_{{\tilde \Delta}_j} \right) \sin \left(\tfrac{d}{4}+\tfrac{1}{2}\sum\limits^3_{j=1}\left({\tilde \Delta}_j-\tfrac{d}{2}\right)\right)\pi\,, 
\end{multline}
so that the dS$_{d+1}$ three-point boundary correlator generated by the vertex \eqref{(a)dscubic} can be obtained from its AdS counterpart by multiplying it with the factor on the second line of the above equation.  
\begin{figure}[t]
    \centering
    \captionsetup{width=0.95\textwidth}
    \includegraphics[width=\textwidth]{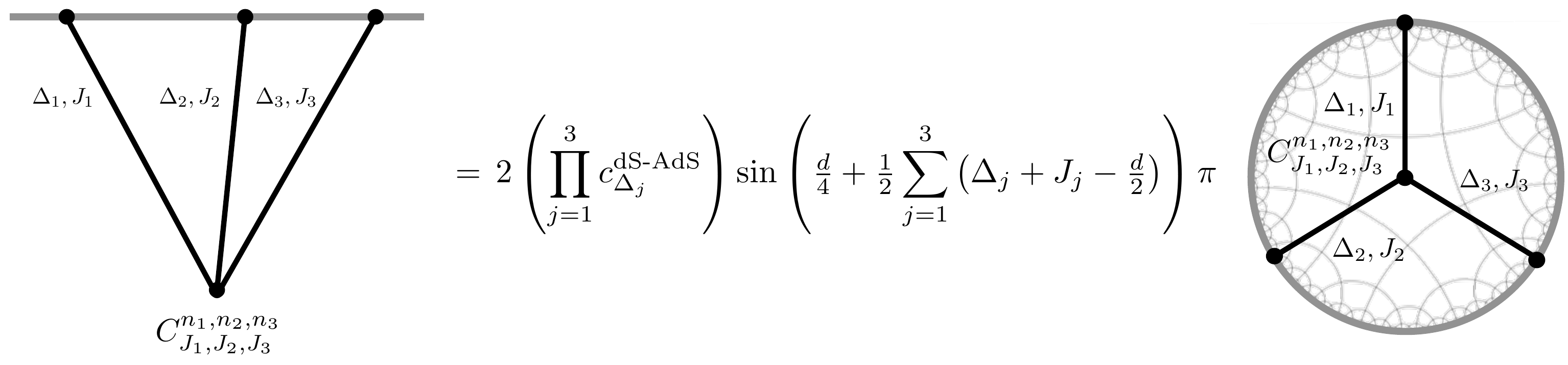}
    \caption{The sinusoidal factor relating spinning three-point diagrams in dS and EAdS depends on the spins of the fields, but it is insensitive to the choice of vertex $C^{n_1,n_2,n_3}_{J_1,J_2,J_3}$ --- it is the same for all vertices involving fields with spins $J_1$-$J_2$-$J_3$.}
    \label{fig::3ptspinningadsds}
\end{figure}

It is interesting to note the effect of the parameters $n_i$, which parameterise the number of derivatives in each vertex. In particular, since both the sine factor in \eqref{spinning3pt0} and the coefficients $c^{\text{dS-AdS}}_{{\tilde \Delta}_j}$ (which are given by the cosine factor \eqref{cseadstods}) are both periodic functions, the effect of the shifts \eqref{delatilde} they induce in the scaling dimensions $\Delta_j$ cancel out at the level of \eqref{spinning3pt0}! We can therefore write (see figure \ref{fig::3ptspinningadsds}.):
\begin{multline}\label{spinning3pt}
\lambda^{\text{AdS}}_{\Delta_1\,\Delta_2\,\Delta_3}  \quad \\\to \quad \lambda^{\text{dS}}_{\Delta_1\,\Delta_2\,\Delta_3}  =  \lambda^{\text{AdS}}_{\Delta_1\,\Delta_2\,\Delta_3}  \times  2\left(\prod\limits^3_{j=1} c^{\text{dS-AdS}}_{\Delta_j} \right) \sin \left(\tfrac{d}{4}+\tfrac{1}{2}\sum\limits^3_{j=1}\left(\Delta_j+J_j-\tfrac{d}{2}\right)\right)\pi\,, 
\end{multline}
which is insensitive to the number of derivatives in a given vertex! 

Finally, let us note that while in the above our use of the differential operator \eqref{fn1n2n3} has been somewhat implicit, it can be used to obtain explicit expressions for spinning boundary three-point functions by evaluating the action of the derivatives on the scalar seed. Such an application is completely mechanical and lends itself to implementation in Mathematica. In position space this was carried out in \cite{Sleight:2016dba,Sleight:2017fpc}. Note that for boundary correlators in momentum space one would first need to determine the Fourier transform of the operators $P_{i+1} \cdot {\cal D}_{P_i}$ and ${\cal D}_{P_i} \cdot {\cal D}_{P_{j}}$. The Fourier transform of $P_{i+1} \cdot {\cal D}_{P_i}$ was given in \cite{Baumann:2019oyu} and similarly one can obtain the Fourier transform of ${\cal D}_{P_i} \cdot {\cal D}_{P_{j}}$.

\subsection{Unitarity}
\label{subsec::contactunitarity}

\begin{figure}[t]
    \centering
    \captionsetup{width=0.95\textwidth}
    \includegraphics[width=0.85\textwidth]{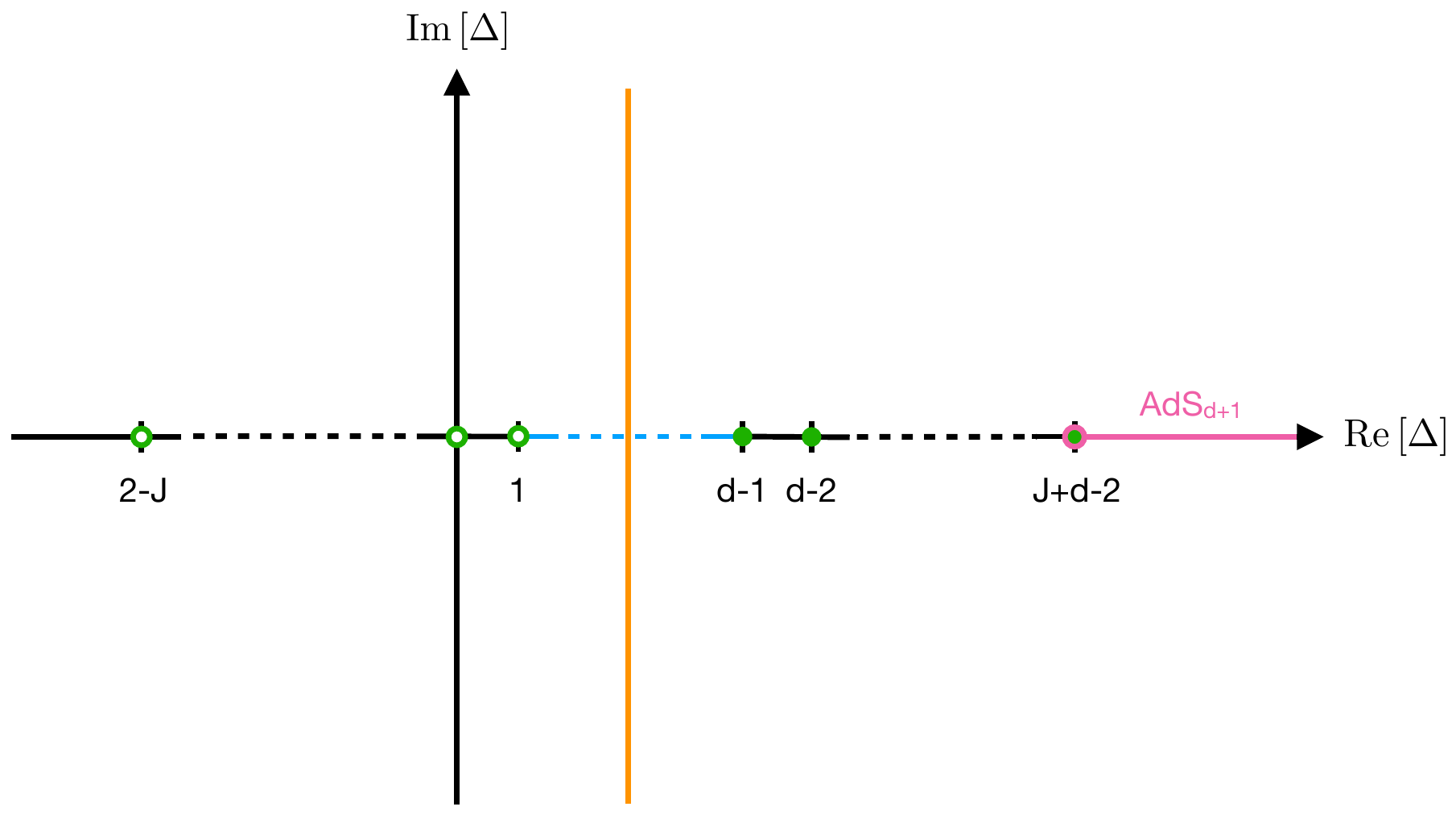}
    \caption{Unitary irreducible spin-$J$ representations of $SO\left(d+1,1\right)$ (green dots, blue line, yellow line) and $SO\left(d,2\right)$ (pink line) in the complex $\Delta$ plane. The yellow line corresponds to the Principal Series and the blue line the complementary series. The green dots denote the discrete series representation, with the hollow green dots marking their shadows ($\mu \to -\mu$). Spin-$J$ unitary irreducible representations of $SO\left(d,2\right)$ only overlap at the massless point, which is the rightmost green dot. Note the symmetry of this figure for unitary irreduble representations of $SO\left(d+1,1\right)$ under $\mu \to -\mu$ (reflection in the yellow line). A similar figure with further details can be found in \cite{Sleight:2019hfp}.}
    \label{fig::unitarity} 
\end{figure}

In the previous sections we reviewed how conformal symmetry constrains boundary three-point functions up to a collection of coefficients $\lambda_{\Delta_1\,\Delta_2\,\Delta_3}$, which does not distinguish between boundary correlators in EAdS and dS. A crucial difference which can be used to differentiate between the two is that in EAdS unitarity is with respect to $SO\left(d,2\right)$ while in dS it is with respect to $SO\left(d+1,1\right)$. This leads to different constraints on the scaling dimensions $\Delta_i$ (see figure \ref{fig::unitarity}) and the three-point (operator product expansion) coefficients $\lambda_{\Delta_1\,\Delta_2\,\Delta_3}$ in these two space-times.

\paragraph{anti-de Sitter.} In anti-de Sitter space unitarity implies a lower bound on the scaling dimensions $\Delta$ \cite{Mack:1975je}: 
\begin{subequations}
\begin{align}
    &\text{Scalars (spin $J=0$):} \qquad \Delta \geq \frac{d}{2}-1,\\
    &\text{Spin $J \geq 1$:} \hspace*{1.475cm} \Delta \geq d-2+J.
\end{align}
\end{subequations}
Unitarity furthermore implies that the three-point coefficients are real, 
\begin{equation}\label{reallambdaads}
\lambda^{\text{AdS}}_{\Delta_1\,\Delta_2\,\Delta_3}\, \in \, \mathbb{R},
\end{equation}
which follows from the reality of conformal correlators under the symmetry group $SO\left(d,2\right)$ (see e.g. \cite{Rattazzi:2008pe}).

\paragraph{de Sitter.}  For nice overviews of the unitary irreducible representations of the de Sitter isometry group and the corresponding quantum fields see e.g. \cite{Joung:2006gj,Joung:2007je,Basile:2016aen}. For scalars $\left(\text{spin }J=0\right)$ these fall into two catgeories:

\begin{itemize}
\item Principal Series: $\Delta = \frac{d}{2}+i\mu$ with $\mu \in \mathbb{R}$. Massive particles.
\item Complementary Series: $\Delta = \frac{d}{2}+i\mu$ with $\mu$ pure imaginary and $|\mu|=\left(0,\tfrac{d}{2}\right)$. Light particles.
\end{itemize}

\noindent For (totally-symmetric) spin $J$ we have (see figure \ref{fig::unitarity}):

\begin{itemize}
\item Principal Series: $\Delta = \frac{d}{2}+i\mu$ with $\mu \in \mathbb{R}$. Massive particles.\
\item Complementary Series: $\Delta = \frac{d}{2}+i\mu$ with $\mu$ pure imaginary and $|\mu|=\left(0,\tfrac{d-2}{2}\right)$. Light particles. 
\item Discrete Series: $\Delta = \frac{d}{2}+i\mu$ with $\mu = \mp \tfrac{i}{2}\left(d-4+2\left(J-r\right)\right)$. (Partially-)massless spin-$J$ particles of depth-$r$. Depth $r=0$ corresponds to massless particles of spin-$J$.
\end{itemize}
In dS Sitter we currently lack an analogue of the constraint \eqref{reallambdaads} on the three-point coefficient in EAdS$_{d+1}$, which is non-perturbative. Working perturbatively, diagram-by-diagram we can use the Mellin-Barnes representation to directly uplift the unitary time evolution encoded by the propagators to the boundary, allowing us to trace how it is imprinted in the boundary correlators that the propagators compute. In particular, we have seen that unitary time evolution from the early time Bunch Davies vacuum in de Sitter emerges from Euclidean AdS via the Wick rotations \eqref{wickinin}, which in the Mellin-Barnes representation is encoded in the phases \eqref{buborel} for external legs and the phases \eqref{dsharmphases} for the internal legs. For each leg in a given diagram, such phases directly uplift to the Mellin-Barnes representation of the corresponding correlator, as we saw explicitly in the previous sections for the three-point boundary correlator in dS$_{d+1}$. Upon summing the $+$ and $-$ branch contributions such phases combined to give a constant sinusoidal factor, which for three-point contact diagrams involving fields of spin $J_1$-$J_2$-$J_3$ reads:
\begin{equation}
 \lambda^{\text{dS}}_{\Delta_1\,\Delta_2\,\Delta_3}  =  \lambda^{\text{AdS}}_{\Delta_1\,\Delta_2\,\Delta_3}  \times  2\left(\prod\limits^3_{j=1} c^{\text{dS-AdS}}_{\Delta_j} \right) \sin \left(\tfrac{d}{4}+\tfrac{1}{2}\sum\limits^3_{j=1}\left(\Delta_j+J_j-\tfrac{d}{2}\right)\right)\pi.
\end{equation}
For $n$-point contact with $n>3$ one can show that:
\begin{subequations}\label{uniacoupdsads}
\begin{align}
\lambda^{\text{dS}}_{
\Delta_1, \ldots , \Delta_n}  &= \lambda^{\text{AdS}}_{
\Delta_1, \ldots , \Delta_n}   \times  2\left(\prod\limits^n_{j=1} c^{\text{dS-AdS}}_{\Delta_j} \right) \sin \left(\tfrac{d(n-2)}{4}-\tfrac{\bar d}{2}+\tfrac{1}{2}\sum\limits^n_{j=1}\left(\Delta_j-\tfrac{d}{2}\right)\right)\pi\,,\\
&= \pm\,  \lambda^{\text{AdS}}_{
\Delta_1, \ldots , \Delta_n}   \times  2\left(\prod\limits^n_{j=1} c^{\text{dS-AdS}}_{\Delta_j} \right) \sin \left(\tfrac{d(n-2)}{4}+\tfrac{1}{2}\sum\limits^n_{j=1}\left(\Delta_j+J_j-\tfrac{d}{2}\right)\right)\pi\,,
\end{align}
\end{subequations}
i.e. up to a sign, which is related to exchanged spin --- see e.g. \eqref{cbedsexch} or section 3.2 of \cite{Sleight:2021iix}.

A consequence of the property \eqref{uniacoupdsads} is the vanishing of dS contact boundary correlators for the values of the boundary dimension $d$, number $n$ of legs, the scaling dimensions $\Delta_i$ and the spins $J_i$ that conspire to give a zero of the sine factor \eqref{uniacoupdsads}, assuming that for such values the Mellin-Barnes representation of the correlator is defined.\footnote{This can be diagnosed by studying the pinching of the integration contours.} This feature has been observed previously in \cite{Sleight:2019mgd}. For example, for contact diagrams involving only conformally coupled scalars (corresponding to $J_i = 0$ and $\Delta_i = \frac{d\pm 1}{2}$ for all $i$), the sine factor \eqref{uniacoupdsads} has a zero when
(section 3.3 of \cite{Sleight:2019mgd}):
\begin{equation}
    n = \pm \left(4m-d\right), \quad m \in \mathbb{N},
\end{equation}
i.e. when $d$ and $n$ are simultaneously odd or simultaneously even, assuming that the Mellin-Barnes representation is defined in each case. For $d=3$, this vanishing for $n$ odd was shown in \cite{Goodhew:2020hob} to indeed follow as a consequence of perturbative unitarity.

It is interesting to note that, in the case that there is a divergence in the Mellin-Barnes representation of boundary correlators \emph{and} there is a zero of the sine factor \eqref{uniacoupdsads}, it can happen that the zero has the effect of cancelling the divergence. Continuing with the above example of contact diagrams involving only conformally coupled scalars, the Mellin-Barnes representation in this case is divergent when:
\begin{equation}
    d(n-2)=n.
\end{equation}
In \cite{Sleight:2019mgd} (section 3.3) this divergence was shown to be cancelled by a corresponding zero of the sine factor \eqref{uniacoupdsads}. In practise this happens because the divergences from the $+$ and $-$ branch contributions (see e.g. the first line of equation \eqref{adstodsscalar}), which as we have see differ only by a phase \eqref{dsads3ptphase}, cancel -- as noted in \cite{Arkani-Hamed:2015bza} for the $d=3$ and $n=3$ case, their equation (5.72). Interestingly, this implies that a diagram contributing to a boundary correlator in dS can be finite even if its EAdS counterpart is divergent. 

\paragraph{Summary.} Above we have seen that while conformal symmetry alone does not distinguish between boundary correlators in EAdS and the Bunch-Davies vacuum of dS, the requirement of unitarity differentiates between them by imposing different constraints on the spectrum of scaling dimensions of states (see figure \ref{fig::unitarity}) and the operator product expansion (OPE) coefficients. While in EAdS unitarity requires the OPE coefficients to be real \cite{Rattazzi:2008pe}, in dS this is not a requirement: At linear order in the coupling they are proportional to the sine factor \eqref{uniacoupdsads}, which is a function of the scaling dimensions and the spins of the particles. As we shall see explicitly later on, higher orders in the coupling then involve sums of such sine factors multiplied together. This also implies the vanishing of diagrams contributing to the perturbative computation of dS boundary for certain collections of particles (and certain boundary dimensions $d$) -- even if its EAdS counterpart is non-zero -- as a consequence of unitary time evolution in dS, which was also observed in \cite{Goodhew:2020hob}. 

Consequently, while any diagram contributing to a boundary correlator in dS can be written as a linear combination of EAdS Witten diagrams \cite{Sleight:2020obc}, since the unitary values of the scaling dimensions in EAdS and dS do not coincide (though they are overlapping, see figure \ref{fig::unitarity}) and the couplings in EAdS and dS differ by sinusoidal factors \eqref{uniacoupdsads}, it should be emphasised that such Witten diagrams are generally not generated by a unitary theory in anti-de Sitter space.

\section{Four-point exchanges}

\label{sec::4ptexch}

In this section we consider four-point tree-level exchanges. In section \ref{subsec::cuttinganddisp} we gave cutting rules to compute the on-shell part of any diagram in (EA)dS, based on the observation that on-shell bulk-to-bulk propagators in (EA)dS decompose into a linear combination \eqref{factproperty} of factorised contributions for the $\Delta^\pm$ modes. The full diagram can then be reconstructed from the dispersion formula \eqref{dispprop}. In the same section this was applied to compute four-point tree-level exchanges of scalar fields. In this section we will show that the same results can be obtained from a boundary perspective.\footnote{It might be useful to note that such statements about ``boundary" and ``bulk" perspectives are somewhat trivial in the Mellin-Barnes representation, which places boundary and bulk on the same footing by trivialising the bulk integrals.} In particular, in section \ref{subsec::bootEAdSexch} we will see how on-shell exchanges in (EA)dS can be Bootstrapped by a combination of factorisation, conformal symmetry and boundary conditions for any collection of internal and external particles. The full exchanges in (EA)dS are then reconstructed using the dispersion formula \eqref{dispprop}, which we argue can be regarded as a general relation between a function and its discontinuity at the level of the Mellin-Barnes representation.

In section \ref{subsec::dsasadsexch} we give the expression for exchanges in dS as a linear combination of EAdS Witten diagrams, which follows from the relations between propagators in (EA)dS reviewed in section \ref{sec::3pt}.

In section \ref{subsec::factunitcbe}, having obtained an expression for the full exchange in dS as a linear combination of AdS exchanges, we reformulate factorisation at the level of the conformal block expansion -- which is inherited from the known conformal block expansion of the constituent AdS exchanges. This allows us to describe factorisation at the level of the full dS exchange, which so far had only been formulated at the level of the individual in-in contributions (which are not observables). We also discuss how unitarity manifests itself in the coefficients of conformal blocks that encode the contribution from the physical exchanged single particle state.

This section gives further technical details on the results presented in section 4 of \cite{Sleight:2020obc} and provides the extension to arbitrary internal and external (integer) spinning fields.

\subsection{Bootstrapping (EA)dS exchanges}
\label{subsec::bootEAdSexch}

Let us first take the exchanged particle to be on-shell, meaning that we are considering the discontinuity \eqref{discs} of the exchange diagram with respect to the internal leg. Particles in (EA)dS$_{d+1}$ are irreducible representations of $SO\left(d+1,1\right)$ which, as we have seen, are labelled by a scaling dimension $\Delta$ and spin $J$. Since the Casimir elements of the algebra take constant values on any irreducible representation, the contribution from the physical exchanged single particle state to the boundary four-point function in, say, the ${\sf s}$-channel is a solution $W$ to the following conformal Casimir equation, which is convenient to formulate in ambient space (reviewed in section \ref{subsec::3ptsaddingspin}) \cite{Dolan:2003hv}:
\begin{align}\label{ccasmir}
    \left(\frac{1}{2} J^{AB}J_{AB} +C_{2}\right) W\left(P_1,P_2,P_3,P_4\right) = 0, \qquad
    C_{2} =-\Delta_+\Delta_-+J\left(J+d-2\right),
\end{align}
with quadratic conformal Casimir $\frac{1}{2} J^{AB}J_{AB}$ and $J_{AB} = J^{(1)}_{AB}+J^{(2)}_{AB}$ are the generators of the $SO\left(d+1,1\right)$ algebra, where $J^{(i)}_{AB}$ acts on the boundary point $P_i$:
\begin{equation}
    i J^{(i)}_{AB} = \left(P_i\right)_{A}\frac{\partial}{\partial \left(P_i\right)^B}-\left(P_i\right)_{B}\frac{\partial}{\partial \left(P_i\right)^A}.
\end{equation}
In the CFT literature the two independent solutions $W_{\Delta_\pm,J}$ to the Casimir equation \eqref{ccasmir} are known as \emph{Conformal Blocks}, which are defined by their limiting behaviour as ${P_{12}\equiv-2P_{1} \cdot P_2 \to 0}$:\footnote{Note that $-2P_1 \cdot P_2 = \left({\bf x}_1-{\bf x}_2\right)^2$.}
\begin{equation}\label{CBs}
    W_{\Delta^\pm,J}\left(P_i\right)\: \sim \: P^{\Delta^\pm-J-\Delta_1-\Delta_2/2}_{12},
\end{equation}
so that the general solution is a linear combination of the two:
\begin{equation}\label{gensol}
    W\left(P_i\right)= \alpha\, W_{\Delta^+,J}\left(P_i\right) + \beta\, W_{\Delta^-,J}\left(P_i\right).  
\end{equation}

The Conformal Casimir equation \eqref{ccasmir} is the boundary counterpart (or ``dual") to the homogeneous wave equation \eqref{adsharm} in (EA)dS$_{d+1}$. In particular (for more details see \cite{Sleight:2017krf}), the $SO\left(d+1,1\right)$ generators acting on an ambient tensor field $\varphi_J\left(X;U\right)$ are given by
\begin{equation}
    iJ_{AB} = X_{A}\frac{\partial}{\partial X^B}-X_{B}\frac{\partial}{\partial X^A}+\Sigma_{AB},
\end{equation}
where $\Sigma_{AB}$ is the spin operator:
\begin{equation}
    \Sigma_{AB} = U_{A}\frac{\partial}{\partial U^B}-U_{B}\frac{\partial}{\partial U^A}.
\end{equation}
Using that the quadratic Casimir is given explicitly by
\begin{equation}
    \frac{1}{2}J^{AB}J_{AB} = U \cdot \frac{\partial}{\partial U} \left(d-2+U \cdot \frac{\partial}{\partial U}\right)+X \cdot \frac{\partial}{\partial X} \left(d+X \cdot \frac{\partial}{\partial X}\right)-X^2 \frac{\partial^2}{\partial X^2},
\end{equation}
and combining it with the expression \eqref{ambpartialcov} for the AdS covariant derivative in terms of ambient partial derivatives, for an ambient field $\varphi_J\left(X;U\right)$ subject to the tangentiality and homogeneity constraints \eqref{THconstr} it follows that the quadratic Casimir equation
\begin{align}
    \left(\frac{1}{2}J^{AB}J_{AB} +C_{2}\right) \varphi_J\left(X;U\right) = 0,
\end{align}
is equivalent to the homogeneous wave equation,
\begin{subequations}
\begin{align}
    \left(\nabla^2-m^2\right)\varphi_J\left(X;U\right) &= 0,\\
    m^2 L^2_{\text{(A)dS}} &= (-)(\Delta^+\Delta^-+J).
\end{align}
\end{subequations}
The conformal blocks \eqref{CBs} are therefore the boundary dual to the on-shell propagators \eqref{disceads} with $\Delta_\pm$ boundary conditions.

What we learn from the above is that on-shell the tree-level exchange of a particle in (EA)dS$_{d+1}$ is fixed, up to normalisation, by \emph{conformal symmetry and boundary conditions}. The conformal Casimir equation \eqref{ccasmir} is in fact equivalent to the homogeneous conformal invariance constraint for exchanges introduced and subsequently applied to dS boundary correlators in the works \cite{Arkani-Hamed:2015bza}, though the connection with the Casimir equation and conformal blocks noted in the companion work \cite{Sleight:2020obc} had not been made explicit. As in the \cite{Arkani-Hamed:2018kmz}, the remaining freedom is fixed by requiring \emph{on-shell factorisation} into a product of the constituent three-point subdiagrams --- i.e. those generated by the vertices that mediate the exchange. 

To implement on-shell factorisation it is useful to consider a special type of solution ${\cal F}_{\mu,J}$ to the conformal Casimir equation \eqref{ccasmir}, which is single-valued in the Euclidean regime (i.e. does not have branch cuts). While the individual conformal blocks \eqref{CBs} do not have this property, ${\cal F}_{\mu,J}$ is the unique linear combination of them which does \cite{Costa:2012cb,Caron-Huot:2017vep} and in the conventions, say, of \cite{Caron-Huot:2017vep} reads\footnote{See \cite{Caron-Huot:2017vep} for the explicit form of the coefficients $K_{\Delta^\pm,J}$.}
\begin{align}\label{CPW}
    {\cal F}_{\mu,J} = \frac{1}{2}\left(W_{
    \Delta^+,J}+\frac{K_{\Delta^-,J}}{K_{\Delta^+,J}}\,W_{
    \Delta^-,J}\right),
\end{align}    
which corresponds to a specific choice of the coefficients $\alpha$ and $\beta$ in \eqref{gensol}. In the literature ${\cal F}_{\mu,J}$ is often referred to as a \emph{Conformal Partial Wave} or a \emph{harmonic function}. It is the boundary dual of the bulk bi-local harmonic functions $\Omega_{\mu,J}$ introduced in section \ref{sec::adsdsdict}, likewise providing an orthogonal basis of Eigenfunctions of the conformal Casimirs. This duality is made manifest by the following integral representation \cite{Hoffmann:2000mx,Dolan:2011dv,SimmonsDuffin:2012uy}:\footnote{To avoid notational clutter, here we leave the dependence of the three-point boundary correlators on the spins of the fields and their cubic couplings implicit.}
\begin{equation}
    {\cal F}_{\mu,J}\left({\bf x}_1,{\bf x}_2;{\bf x}_3,{\bf x}_4\right) \propto \int d^d{\bf x}^\prime\, F_{\Delta_1, \Delta_2, \Delta^+}\left({\bf x}_1, {\bf x}_2, {\bf x}^\prime \right)F_{\Delta^-, \Delta_3, \Delta_4}\left({\bf x}^\prime, {\bf x}_3, {\bf x}_4 \right),
\end{equation}
which is (up to normalisation) an integrated product of three-point boundary correlators of the exchanged field, one with $\Delta^+$ and the other with $\Delta^-$ boundary conditions. It is the boundary counterpart of the split representation \eqref{splitrep} of the bulk harmonic functions. Likewise, in Fourier space this is simply the product of the three-point boundary correlators:
\begin{equation}\label{cpwfsprod}
    {\cal F}_{\mu,J}\left({\bf k}_1,{\bf k}_2,{\bf p};-{\bf p},{\bf k}_3,{\bf k}_4\right) \propto  F_{\Delta_1, \Delta_2, \Delta^+}\left({\bf k}_1, {\bf k}_2, {\bf p} \right)F_{\Delta^-, \Delta_3, \Delta_4}\left(-{\bf p}, {\bf k}_3, {\bf k}_4 \right).
\end{equation}

The proportionality constant can be fixed by requiring that \eqref{cpwfsprod} is consistent with factorisation of the exchange in (EA)dS in the case that the boundary condition on the exchanged field produces the conformal partial wave \eqref{CPW}. More general boundary conditions can be obtained by employing the Mellin-Barnes representation and multiplying by the projectors \eqref{projectors}. This is explained in the following, starting with the exchange in EAdS.

\paragraph{On-shell exchange in EAdS.} Let us first consider the exchange of a particle in EAdS$_{d+1}$. Suppose that the particle is subject to the linear combination of $\Delta^+$ and $\Delta^-$ boundary conditions such that the on-shell exchange is given (up to normalisation) by the conformal partial wave \eqref{CPW}. As stated above, the normalisation is fixed by requiring on-shell factorisation into the constituent three-point boundary correlators generated by the cubic vertices. Given the representation \eqref{cpwfsprod} of the Conformal Partial Wave in Fourier space, this fixes the on-shell exchange generated by cubic vertices $C^{n_1,n_2,n}_{J_1,J_2,J}$ and $C^{{\bar n},n_3,n_4}_{J,J_3,J_4}$, which is the discontinuity \eqref{discs} of the full exchange, to be
\begin{subequations}
\begin{align}\label{onshellexhccpw}
   \text{Disc}_{{\sf s}}\left[ {\cal A}^{\text{EAdS}}_{\text{CPW}}\left({\bf k}_1,{\bf k}_2,{\bf p};-{\bf p},{\bf k}_3,{\bf k}_4\right)\right] &={\cal F}^{\text{EAdS}}_{\mu,J}\left({\bf k}_1,{\bf k}_2,{\bf p};-{\bf p},{\bf k}_3,{\bf k}_4\right),\\ \label{CPWAdS}
   {\cal F}^{\text{EAdS}}_{\mu,J}\left({\bf k}_1,{\bf k}_2,{\bf p};-{\bf p},{\bf k}_3,{\bf k}_4\right)&=
   F^{\text{AdS}}_{\Delta_1,\Delta_2,\Delta^+; J_1, J_2, J; n_1, n_2, n}\left({\bf k}_1,{\bf k}_2,{\bf p}\right)\\ &\hspace*{2cm}\times F^{\text{AdS}}_{\Delta^-,\Delta_3,\Delta_4; J, J_3, J_4; {\bar n}, n_3, n_4}\left(-{\bf p},{\bf k}_3,{\bf k}_4\right), \nonumber
\end{align}
\end{subequations}
where ${\cal F}^{\text{EAdS}}_{\mu,J}$ is the Conformal Partial Wave \eqref{cpwfsprod} normalised by the three-point boundary correlators \eqref{spin3ptfromscse}. The subscript CPW in \eqref{onshellexhccpw} simply denotes the fact that the exchange field is subject to the linear combination of $\Delta^+$ and $\Delta^-$ boundary conditions that corresponds to the Conformal Partial Wave \eqref{CPW}.

To obtain the on-shell exchange for a general linear combination of $\Delta^+$ and $\Delta^-$ boundary conditions on the exchanged field it is useful to employ the Mellin-Barnes representation. The Mellin-Barnes representation of a Conformal Partial Wave is inherited from that of its constituent three-point boundary correlators described in section \ref{sec::3pt}. In particular, 
\begin{multline}
    {\cal F}^{\text{EAdS}}_{\mu,J}\left({\bf k}_1,{\bf k}_2,{\bf p};-{\bf p},{\bf k}_3,{\bf k}_4\right) \\= \int^{+i\infty}_{-i\infty} \left[ds\right]_4 \left[du\,d{\bar u}\right] {\cal F}^{\text{EAdS}}_{\mu,J}\left(s_1,{\bf k}_1,s_2,{\bf k}_2,u,{\bf p};{\bar u},-{\bf p},s_3,{\bf k}_3,s_4,{\bf k}_4\right), 
\end{multline}
where 
\begin{multline}\label{EAdScpw}
  \hspace*{-0.5cm}  {\cal F}^{\text{EAdS}}_{\mu,J}\left(s_1,{\bf k}_1,s_2,{\bf k}_2,u,{\bf p};{\bar u},-{\bf p},s_3,{\bf k}_3,s_4,{\bf k}_4\right) = F^{\text{AdS}}_{\Delta_1,\Delta_2,\Delta^+; J_1, J_2, J; n_1, n_2, n}\left(s_1,{\bf k}_1,s_2,{\bf k}_2,u,{\bf p}\right)\\ \times F^{\text{AdS}}_{\Delta^-,\Delta_3,\Delta_4; J, J_3, J_4; {\bar n}, n_3, n_4}\left({\bar u},-{\bf p},s_3,{\bf k}_3,s_4,{\bf k}_4\right).
\end{multline}
Note that, as in section \ref{sec::adsdsdict}, each external leg is described by an external Mellin variable $s_i$ and the internal leg is described by a pair of internal Mellin variables $u$, ${\bar u}$. 
To implement the $\Delta^\pm$ boundary condition on the exchanged particle, at the level of the Mellin-Barnes representation we can simply take the projectors \eqref{projectors} --- which in the bulk act on the bulk harmonic function $\Omega_{\mu,J}$ --- and uplift them  to the boundary, where they act on the boundary harmonic function ${\cal F }_{\mu,J}$, projecting onto the $\Delta^\pm$ boundary conditions. The on-shell exchange of a particle in EAdS subject to a generic linear combination of $\Delta^+$ and $\Delta^-$ boundary conditions therefore reads \cite{Sleight:2020obc} 
\begin{multline}\label{disceadsexch}
     \text{Disc}_{{\sf s}}\left[ {\cal A}^{\text{EAdS}}_{\alpha \Delta^++\beta \Delta^-,J}\left(s_1,{\bf k}_1,s_2,{\bf k}_2,u,{\bf p};{\bar u}, -{\bf p},s_3,{\bf k}_3,s_4,{\bf k}_4\right)\right] = \underbrace{\left(\alpha\, \omega_{\Delta^+}\left(u,{\bar u}\right)+\beta\, \omega_{\Delta^-}\left(u,{\bar u}\right)\right)}_{\text{boundary conditions}}\\ \times \frac{\Gamma\left(1+i\mu\right)\Gamma\left(1-i\mu\right)}{\pi}\underbrace{{\cal F}^{\text{EAdS}}_{\mu,J}\left(s_1,{\bf k}_1,s_2,{\bf k}_2,u,{\bf p};{\bar u}, -{\bf p},s_3,{\bf k}_3,s_4,{\bf k}_4\right)}_{\text{conformal symmetry and factorisation}},
\end{multline}
which was fixed by a combination of \emph{conformal symmetry}, \emph{factorisation} and \emph{boundary conditions}. Note that this matches the result \eqref{discadsexch0} obtained in section \ref{subsec::cuttinganddisp} using the cutting rules.

\begin{figure}[t]
    \centering
    \captionsetup{width=0.95\textwidth}
    \includegraphics[width=\textwidth]{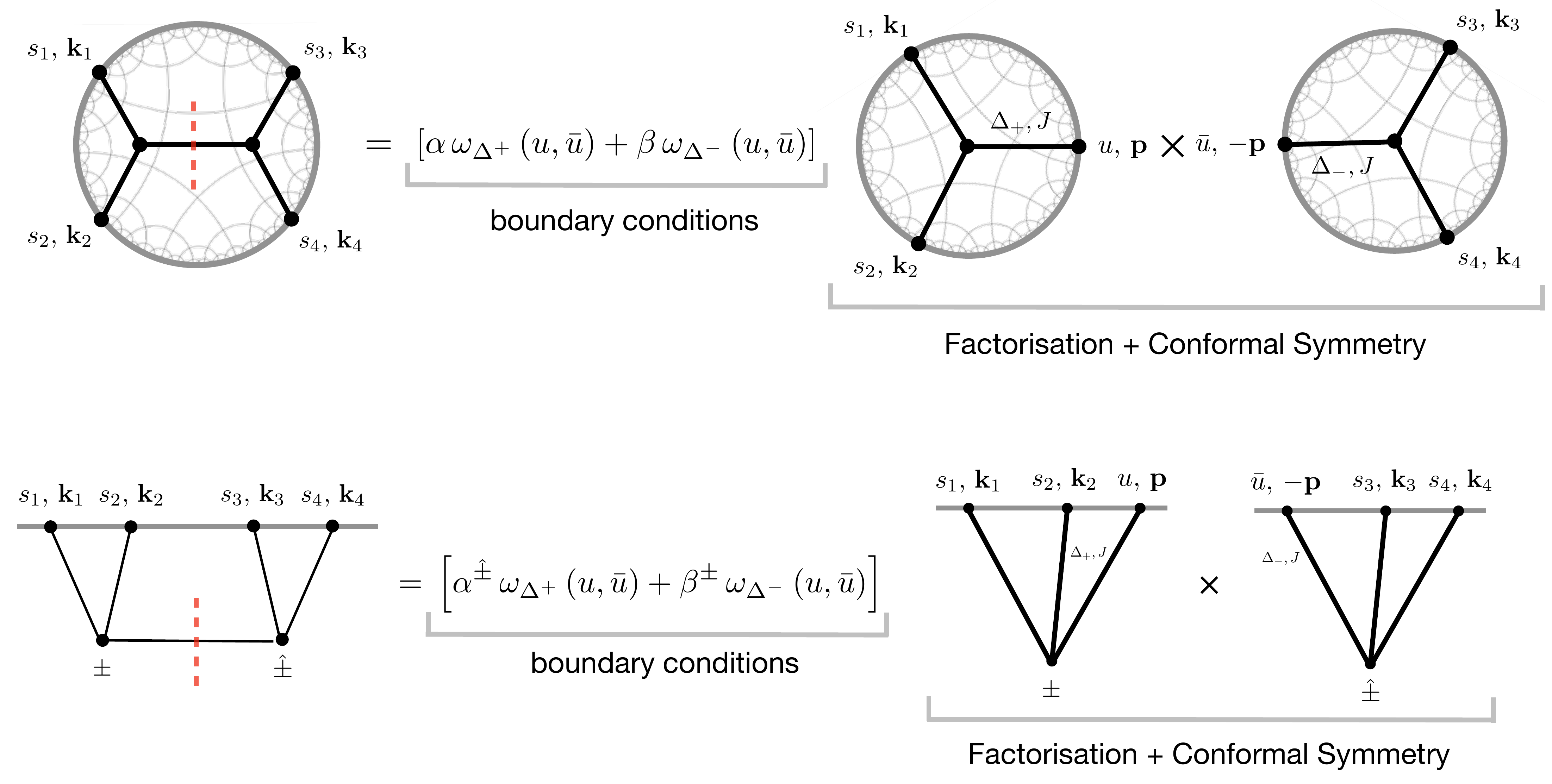}
    \caption{Exchanges both in EAdS and dS are determined by a combination of factorisation + conformal symmetry (which is ensured by an appropriately normalised conformal partial wave \eqref{cpwfsprod}) and boundary conditions which is implemented in the Mellin-Barnes representation by the projectors $\omega_{\Delta^\pm}\left(u,{\bar u}\right)$.}
    \label{fig::cuteachadsds}
\end{figure}

\paragraph{On-shell exchange in dS.} In dS the only difference with respect to the EAdS case considered above is that, guided by the properties of dS in-in propagators reviewed in section \ref{subsec::cuttinganddisp}, we require factorisation at the level of each in-in contribution. In particular, the exchange of a particle in dS decomposes as
\begin{equation}
    {\cal A}^{\text{dS}}_{\mu,J}\left({\bf k}_1,{\bf k}_2,{\bf p};-{\bf p},{\bf k}_3,{\bf k}_4\right)= \sum\limits_{\pm \, {\hat \pm}}\,\left(\pm i\right)\left({\hat \pm} i\right)\, {\cal A}^{\pm, {\hat \pm}}_{\mu,J}\left({\bf k}_1,{\bf k}_2,{\bf p};-{\bf p},{\bf k}_3,{\bf k}_4\right),
\end{equation}
where ${\cal A}^{\pm, {\hat \pm}}_{\mu,J}$ is the contribution from the $\pm \, {\hat \pm}$ branches of the in-in contour. We can now proceed to fix the on-shell exchange following the same steps as for the EAdS exchange above at the level of each in-in contribution, where the boundary condition is given by the requirement that the vacuum at early times is Bunch Davies -- which is selected by the coefficients \eqref{coeffsab} of the projectors $\omega_{\Delta^\pm}\left(u,{\bar u}\right)$.

In particular, \emph{conformal symmetry}, \emph{factorisation} and \emph{Bunch Davies vacuum} fixes the on-shell exchange in dS$_{d+1}$ to be \cite{Sleight:2020obc}
\begin{multline}\label{discbddspm2}
    \text{Disc}_{{\sf s}}\left[{\cal A}^{\pm, {\hat \pm}}_{\mu,J}\left(s_1,{\bf k}_1,s_2,{\bf k}_2,u,{\bf p};{\bar u}, -{\bf p},s_3,{\bf k}_3,s_4,{\bf k}_4\right)\right] = \underbrace{\left(\alpha^{{\hat \pm}}\, \omega_{\Delta^+}\left(u,{\bar u}\right)+\beta^{\pm}\, \omega_{\Delta^-}\left(u,{\bar u}\right)\right)}_{\text{boundary conditions}}\\ \times \frac{\Gamma\left(1+i\mu\right)\Gamma\left(1-i\mu\right)}{\pi}\underbrace{{\cal F}^{\,\pm, {\hat \pm}}_{\mu,J}\left(s_1,{\bf k}_1,s_2,{\bf k}_2,u,{\bf p};{\bar u}, -{\bf p},s_3,{\bf k}_3,s_4,{\bf k}_4\right)}_{\text{conformal symmetry and factorisation}},
\end{multline}
where ${\cal F}^{\,\pm, {\hat \pm}}_{\mu, J}$ is the dS-normalised conformal partial wave \eqref{cpwfsprod} on the $\pm\, {\hat \pm}$ branch:
\begin{multline}\label{pmdscpw}
    {\cal F}^{\pm, {\hat \pm}}_{\mu,J}\left(s_1,{\bf k}_1,s_2,{\bf k}_2,u,{\bf p};{\bar u}, -{\bf p},s_3,{\bf k}_3,s_4,{\bf k}_4\right) = F^{\pm}_{\Delta_1,\Delta_2,\Delta^+; J_1, J_2, J; n_1, n_2, n}\left(s_1,{\bf k}_1,s_2,{\bf k}_2,u,{\bf p}\right)\\ \times F^{{\hat \pm}}_{\Delta^-,\Delta_3,\Delta_4; J, J_3, J_4; {\bar n}, n_3, n_4}\left({\bar u},-{\bf p},s_3,{\bf k}_3,s_4,{\bf k}_4\right),
\end{multline}
and the constituent three-point boundary correlators are given by \eqref{spin3ptfromscse}. Note that this matches the result \eqref{discdsexch0} obtained in section \ref{subsec::cuttinganddisp} using the cutting rules.

\paragraph{Full exchange in (EA)dS.}

In the above we determined the on-shell exchange in EAdS \eqref{disceadsexch} and in the Bunch-Davies vacuum of dS \eqref{discbddspm2} by solving the conformal Casimir equation \eqref{ccasmir} and imposing factorisation and boundary conditions. The full exchange satisfies an \emph{inhomogeneous} confomal invariance constraint, which is the conformal Casimir equation \eqref{ccasmir} with a contact term source on the r.h.s. and is what enforces \emph{locality} of the solution.\footnote{The analogous constraint in flat space is what ensures tree exchange amplitudes have simple poles in the Mandelstam variables.} This is equivalent to the inhomogeneous conformal invariance constraint introduced in \cite{Arkani-Hamed:2015bza,Arkani-Hamed:2018kmz} for dS exchanges, which was solved directly in \cite{Arkani-Hamed:2018kmz} using a series expansion. It is furthermore the boundary counterpart of the inhomogeneous wave equation \eqref{bubuadseom} satisfied by the corresponding bulk-to-bulk propagators. 

As described in section \ref{subsec::cuttinganddisp}, alternatively one can obtain the full exchange from its on-shell part using the dispersion formula \eqref{dispersionadsdsexch}. This simply amounts to multiplying the on-shell exchange by the factor $\csc\left(\pi\left(u+{\bar u}\right)\right)$ at the level of the Mellin-Barnes representation. Before proceeding, since this dispersion formula was derived from the properties of (EA)dS bulk-to-bulk propagators, it is useful to first understand how it can be obtained from a boundary perspective. Let us suppose that that we have a function $f\left(z\right)$ which has singularities on the negative real axis. The standard dispersion relation argument implies:
\begin{align}
    f(a)=\oint \frac{dz}{2\pi i}\,\frac{f(z)}{z-a}=\int \frac{dz}{\pi }\,\frac1{z-a}\,\text{Disc}\,f(z) \,,
\end{align}
where we have assumed that the integral vanishes at infinity and where the last integral is assumed to be over the singularities of $f(z)$ with $\text{Disc}$ defined in \eqref{discs}. If one assumes the Mellin-Barnes representation of $\text{Disc}f(y)$ to be known,
\begin{align}
   \text{Disc}\, f(y)=\int_{-i\infty}^{+i\infty}\frac{ds}{2\pi i}\,\mathcal{M}(s)\,(-y)^{-s},\qquad y<0,
\end{align}
by exchanging the integral in $s$ and in $y$ we obtain:
\begin{align}
    f(z)=-\int_{-i\infty}^{+i\infty}\frac{ds}{2\pi i}\,\mathcal{M}(s)\int_{0}^{\infty} \frac{dy}{\pi }\,\frac{y^{-s}}{y+z}=\int_{-i\infty}^{+i\infty}\frac{ds}{2\pi i}\,\mathcal{M}(s)\csc(\pi s)\,z^{-s}\,.
\end{align}
In other words, at the level of the Mellin-Barnes representation, reconstructing a function from its discontinuity \eqref{discs} amounts to multiplying the discontinuity by a cosecant factor, with possible ambiguities arising from the choice in integration contour passing on the right or left of the poles present in the analyticity strip.

An example of the above Mellin space inversion is given by the function
\begin{align}
    f(z)=-\log(1-z^2)\,.
\end{align}
The Mellin transform of the above function can be easily evaluated and reads:
\begin{align}
    f(z)=-\int_{c-i\infty}^{c+i\infty}\frac{ds}{2\pi i}\,\frac{\pi}{s}\left(\cot \left(\frac{\pi  s}{2}\right)-i\right)\,z^{-s}\,,
\end{align}
whose strip of analyticity is $-2<c<0$. Formally the above integral is exponentially suppressed for $s=Re^{i\theta}$ and $R\to\infty$ only when
\begin{align}
    \arg (z) \sin (\theta )-\cos (\theta ) \log (\left| z\right| )<0\,.
\end{align}
By bending the integration contour at infinity towards the negative real axis the integral will remain exponentially suppressed also for $\arg{z}=\pm i\pi$. We can evaluate the discontinuity of the above function at the Mellin integral splitting the integrals into two terms:
\begin{multline}
    \text{Disc}f(y)=-\int_{c-i\infty}^{c+i\infty}\frac{ds}{2\pi i}\,\frac{\pi}{s}\left(1+e^{-i \pi  s}\right)\,y^{-s}\\=-\pi\int_{c-i\infty}^{c+i\infty}\frac{ds}{2\pi i}\,\frac{y^{-s}}{s}-\pi\int_{c-i\infty}^{c+i\infty}\frac{ds}{2\pi i}\,\frac{(-y)^{-s}}{s}=\pi\theta(|y|-1)\,,
\end{multline}
where the integration contour passes on the left of the pole $s\sim0$.
Note that in order to evaluate the second integral we assumed $-y>0$. Its now obvious that from the knowledge of the discontinuity of $f(z)$ and its Mellin representation, under suitable assumptions about the convergence of the corresponding Mellin integrals, we can obtain the original function Mellin representation multiplying by $\csc(\pi s)$ assuming that the function we are reconstructing is well-behaved at infinity. In particular we have in the above example:
\begin{align}
    \pi\theta(|y|-1)\rightarrow -\frac{\pi}{s}\left(1+e^{-i \pi  s}\right)\rightarrow -\frac{\pi\csc(\pi s)}{s}\left(1+e^{-i \pi  s}\right)\rightarrow -\log(1-z^2)\,,
\end{align}
where in the first step the integration contour is chosen to pass on the left of the pole $s\sim 0$ and the function can easily be obtained by closing the contour on the negative axis. Note that the choice of integration contour for the $\csc(\pi s)$ is ambiguous, as already stressed in \cite{Sleight:2019hfp}. However, there is a unique choice for which the reconstructed function is well behaved for $z\to\infty$. This also allows to unambiguously identify the polynomial ambiguity which, being singular at infinity, cannot be reconstructed by the dispersion relation argument.

The above discussion makes clear how the expression \eqref{dispprop} we have obtained for the bulk-to-bulk propagators in terms of the $\csc$ is actually a generic relation between a function and its discontinuity at the level of the Mellin-Barnes representation.

Coming back to the exchange diagrams, by imposing conformal symmetry, factorisation, boundary conditions and \emph{locality}, the full exchange in EAdS for a general linear combination of $\Delta^\pm$ boundary conditions on the exchanged field is then reconstructed from its discontinuity \eqref{disceadsexch} by multiplying it with $\csc\left(\pi\left(u+{\bar u}\right)\right)$  \cite{Sleight:2020obc}:
\begin{align}\label{fullAdSexchMB}
  &  {\cal A}^{\text{EAdS}}_{\alpha \Delta^++\beta \Delta^-,J}\left(s_1,{\bf k}_1,s_2,{\bf k}_2,u,{\bf p};{\bar u}, -{\bf p},s_3,{\bf k}_3,s_4,{\bf k}_4\right) = \underbrace{\csc\left(\pi\left(u+{\bar u}\right)\right)}_{\text{contact terms}} \\ & \hspace*{4cm}\times \underbrace{\text{Disc}_{{\sf s}}\left[{\cal A}^{\text{EAdS}}_{\alpha \Delta^++\beta \Delta^-,J}\left(s_1,{\bf k}_1,s_2,{\bf k}_2,u,{\bf p};{\bar u}, -{\bf p},s_3,{\bf k}_3,s_4,{\bf k}_4\right)\right]}_{\text{on-shell exchange}},\nonumber
\end{align}
and likewise for the exchange in the Bunch-Davies vacuum of dS we have \cite{Sleight:2020obc}
\begin{align}\label{dsexchBD}
& {\cal A}^{\text{dS}}_{\mu,J}\left(s_1,{\bf k}_1,s_2,{\bf k}_2,u,{\bf p};{\bar u}, -{\bf p},s_3,{\bf k}_3,s_4,{\bf k}_4\right)=\underbrace{\csc\left(\pi\left(u+{\bar u}\right)\right)}_{\text{contact terms}} \\ &
 \hspace*{3cm} \times \underbrace{\sum\limits_{\pm\,{\hat \pm}} \,\left(\pm i\right)\left({\hat \pm} i\right)\,  \text{Disc}_{{\sf s}}\left[{\cal A}^{\pm, {\hat \pm}}_{\mu,J}\left(s_1,{\bf k}_1,s_2,{\bf k}_2,u,{\bf p};{\bar u}, -{\bf p},s_3,{\bf k}_3,s_4,{\bf k}_4\right)\right]}_{
 \text{on-shell exchange}}. \nonumber
\end{align}

Various properties of the above Mellin-Barnes representation for (EA)dS exchanges were described in \cite{Sleight:2019mgd,Sleight:2019hfp}.

\subsection{dS exchanges as a linear combination of EAdS exchanges}
\label{subsec::dsasadsexch}

\begin{figure}[t]
    \centering
    \captionsetup{width=0.95\textwidth}
    \includegraphics[width=0.8\textwidth]{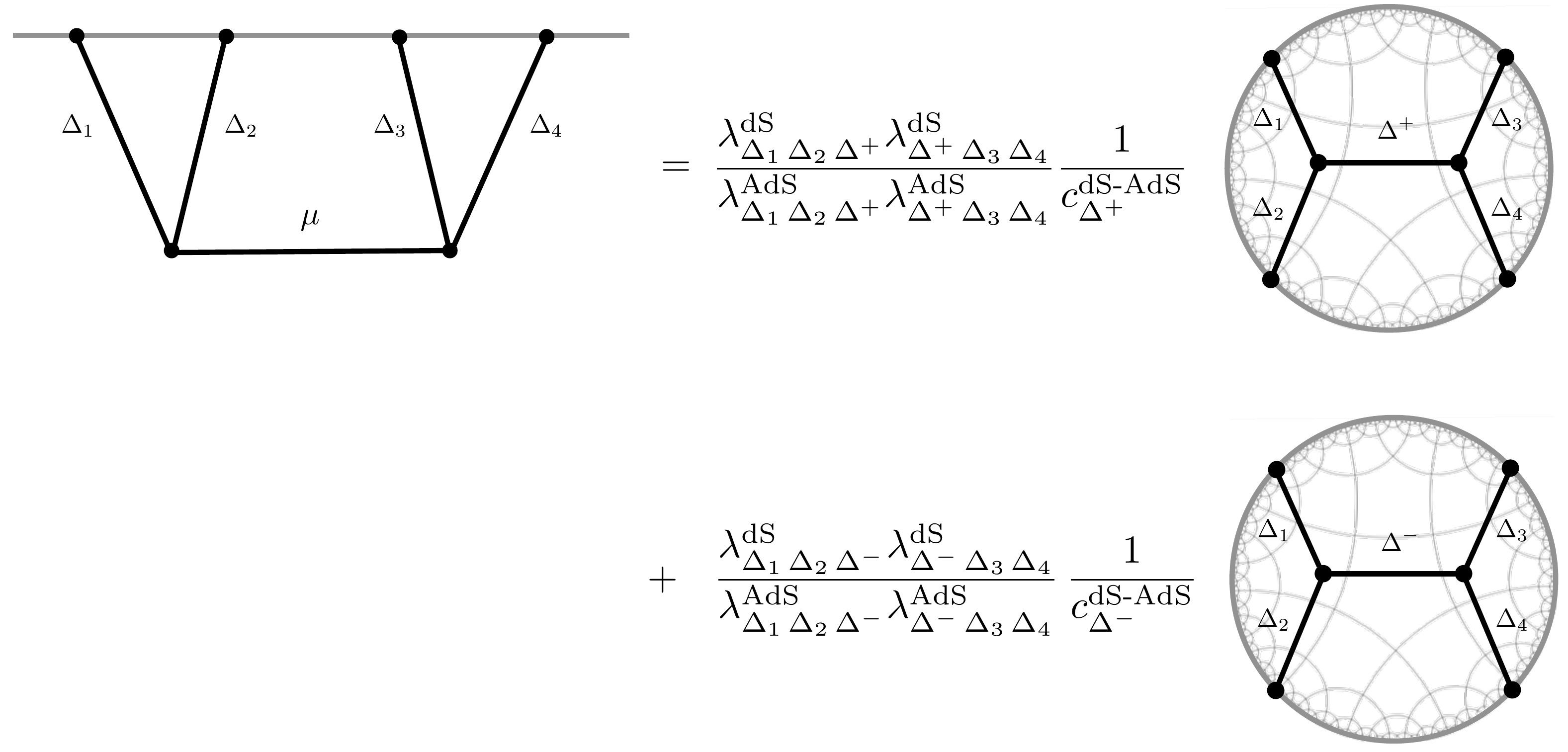}
    \caption{An exchange in the Bunch-Davies vacuum of dS can be expressed as a linear combination of EAdS exchanges with $\Delta^\pm$ boundary conditions. The factors $ \lambda^{\text{dS}}_{\Delta_{1,3}\,\Delta_{2,4}\,\Delta^\pm}/\lambda^{\text{AdS}}_{\Delta_{1,3}\,\Delta_{2,4}\,\Delta^\pm}$ account for the change in three-point coefficient from EAdS to dS and $c^{\text{dS-AdS}}_{\Delta^\pm}$ for the change in two-point coefficient for the exchanged field. Notice that this expression is symmetric under $\Delta^+ \leftrightarrow \Delta^-$.}
    \label{fig::exchdsasads}
\end{figure}

In section \ref{subsec::props(EA)dS} we reviewed the observation that diagrams contributing to a boundary correlator in the Bunch-Davies vacuum of dS$_{d+1}$ can be expressed as a linear combination of Witten diagrams in EAdS$_{d+1}$, which followed as an immediate consequence of the fact that in-in propagators in the Bunch-Davies vacuum of dS$_{d+1}$ are given by a specific linear combination of Wick rotated bulk-to-bulk propagators in EAdS$_{d+1}$ \cite{Sleight:2020obc}. While this statement followed from the properties of bulk-to-bulk propagators, it can also be understood from a purely boundary perspective. To see this it is sufficient to note that, in solving for the EAdS exchange and the $\pm {\hat \pm}$ contribution to the dS exchange in the previous section, the only difference between the two was the choice of normalisation for the Conformal Partial Wave \eqref{cpwfsprod} -- which is chosen to be consistent with on-shell factorisation into the constituent (EA)dS three-point boundary correlators. In section \ref{sec::3pt} we reviewed the fact that $\pm$ three-point boundary correlators in the Bunch-Davies vacuum of dS$_{d+1}$ differ from their EAdS Witten diagram counterpart by a constant phase, see equation \eqref{dsads3ptphase}. This in turn implies that the EAdS-normalised and $\pm\, {\hat \pm}$ dS-normalised conformal partial waves \eqref{EAdScpw} and \eqref{pmdscpw} are related as follows:
\begin{multline}
  {\cal F}^{\pm\, {\hat \pm}}_{\mu,J}=  c^{\text{dS-AdS}}_{ \Delta^+}c^{\text{dS-AdS}}_{\Delta^-}\left(\prod\limits^4_{i=1} c^{\text{dS-AdS}}_{\Delta_i} \right)\\ \times \,e^{\mp \left(\tfrac{-d+\Delta_1+\Delta_2+\Delta^++J_1+J_2+J}{2}\right)\pi i}e^{{\hat \mp} \left(\tfrac{-d+\Delta^-+\Delta_3+\Delta_4+J+J_3+J_4}{2}\right)\pi i}\,{\cal F}^{\text{EAdS}}_{\mu,J}.
\end{multline}
This gives the $\pm\, {\hat \pm}$ contribution to the dS exchange \eqref{dsexchBD} as the following linear combination of EAdS exchanges with $\Delta^\pm$ boundary conditions:
\begin{multline}\label{dSasAdSexchpm}
   \hspace*{-0.75cm} {\cal A}^{\pm\, {\hat \pm}}_{\mu,J}=\left(\prod\limits^4_{i=1} c^{\text{dS-AdS}}_{\Delta_i} \right) \left[c^{\text{dS-AdS}}_{\Delta^+}e^{\mp \left(\tfrac{-d+\Delta_1+\Delta_2+\Delta^++J_1+J_2+J}{2}\right)\pi i}e^{{\hat \mp} \left(\tfrac{-d+\Delta^++\Delta_3+\Delta_4+J+J_3+J_4}{2}\right)\pi i} {\cal A}^{\text{EAdS}}_{\Delta^+,J}\,\right.\\\left.+\,\left(\Delta^+ \to \Delta^-\right)\right],
\end{multline}
where we used that in the Bunch-Davies vacuum the coefficients $\alpha^{{\hat \pm}}$ and $\beta^{\pm}$ onto the $\Delta^\pm$ modes are given by equation \eqref{coeffsab}. Summing the contributions from all branches of the in-in contour the full dS exchange reads (see figure \ref{fig::exchdsasads}):\footnote{Note that the term with ${\cal A}^{\text{EAdS}}_{\Delta^-,J}$ is obtained from that with ${\cal A}^{\text{EAdS}}_{\Delta^+,J}$ on the left by simply interchanging $\Delta^+ \leftrightarrow \Delta^-$.}
\begin{shaded}
\begin{equation}\label{dSasAdSexch}
    {\cal A}^{\text{dS}}_{\mu,J} =  \frac{\lambda^{\text{dS}}_{\Delta_1\,\Delta_2\,\Delta^+}\lambda^{\text{dS}}_{\Delta^+\,\Delta_3\,\Delta_4}}{\lambda^{\text{AdS}}_{\Delta_1\,\Delta_2\,\Delta^+}\lambda^{\text{AdS}}_{\Delta^+\,\Delta_3\,\Delta_4}}\frac{1}{c^{\text{dS-AdS}}_{\Delta^+}}\,{\cal A}^{\text{EAdS}}_{\Delta^+,J}+\frac{\lambda^{\text{dS}}_{\Delta_1\,\Delta_2\,\Delta^-}\lambda^{\text{dS}}_{\Delta^-\,\Delta_3\,\Delta_4}}{\lambda^{\text{AdS}}_{\Delta_1\,\Delta_2\,\Delta^-}\lambda^{\text{AdS}}_{\Delta^-\,\Delta_3\,\Delta_4}}\,\frac{1}{c^{\text{dS-AdS}}_{\Delta^-}}{\cal A}^{\text{EAdS}}_{\Delta^-,J},
\end{equation}
\end{shaded}
\noindent where the constant phases that weight each contribution \eqref{dSasAdSexchpm} from the in-in contour combine to give the constant sinusoidal factors \eqref{spinning3pt} that take us from the AdS three-point coefficients $\lambda^{\text{AdS}}_{\Delta_1\,\Delta_2\,\Delta^\pm}$ and $\lambda^{\text{AdS}}_{\Delta^\pm\,\Delta_3\,\Delta_4}$ to the dS three-point coefficients $\lambda^{\text{dS}}_{\Delta_1\,\Delta_2\,\Delta^\pm}$ and $\lambda^{\text{dS}}_{\Delta^\pm\,\Delta_3\,\Delta_4}$: 
\begin{multline}
    \frac{\lambda^{\text{dS}}_{\Delta_{1,3}\,\Delta_{2,4}\,\Delta^\pm}}{  \lambda^{\text{AdS}}_{\Delta_{1,3}\,\Delta_{2,4}\,\Delta^\pm}}= 2\,c^{\text{dS-AdS}}_{ \Delta_{1,3}}\,c^{\text{dS-AdS}}_{ \Delta_{2,4}}\,c^{\text{dS-AdS}}_{ \Delta^\pm}\\ \times \sin \left(\frac{-d+\Delta_{1,3}+\Delta_{2,4}+\Delta^\pm+J_{1,3}+J_{2,4}+J}{2}\right)\pi\,.
\end{multline}
Likewise, the factors $c^{\text{dS-AdS}}_{\Delta^\pm}$ account for the change in two-point function coefficient from EAdS to dS. The result \eqref{dSasAdSexch} gives the extension of the results presented in section 4 of \cite{Sleight:2020obc} to any cubic coupling of the exchanged spin-$J$ field to external fields of arbitrary integer spins $J_{1,2,3,4}$.

\subsection{Factorisation in dS revisited, unitarity and the conformal block expansion}

\label{subsec::factunitcbe}

In the previous sections we have seen how diagrams contributing to boundary correlators in (EA)dS factorise when an internal leg is placed on-shell. This follows from the properties of bulk-to-bulk propagators described in section \ref{subsec::cuttinganddisp} and in the dS case, in particular, is at the level of each \emph{individual} contribution to the in-in contour. It is compelling to understand such properties at the level of the full diagram, since the in-in formalism only constitutes an auxiliary step to obtain the observable. Given the general expression \eqref{dSasAdSexch} for the full exchange in dS, let us explore this in the following.

It is instructive to study factorisation at the level of the conformal block \eqref{gensol} expansion. The conformal blocks  $W_{\Delta^\pm,J}$ encode the contribution from the exchanged $\Delta^\pm$ mode in the ${\sf s}$-channel. In AdS this is well understood \cite{Heemskerk:2009pn,ElShowk:2011ag,Alday:2017gde,Meltzer:2019nbs}, where the on-shell exchange for the $\Delta^\pm$ boundary condition is given by\footnote{Note that in momentum space this amounts to taking the discontinuity \eqref{discs}.}
\begin{equation}\label{osexchcb}
    {\cal A}^{\text{EAdS}}_{\Delta^\pm,\,J}\Bigg|_{\text{on-shell}} = \frac{\lambda^{\text{AdS}}_{\Delta_1\,\Delta_2\, \Delta^\pm}\,\lambda^{\text{AdS}}_{ \Delta^\pm\, \Delta_3\,\Delta_4}}{C^{\text{AdS}}_{\Delta^\pm,J}}\,W_{\Delta^\pm,J},
\end{equation}
where the coefficient is factorised into the three-point coefficients $\lambda^{\text{AdS}}_{\Delta_1\,\Delta_2\, \Delta^\pm}$ and $\lambda^{\text{AdS}}_{ \Delta^\pm\, \Delta_3\,\Delta_4}$ of the contact diagrams generated by the cubic vertices that mediate the $\Delta^\pm$ exchange and $C^{\text{AdS}}_{\Delta,J}$ is the coefficient of the tree-level two-point function in AdS \eqref{normbubo}. This is how the factorisation illustrated in figure \ref{fig::exchdiscfact}  manifests itself at the level of the conformal block expansion. In particular, note that \eqref{osexchcb} can be obtained from the Mellin-Barnes representation of the on-shell exchange \eqref{disceadsexch} using that  
\begin{multline}
    \omega_{\Delta^\pm}\left(u,{\bar u}\right)\,{\cal F}_{\mu,J}\left(s_1,{\bf k}_1,s_2,{\bf k}_2,u,{\bf p};{\bar u}, -{\bf p},s_3,{\bf k}_3,s_4,{\bf k}_4\right)\\ \propto \, W_{\Delta^\pm,J}\left(s_1,{\bf k}_1,s_2,{\bf k}_2,u,{\bf p};{\bar u}, -{\bf p},s_3,{\bf k}_3,s_4,{\bf k}_4\right),
\end{multline}
where the projectors $\omega_{\Delta^\pm}\left(u,{\bar u}\right)$ select the $\Delta^\pm$ mode, as they did at the level of the propagators in section \ref{subsec::props(EA)dS}.

Using the expression \eqref{dSasAdSexch} for the full dS exchange as a linear combination of EAdS exchanges, we can write the analogous statement for dS exchanges in the Bunch-Davies vacuum:
 \begin{equation}\label{osexchds}
     {\cal A}^{\text{dS}}_{\mu,\,J}\Bigg|_{\text{on-shell}} = \frac{\lambda^{\text{dS}}_{\Delta_1\,\Delta_2\, \Delta^+}\,\lambda^{\text{dS}}_{ \Delta^+\, \Delta_3\,\Delta_4}}{C^{\text{dS}}_{\Delta^+,J}}\,W_{\Delta^+,J}+\frac{\lambda^{\text{dS}}_{\Delta_1\,\Delta_2\, \Delta^-}\,\lambda^{\text{dS}}_{ \Delta^-\, \Delta_3\,\Delta_4}}{C^{\text{dS}}_{\Delta^-,J}}\,W_{\Delta^-,J},
 \end{equation}
 which is a linear combination of conformal blocks for the exchanged $\Delta^\pm$ modes, where each coefficient is factorised into the three-point coefficients $\lambda^{\text{dS}}_{\Delta_1\,\Delta_2\, \Delta^\pm}$ and $\lambda^{\text{dS}}_{ \Delta^\pm\, \Delta_3\,\Delta_4}$ of the contact diagrams generated by the cubic vertices that mediate the $\Delta^\pm$ exchange. This is factorisation at the level of the full dS exchange, as opposed to factorisation at the level of each in-in contribution as considered previously. 
 
 Let us note that unitarity in AdS constrains the coefficients of the conformal block encoding the on-shell exchange in \eqref{osexchcb} to be real. This follows from the fact that unitarity in AdS requires $\lambda^{\text{AdS}}_{\Delta_1 \Delta_2 \Delta_2} \in \mathbb{R}$ \cite{Rattazzi:2008pe}. In the case of equal external fields this furthermore implies that the coefficient is positive. As reviewed in section \ref{subsec::contactunitarity}, in dS the three-point coefficients are no longer required to be real and at linear order in the coupling (as is the case for the exchange above) unitary time evolution is encoded in the sinusoidal factors \eqref{uniacoupdsads}. In particular,
 \begin{multline}
   \hspace*{-0.5cm}  \frac{\lambda^{\text{dS}}_{\Delta_1\,\Delta_2\, \Delta^\pm}\,\lambda^{\text{dS}}_{ \Delta^\pm\, \Delta_3\,\Delta_4}}{C^{\text{dS}}_{\Delta^\pm,J}} = 4\left(\prod\limits^4_{j=1} c^{\text{dS-AdS}}_{\Delta_j} \right)c^{\text{dS-AdS}}_{ \Delta^\pm}\,\\ \times \sin \left(\left(\tfrac{\Delta_1+\Delta_2+\Delta^\pm+J_1+J_2+J-d}{2}\right)\pi\right)\,\sin \left(\left(\tfrac{\Delta_3+\Delta_4+\Delta^\pm+J_3+J_4+J-d}{2}\right)\pi\right) \frac{\lambda^{\text{AdS}}_{\Delta_1\,\Delta_2\, \Delta^\pm}\,\lambda^{\text{AdS}}_{ \Delta^\pm\, \Delta_3\,\Delta_4}}{C^{\text{AdS}}_{\Delta^\pm,J}}.
 \end{multline}
 Note that, as for the contact diagrams discussed in section \ref{subsec::contactunitarity}, such sine factors seem to open up the the possibility that dS exchange diagrams vanish for certain values of the scaling dimensions, spins and the boundary dimension $d$. It would be interesting to understand if/how such products of sine factors \eqref{spinning3pt} could come out of the non-perturbative unitarity constraints very recently proposed in \cite{Hogervorst:2021uvp,DiPietro:2021sjt}.
 
Above we have given the on-shell exchange in dS in terms of conformal blocks, which encode the contribution from the physical exchanged single particle state. It is important to appreciate that the full exchange in dS also admits a conformal block expansion (\cite{Sleight:2020obc}, section 5) which, via the identity \eqref{dSasAdSexch}, can be obtained from the known conformal block decomposition of the exchanges in EAdS. The latter takes the form:
\begin{multline}\label{adsfullcbd}
    {\cal A}^{\text{EAdS}}_{\Delta^\pm,\,J} = \frac{\lambda^{\text{AdS}}_{\Delta_1\,\Delta_2\, \Delta^\pm}\,\lambda^{\text{AdS}}_{ \Delta^\pm\, \Delta_3\,\Delta_4}}{C^{\text{AdS}}_{\Delta^\pm,J}}\,W_{\Delta^\pm,J}\\+\sum\limits^J_{J^\prime=0}\sum\limits^
    \infty_{n=0}\,{}^{(\sf s)}\,a_{n,J^
    \prime}W_{\Delta_1+\Delta_2+2n+J^\prime,J^\prime}+\sum\limits^J_{J^\prime=0}\sum\limits^
    \infty_{n=0}\,{}^{(\sf s)}\,b_{n,J^
    \prime}W_{\Delta_3+\Delta_4+2n+J^\prime,J^\prime},
\end{multline}
where the on-shell exchanges \eqref{osexchcb} are completed by an infinite tail of conformal blocks corresponding to the exchange of double-trace operators with scaling dimensions $\Delta_{1,3}+\Delta_{2,4}+2n+J^\prime$, which encode the contact terms of the exchange. The conformal block coefficients ${}^{(\sf s)}\,a_{n,J^
    \prime}$ and ${}^{(\sf s)}\,b_{n,J^
    \prime}$ are the same for both the $\Delta^\pm$ EAdS exchanges and have been computed e.g. in \cite{El-Showk:2011yvt,Alday:2017gde,Gopakumar:2016cpb,Zhou:2018sfz,Gopakumar:2018xqi,Sleight:2019ive} using a variety of techniques. Using the identity 
    \eqref{dSasAdSexch} we can then immediately write down the conformal block expansion of the full dS exchange:
\begin{multline}\label{cbedsexch}
    {\cal A}^{\text{dS}}_{\mu,J}= \frac{\lambda^{\text{dS}}_{\Delta_1\,\Delta_2\, \Delta^+}\,\lambda^{\text{dS}}_{ \Delta^+\, \Delta_3\,\Delta_4}}{C^{\text{dS}}_{\Delta^+,J}}\,W_{\Delta^+,J}+\frac{\lambda^{\text{dS}}_{\Delta_1\,\Delta_2\, \Delta^-}\,\lambda^{\text{dS}}_{ \Delta^-\, \Delta_3\,\Delta_4}}{C^{\text{dS}}_{\Delta^-,J}}\,W_{\Delta^-,J}\\
  +  2\left(\prod\limits^4_{i=1} c^{\text{dS-AdS}}_{\Delta_i}\right)\sin\left(\left(\tfrac{-d+2J+\Delta_1+\Delta_2+\Delta_3+\Delta_4+J_1+J_2+J_3+J_4}{2}\right)\pi\right)\\
  \times \left[\sum\limits^J_{J^\prime=0}\sum\limits^
    \infty_{n=0}\,{}^{(\sf s)}\,a_{n,J^
    \prime}W_{\Delta_1+\Delta_2+2n+J^\prime,J^\prime}+\sum\limits^J_{J^\prime=0}\sum\limits^
    \infty_{n=0}\,{}^{(\sf s)}\,b_{n,J^
    \prime}W_{\Delta_3+\Delta_4+2n+J^\prime,J^\prime}\right],
\end{multline}
where we used that 
\begin{multline}\label{3ptto4pttriid}
    \frac{\lambda^{\text{dS}}_{\Delta_1\,\Delta_2\,\Delta^+}\lambda^{\text{dS}}_{\Delta^+\,\Delta_3\,\Delta_4}}{\lambda^{\text{AdS}}_{\Delta_1\,\Delta_2\,\Delta^+}\lambda^{\text{AdS}}_{\Delta^+\,\Delta_3\,\Delta_4}}\frac{1}{c^{\text{dS-AdS}}_{\Delta^+}}\,+\frac{\lambda^{\text{dS}}_{\Delta_1\,\Delta_2\,\Delta^-}\lambda^{\text{dS}}_{\Delta^-\,\Delta_3\,\Delta_4}}{\lambda^{\text{AdS}}_{\Delta_1\,\Delta_2\,\Delta^-}\lambda^{\text{AdS}}_{\Delta^-\,\Delta_3\,\Delta_4}}\,\frac{1}{c^{\text{dS-AdS}}_{\Delta^-}}\\ =  2\left(\prod\limits^4_{i=1} c^{\text{dS-AdS}}_{\Delta_i}\right)\sin\left(\left(\tfrac{-d+2J+\Delta_1+\Delta_2+\Delta_3+\Delta_4+J_1+J_2+J_3+J_4}{2}\right)\pi\right).
\end{multline}
Notice that the additional terms in the conformal block expansion of the full exchange \eqref{cbedsexch} with respect to the on-shell part \eqref{osexchds} are given by the contact terms in the EAdS exchanges \eqref{osexchds} multiplied by the sinusoidal factor \eqref{uniacoupdsads}. This is consistent with the interpretation of such additional terms as contact terms in the dS exchange, where the sinusoidal factor \eqref{uniacoupdsads} corrects for the change in coefficient of contact diagrams as we go from EAdS to dS, as described in section \ref{sec::3pt}. 

In a similar spirit, via the identity \eqref{dSasAdSexch} one can obtain the conformal block decomposition of the dS exchanges in the crossed channels, which is inherited from the known \cite{El-Showk:2011yvt,Gopakumar:2016cpb,Alday:2017gde,Sleight:2018ryu,Liu:2018jhs,Zhou:2018sfz,Albayrak:2019gnz,Gopakumar:2018xqi,Sleight:2019ive} crossing decompositions of EAdS exchanges. This was described in \cite{Sleight:2020obc} (section 5) and will not be discussed further here.

\section{dS to AdS rules}
\label{sec::generalalg}

In section \ref{sec::adsdsdict} have seen that dS diagrams in the Bunch-Davies vacuum can be expressed a linear combination of EAdS Witten diagrams for the $\Delta^\pm$ boundary conditions, which followed from the fact that dS in-in propagators can be expressed as a linear combination of analytically continued $\Delta^\pm$ EAdS propagators \cite{Sleight:2020obc}. In the previous sections this has been applied to contact diagrams and four-point exchanges at tree level, extending previous results to include arbitrary spinning fields. The precise relative coefficients of the EAdS Witten diagrams that give rise to the corresponding diagram in dS are, however, not manifest from such properties of the dS in-in propagators: It is necessary to run the machinery of the Schwinger-Keldysh formalism as an auxiliary step, adding each contribution from the in-in contour. In this section we therefore provide a set of rules which allow us to immediately write down the precise linear combination to EAdS Witten diagrams that give the corresponding diagram in the Bunch-Davies vacuum of dS, bypassing the Schwinger-Keldysh formalism. We will see that such rules can be understood from a boundary/bootstrap perspective, with the coefficients weighting the EAS Witten diagrams following from requirement of consistent factorisation and boundary conditions of the diagram in dS. From this Bootstrap perspective, the statement that dS diagrams can be expressed a linear combination of EAdS Witten diagrams is also a statement about the analyticity properties of dS diagrams in the Bunch-Davies vacuum -- which are being inherited from those their constituent EAdS Witten diagrams. This observation was also made in the companion paper \cite{Sleight:2020obc} and is elaborated on in section \ref{sec::analyticity}.

To establish such a set of dS to EAdS Witten diagram rules, it is instructive to first write down how in practise each in-in diagram contribution to a full diagram in dS gets decomposed into a linear combination of corresponding EAdS Witten diagrams, which we can do using the properties of (EA)dS propagators reviewed in section \ref{sec::adsdsdict}:
\begin{enumerate}
    \item For a given in-in diagram, where each vertex is sitting either on the $+$ or $-$ branch of the in-in contour, decompose each internal leg into its individual $\Delta^+$ and $\Delta^-$ mode contributions:
    \begin{subequations}
    \begin{align}
   \hspace*{-0.25cm} \Pi^{\pm,\, {\hat \pm}}_{\mu,J}\left(u,{\bf p};{\bar u},-{\bf p}\right) &= \alpha^{{\hat \pm}}\,\Pi^{\pm,\, {\hat \pm}}_{\Delta^+,J}\left(u,{\bf p};{\bar u},-{\bf p}\right)+\beta^{\pm}\,\Pi^{\pm,\, {\hat \pm}}_{\Delta^-,J}\left(u,{\bf p};{\bar u},-{\bf p}\right),\\
     \hspace*{-0.25cm} \Pi^{\pm,\, {\hat \pm}}_{\Delta,J}\left(u,{\bf p};{\bar u},-{\bf p}\right)&=\csc\left(\pi\left(u+{\bar u}\right)\right) \omega_{\Delta}\left(u,{\bar u}\right) \Gamma\left(i\mu\right)\Gamma\left(-i\mu\right) \Omega^{\pm,\, {\hat \pm}}_{\mu,J}\left(u,{\bf p};{\bar u},-{\bf p}\right).
\end{align}
 \end{subequations}
This can be represented diagrammatically as:
\begin{equation}
    \includegraphics[width=0.9\textwidth]{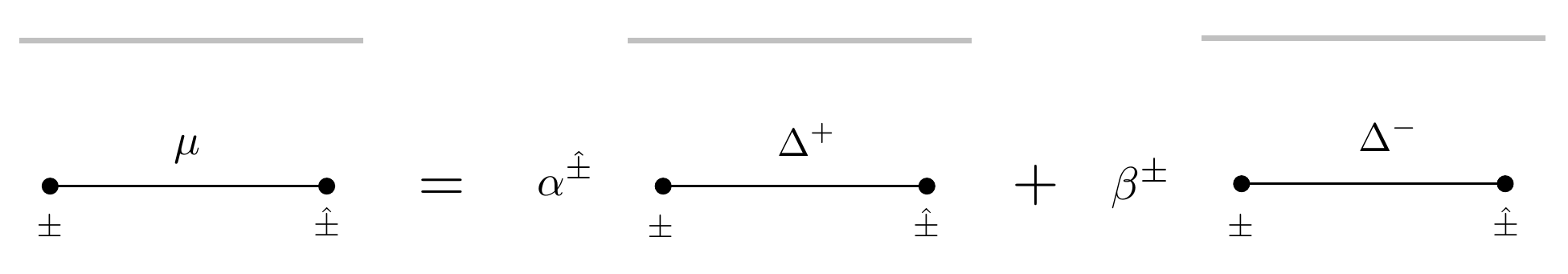}.
\end{equation}
\item At the level of the Mellin-Barnes representation, the $\Delta^\pm$ mode contributions to each dS in-in propagator differ from the corresponding $\Delta^\pm$ bulk-to-bulk propagator in EAdS by the following phase:\footnote{Note that is simply the analytic continuation \eqref{analcontbubu} of the $\Delta^\pm$ EAdS propagators to the $\pm\, {\hat \pm}$ branches of the in-in contour in dS according to the Wick rotations \eqref{wickinin}.} 
\begin{subequations}
\begin{align}
    \alpha^{{\hat \pm}}\,\Pi^{\pm,\, {\hat \pm}}_{\Delta^+,J}\left(u,{\bf p};{\bar u},-{\bf p}\right) &= c^{\text{dS-AdS}}_{\Delta^+}\, e^{\mp \left(u+\frac{i\mu}{2}\right)\pi i}e^{{\hat \mp} \left({\bar u}+\frac{i\mu}{2}\right)\pi i}\,\Pi^{\text{AdS}}_{\Delta^+,J}\left(u,{\bf p};{\bar u},-{\bf p}\right),\\
    \beta^{\pm}\,\Pi^{\pm,\, {\hat \pm}}_{\Delta^-,J}\left(u,{\bf p};{\bar u},-{\bf p}\right) &= c^{\text{dS-AdS}}_{\Delta^-}\,e^{\mp \left(u-\frac{i\mu}{2}\right)\pi i}e^{{\hat \mp} \left({\bar u}-\frac{i\mu}{2}\right)\pi i}\,\Pi^{\text{AdS}}_{\Delta^-,J}\left(u,{\bf p};{\bar u},-{\bf p}\right),
\end{align}
\end{subequations}
where $\Delta^\pm=\frac{d}{2}\pm i\mu$. Diagrammatically we can therefore make the following replacements for each $\Delta^\pm$ contribution to a given internal leg:
\begin{equation}
    \includegraphics[width=0.8\textwidth]{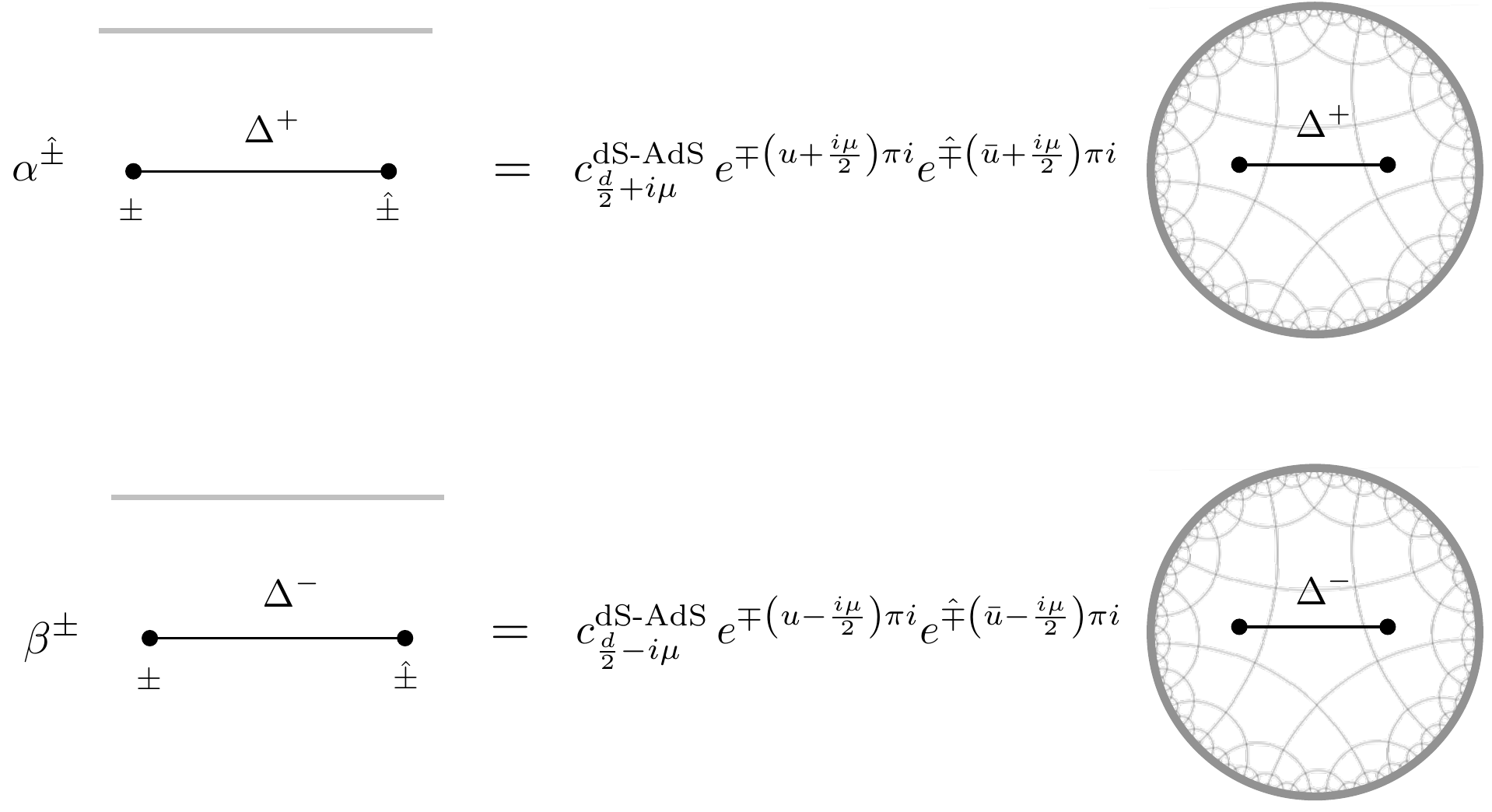}.
\end{equation}
\item Steps 1 and 2 above express each internal leg in a dS in-in diagram as a linear combination of internal legs in EAdS with $\Delta^\pm$ boundary conditions. Similarly, for each external leg we apply the identity \eqref{buborel} which expresses it as a phase multiplying the corresponding EAdS bulk-to-boundary propagator:
\begin{equation}\label{analcontbubufig}
    \includegraphics[width=0.8\textwidth]{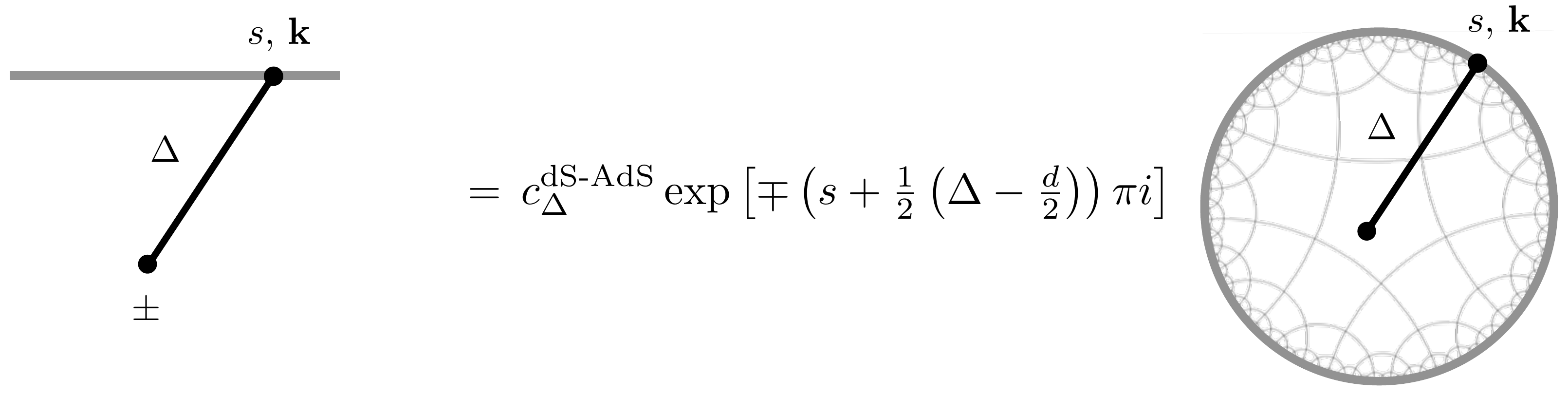}.
\end{equation}

\item The above steps reduce a given in-in diagram in dS to a linear combination of diagrams in EAdS with definite $\Delta^\pm$ boundary condition, where each vertex is dressed by the following phase: 
\begin{equation}\label{fig::vertexdstoAdS}
    \includegraphics[width=0.8\textwidth]{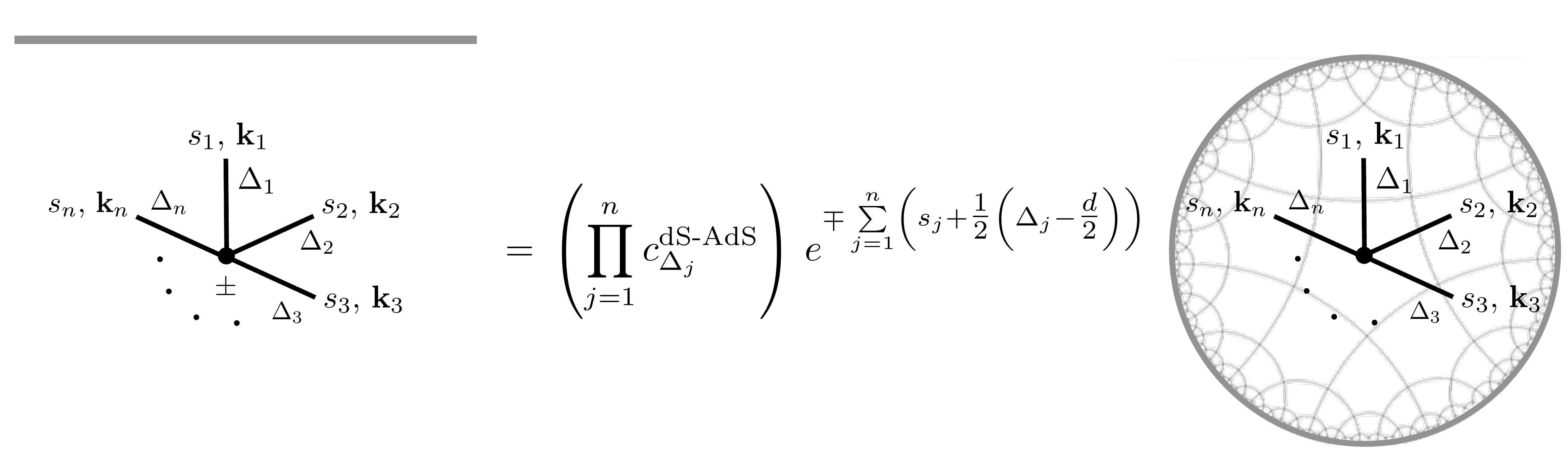}.
\end{equation}

Recall that at each vertex there is a Dirac delta function \eqref{ddmb} which constrains the sum of Mellin variables associated to the vertex to be a constant, by virtue of scale symmetry. See section \ref{sec::3pt}. For each vertex the phase \eqref{fig::vertexdstoAdS} is therefore constant and reads:
\begin{equation}\label{fig::vertexdstoAdSconst}
   \hspace*{-0.5cm} \includegraphics[width=0.9\textwidth]{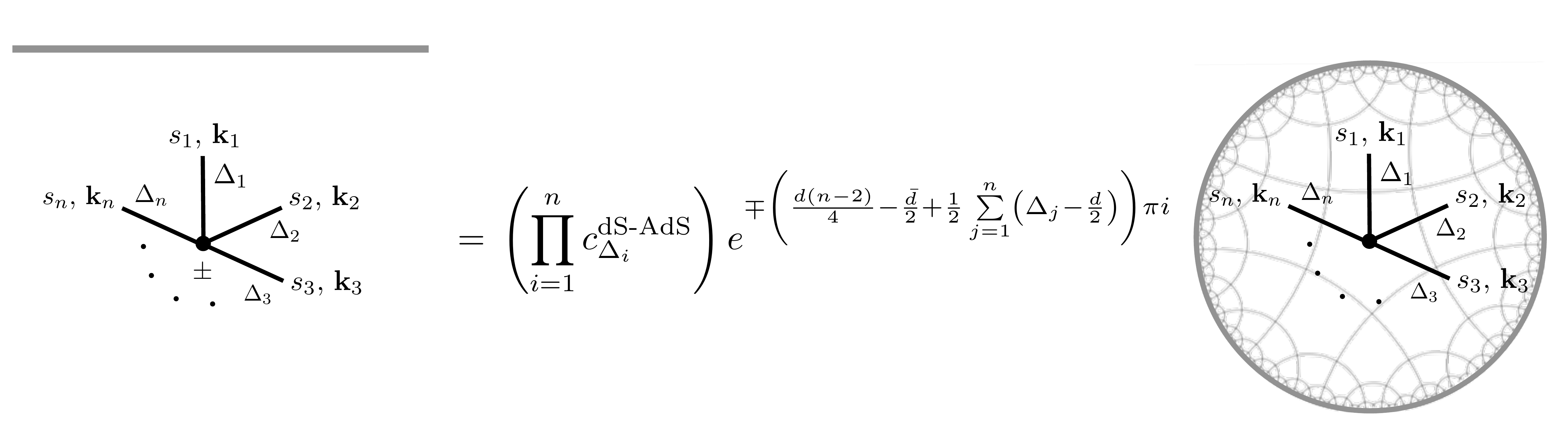}.
\end{equation}
\item To obtain the full dS diagram one sums over each in-in contribution above after multiplying each vertex by $\left(\pm i\right)$, where the $\pm$ refers to the branch of the in-in contour that the vertex is sitting on. In the full dS diagram the constant phases \eqref{fig::vertexdstoAdS} that dress each in-in contribution combine to give the sinusoidal factor \eqref{npointsine} that converts a contact diagram in EAdS to a contact boundary correlator in dS, as detailed in section \ref{sec::3pt}. Diagrammatically we can represent this as
\begin{equation}
    \includegraphics[width=0.65\textwidth]{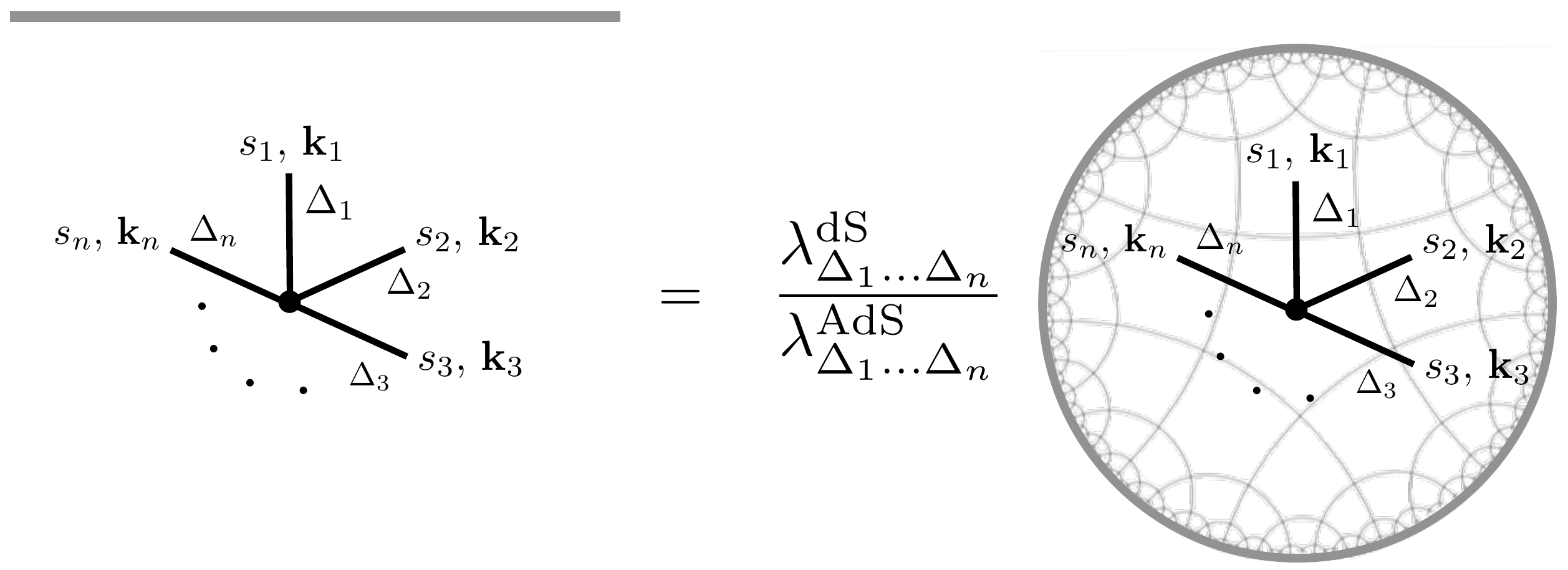},
\end{equation}
where in this diagram we have summed over all in-in contributions and:
\begin{equation}\label{sinereviw}
    \lambda^{\text{dS}}_{
\Delta_1, \ldots , \Delta_n}/\lambda^{\text{AdS}}_{
\Delta_1, \ldots , \Delta_n} =  2\left(\prod\limits^n_{i=1} c^{\text{dS-AdS}}_{\Delta_i} \right)\sin \left(\tfrac{d(n-2)}{4}-\tfrac{\bar d}{2}+\tfrac{1}{2}\sum\limits^n_{i=1}\left(\Delta_i-\tfrac{d}{2}\right)\right)\pi.
\end{equation}
 Note that for each internal line we must divide by the factor $c^{\text{dS-AdS}}_{\Delta}$, as in equation \eqref{dSasAdSexch} for the example of the exchange. This is because internal lines connect two vertices and the relation \eqref{analcontbubufig} between internal lines in dS and dS only contains a single factor of $c^{\text{dS-AdS}}_{\Delta}$. 
\end{enumerate}

In summary, the above steps give a general algorithm to decompose any diagram contributing to a boundary correlator in the Bunch-Davies vacuum of dS as a linear combination of corresponding EAdS Witten diagrams, using the in-in formalism as an auxiliary step. While this was clear from the properties of dS propagators reviewed in section \ref{subsec::props(EA)dS}, what we learn is that each dS diagram decomposes as a sum of EAdS diagrams over all possible combinations of $\Delta^\pm$ boundary conditions on the internal legs in a manner that is symmetric under $\Delta^+ \leftrightarrow \Delta^-$ where, in particular, each vertex is multiplied by the sinusoidal factor \eqref{sinereviw} that accounts for the change in the coefficient of contact subdiagrams as we move from AdS to dS. 

These observations naturally suggest a set of rules which we can apply to immediately write down the decomposition of a dS diagram in the Bunch-Davies vacuum as a linear combination of EAdS Witten diagrams --- without making any reference to the in-in formalism! These are as follows: 

\paragraph{dS to EAdS rules.}

\begin{enumerate}
    \item For a given dS diagram, first draw the same diagram in EAdS: Each external line connected to the boundary of dS becomes an external line connected to the boundary of AdS. Each internal line in dS becomes an internal line in AdS. Accordingly, each vertex in dS is translated to the same vertex in AdS. For example, if we were considering the exchange diagrams of section \ref{sec::4ptexch} this would read:
\begin{equation}
    \includegraphics[width=0.5\textwidth]{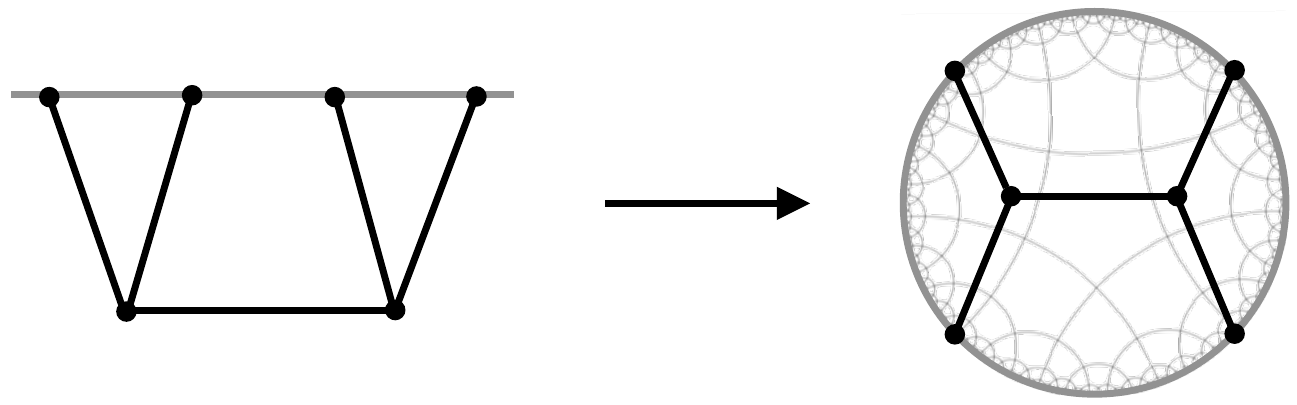},
\end{equation} 
\item Pick an internal line of the AdS diagram and assign to it the $\Delta^+$ related to the mass of the corresponding particle in dS. Divide by $c^{\text{dS-AdS}}_{\Delta^+}$. Proceed to the next internal leg and follow the same procedure until all internal legs have been exhausted. Continuing the example of the exchange diagram above this is simply:
\begin{equation}
    \includegraphics[width=0.9\textwidth]{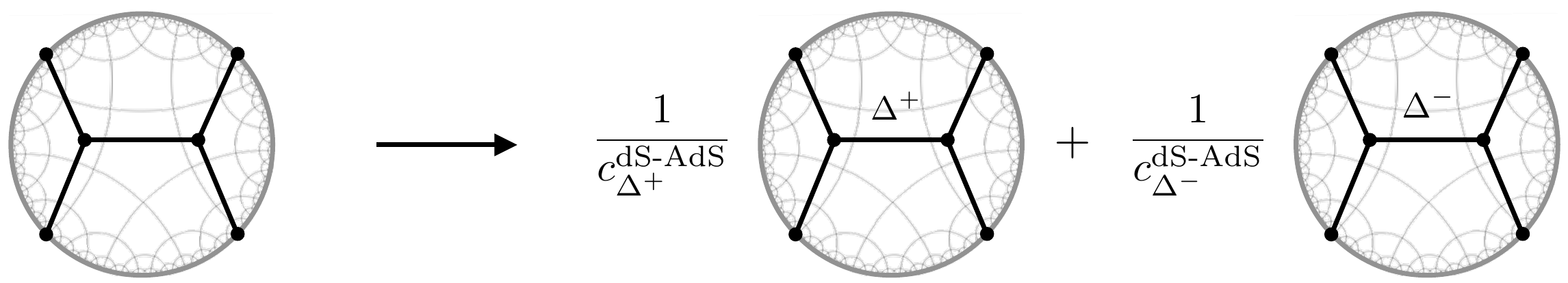},
\end{equation}

\item Finally, for each vertex multiply by the sinusodial factor \eqref{sinereviw} that converts the contact diagram the vertex generates in EAdS to a contact diagram in dS. For the example of the exchange this gives:
\begin{equation}\nonumber
\hspace*{-0.5cm}    \includegraphics[width=0.75\textwidth]{exchdsasads.pdf}
\end{equation}
which recovers the result \eqref{dSasAdSexch}.
\end{enumerate}

From a bootstrap perspective the above rules can be understood as follows: The analyticity properties of diagrams contributing to dS boundary correlators in the Bunch-Davies vacuum are inherited from the corresponding Witten diagrams with $\Delta^\pm$ boundary conditions. The assumption that we are in the Bunch-Davies vacuum at early times requires that the linear combination of such EAdS Witten diagrams is symmetric under $\Delta^+ \leftrightarrow \Delta^-$ for each internal leg, with the precise relative coefficients fixed by factorisation -- this is ensured by the fact that all contact subdiagrams are contact diagrams in dS, which is implemented by multiplying by the ratio for each vertex (rule \# 3 above).

To illustrate how these rules are work beyond the tree level exchanges and contact diagrams considered so far, let us give some examples. For concreteness we will focus on diagrams generated by scalar fields with non-derivative interactions though, as is clear from the above, it is trivial to write down the corresponding expressions for spinning fields and derivative interactions. In sections \ref{subsecc::candy4pt}, \ref{subsec::exbox4pt} and \ref{subsec::2ptbubbbleex} we consider one-loop examples of the four-point candy diagram, four-point box diagram and the two-point bubble diagram respectively. In section \ref{subsec::propprods} we derive some useful identities to compute higher loop diagrams in (EA)dS of the ``beach ball"-type i.e. formed by taking powers of bulk-to-bulk propagators.

\subsection{Example: Four-point candy diagram} 
\label{subsecc::candy4pt}

Consider the four-point candy diagram generated by the following non-derivative quartic vertices of scalar fields:
\begin{equation}
   g_1\, \sigma_1 \sigma_2 \phi_1 \phi_2, \qquad  g_2\,\sigma_3 \sigma_4 \phi_1 \phi_2, \label{candyquarticvertices}
\end{equation}
where we take the $\sigma_i$ as the external fields with scaling dimensions $\delta_i$ and the $\phi_{1,2}$ exchanged. Following the rules given above, the four-point candy diagram in dS$_{d+1}$ is given by the following linear combination of the corresponding candy Witten diagrams in EAdS$_{d+1}$:
\begin{equation}
    \includegraphics[width=0.9\textwidth]{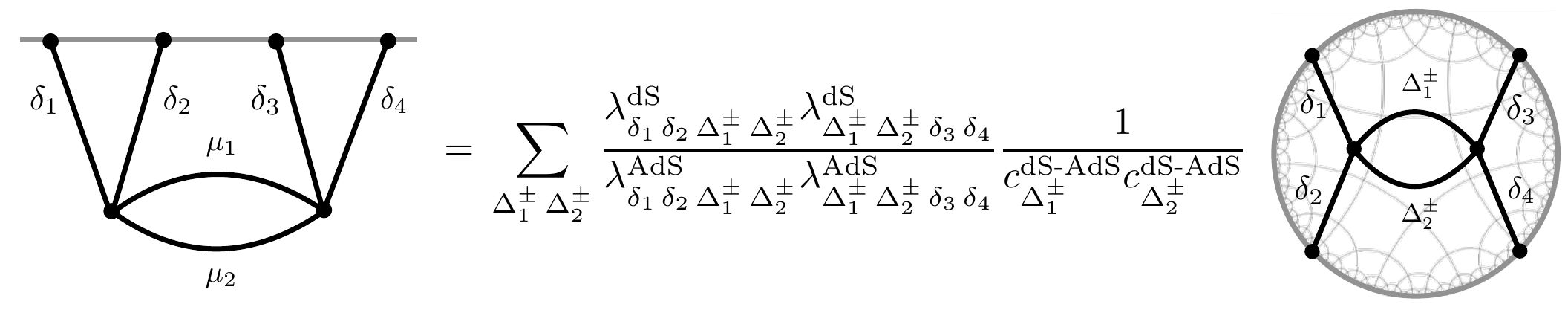},
\end{equation} 
where the $\Delta^{\pm}_{1,2}=\frac{d}{2}\pm i\mu_{1,2}$ modes are those of fields $\phi_{1,2}$. For each term in the sum, the ratios \eqref{sinereviw}
\begin{align}
 \hspace*{-0.5cm}  \lambda^{\text{dS}}_{\delta_a\,\delta_b\,\Delta^\pm_1\,\Delta^\pm_2} &= 2\,  c^{\text{dS-AdS}}_{\delta_a}c^{\text{dS-AdS}}_{\delta_b}c^{\text{dS-AdS}}_{\Delta^\pm_1}c^{\text{dS-AdS}}_{\Delta^\pm_2} \sin \left(\tfrac{\Delta^\pm_1+\Delta^\pm_2+\delta_a+\delta_b-d}{2}\right)\pi \, \times  \, \lambda^{\text{AdS}}_{\delta_a\,\delta_b\,\Delta^\pm_1\,\Delta^\pm_2},
\end{align}
take us from the quartic contact Witten diagrams generated by the vertices \eqref{candyquarticvertices} in EAdS to their counterparts in dS, as described in section \ref{subsec::npt}. Candy Witten diagrams in EAdS have been considered in various works e.g. \cite{Aharony:2016dwx,Cardona:2017tsw,Yuan:2018qva,Bertan:2018khc,Bertan:2018afl,Ghosh:2019lsx,Shyani:2019wed,Carmi:2019ocp,Meltzer:2019nbs,Meltzer:2019nbs,Carmi:2021dsn} using a variety of techniques. It can also be computed in the Mellin-Barnes representation, for instance using the cutting rules and dispersion formula given in section \ref{subsec::cuttinganddisp}. In particular, placing both internal legs on-shell the diagram factorises into a product of quartic contact Witten diagrams generated by the vertices \eqref{candyquarticvertices}: 
\begin{equation}\label{fig::candycut}
    \includegraphics[width=0.8\textwidth]{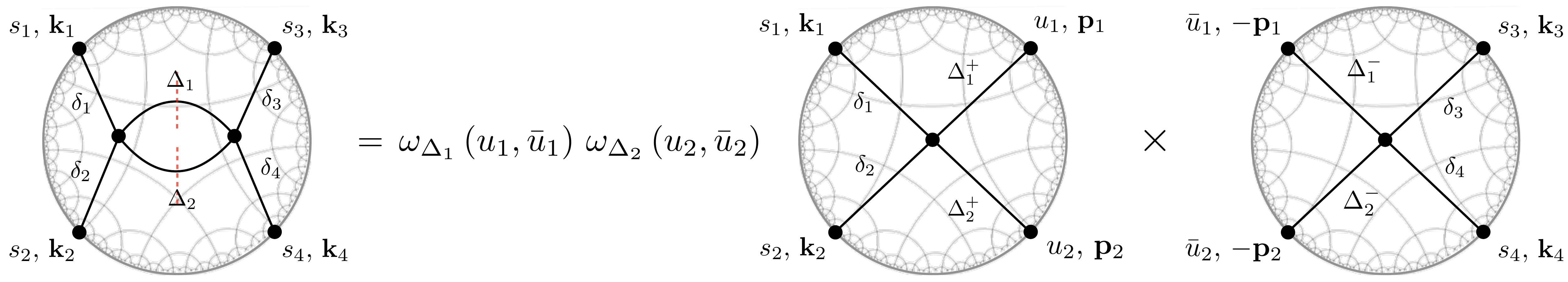}.
\end{equation} 
The Mellin-Barnes representation for such four-point contact Witten diagram factors was given in equation \eqref{nptcontMB}. The full diagram is then obtained by applying the dispersion formula \eqref{dispprop} to each internal leg:
\begin{equation}\label{fig::candydisp}
    \includegraphics[width=0.65\textwidth]{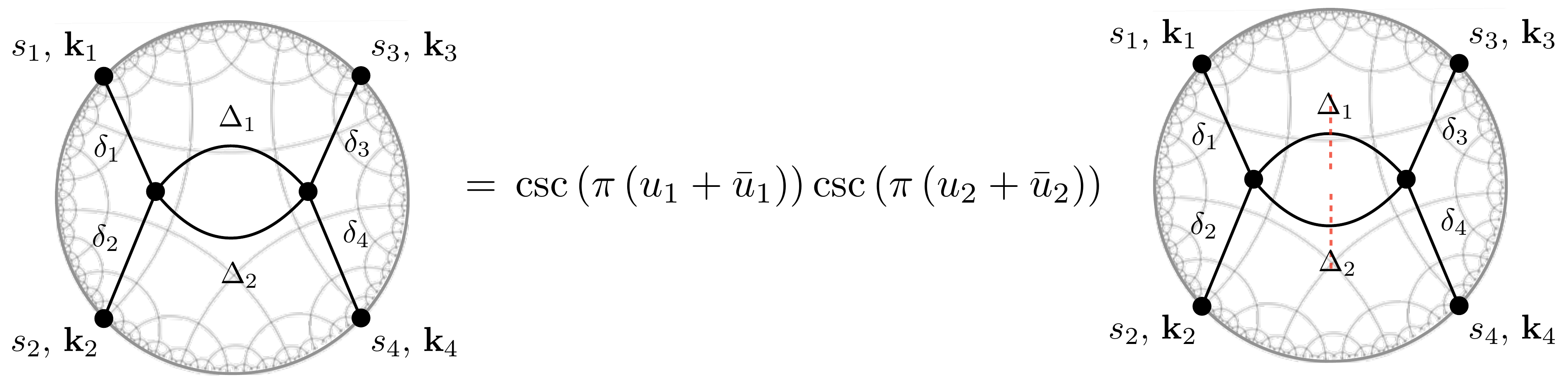}.
\end{equation} 

\subsection{Example: Four-point box diagram}
\label{subsec::exbox4pt}

Similarly we can consider the four-point box diagram generated by the non-derivative cubic vertices:
\begin{equation}\label{boxcubicvertices}
   g_1\, \sigma_1 \phi_3 \phi_4, \qquad  g_2\,\sigma_2 \phi_1 \phi_4,   \qquad  g_3\,\sigma_3 \phi_1 \phi_2, \qquad  g_2\,\sigma_4 \phi_2 \phi_3,
\end{equation}
taking the $\sigma_{1,2,3,4}$ as external scalar fields as before and $\phi_{1,2,3,4}$ exchanged. Applying the rules given above, the four-point box diagram in dS$_{d+1}$ is given by the following linear combination of the corresponding box Witten diagrams in EAdS$_{d+1}$:
\begin{equation}
    \includegraphics[width=\textwidth]{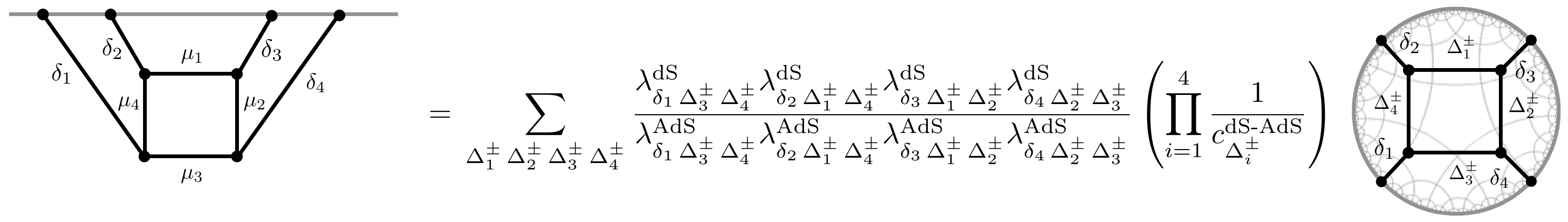},
\end{equation} 
where the $\Delta^{\pm}_{1,2,3,4}=\frac{d}{2}\pm i\mu_{1,2,3,4}$. In this case, the ratios 
\begin{align}
 \lambda^{\text{dS}}_{\delta_a\,\Delta^\pm_b\,\Delta^\pm_c} &= 2\,  c^{\text{dS-AdS}}_{\delta_a}c^{\text{dS-AdS}}_{\Delta^\pm_b}c^{\text{dS-AdS}}_{\Delta^\pm_c} \sin \left(\tfrac{\delta_a+\Delta^\pm_b+\Delta^\pm_c-d}{2}\right)\pi \, \times  \, \lambda^{\text{AdS}}_{\delta_a\,\Delta^\pm_b\,\Delta^\pm_c},
\end{align}
take us from the three-point contact Witten diagrams generated by the cubic vertices \eqref{boxcubicvertices} in EAdS to their counterparts in dS. Box Witten diagrams in EAdS have been considered using various techniques in e.g. \cite{Aharony:2016dwx,Yuan:2018qva,Carmi:2019ocp,Meltzer:2019nbs,Albayrak:2020bso,Carmi:2021dsn}. In the Mellin-Barnes representation, using the cutting rules outlined in section \ref{subsec::cuttinganddisp}, if we put two of the internal legs on-shell the diagram factorises into a product of four-point exchanges:
\begin{equation}\label{fig::boxcut}
    \includegraphics[width=0.75\textwidth]{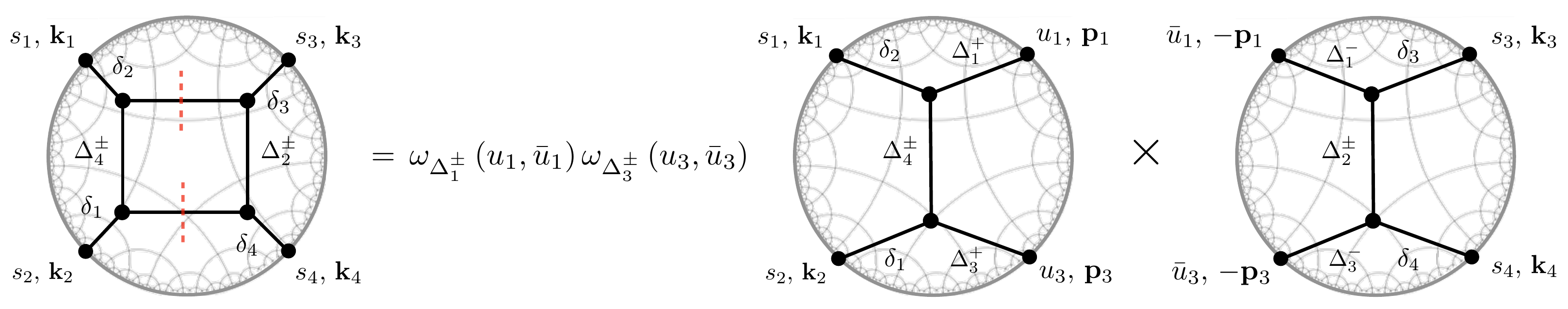},
\end{equation} 
whose Mellin-Barnes representation was given in \eqref{fullAdSexchMB}. As for the candy diagram considered above, the full diagram can be reconstructed applying the dispersion formula \eqref{dispprop}, which amounts to multiplying by the cosecent factors $\csc\left(\pi \left(u_1+{\bar u}_1\right)\right)\csc\left(\pi \left(u_2+{\bar u}_2\right)\right)$.

\subsection{Example: Two-point bubble diagrams}
\label{subsec::2ptbubbbleex}

It is also interesting to consider the effect of loop corrections to boundary two-point functions in de Sitter. See e.g. \cite{Marolf:2010zp,Krotov:2010ma}.

Conformal symmetry constrains boundary two-point functions of scalar fields to take the following form:
\begin{subequations}\label{anomdimdef}
\begin{align}
    \langle {\cal O}\left({\bf x}_1\right){\cal O}_1\left({\bf x}_2\right) \rangle &= \frac{C}{\left({\bf x}^2_{12}\right)^{\Delta+\gamma}}\\
    &=\frac{C}{\left({\bf x}^2_{12}\right)^{\Delta}}\left(1-\gamma\,\log\left({\bf x}^2_{12}\right)+\ldots\right),
\end{align}
\end{subequations}
where $\Delta$ is the tree-level scaling dimension and the anomalous dimension $\gamma$ is induced by loop corrections, which we see from the second line can be read off from logarithmic terms generated by loop diagrams.

Let us consider the one-loop two-point bubble diagram generated by the following cubic vertex of scalar fields $\phi_{1,2}$ and $\phi$:
\begin{equation}\label{cubic12phi}
    g\, \phi_1 \phi_2 \phi.
\end{equation}
Applying the rules outlined at the beginning of this section, the bubble diagram in dS$_{d+1}$ is given by the following linear combination of one-loop bubble Witten diagrams in EAdS$_{d+1}$:
\begin{equation}
    \includegraphics[width=0.8\textwidth]{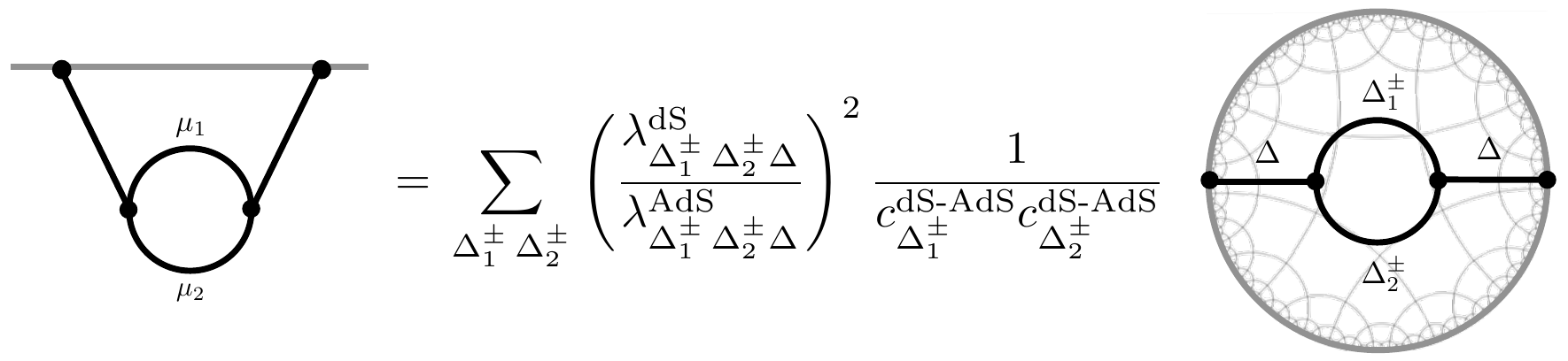},
\end{equation}
where
\begin{equation}
     \lambda^{\text{dS}}_{\Delta^\pm_1\,\Delta^\pm_2\,\Delta} = 2\,  c^{\text{dS-AdS}}_{\Delta}c^{\text{dS-AdS}}_{\Delta^\pm_1}c^{\text{dS-AdS}}_{\Delta^\pm_2} \sin \left(\tfrac{\Delta+\Delta^\pm_1+\Delta^\pm_2-d}{2}\right)\pi \, \times  \, \lambda^{\text{AdS}}_{\Delta^\pm_1\,\Delta^\pm_2\,\Delta},
\end{equation}
accounts for the change in the coefficient of the contact diagram generated by the vertex \eqref{cubic12phi} as we move from EAdS to dS. Accordingly, the corresponding anomalous dimensions $\gamma^{\text{dS}}_{\mu_1\,\mu_2}$ and $\gamma^{\text{AdS}}_{\Delta^\pm_1,\,\Delta^{\hat \pm}_2}$ which, as reviewed above, are given by the coefficients of the logarithmic terms in ${\cal B}^{\text{dS}}_{\mu_1,\,\mu_2}$ and ${\cal B}^{\text{AdS}}_{\Delta^\pm_1,\,\Delta^{\hat \pm}_2}$ respectively, are related via:
\begin{equation}\label{gammadsasads}
   \gamma^{\text{dS}}_{\mu_1\,\mu_2} =\,\frac{1}{c^{\text{dS-AdS}}_{\Delta}}\sum\limits_{\Delta^\pm_1\,\Delta^\pm_2} \left(\frac{\lambda^{\text{dS}}_{\Delta^\pm_1\,\Delta^\pm_2\Delta}}{\lambda^{\text{AdS}}_{\Delta^\pm_1\,\Delta^\pm_2\Delta}}\right)^2\frac{1}{c^{\text{dS-AdS}}_{\Delta^\pm_1}c^{\text{dS-AdS}}_{\Delta^\pm_2}}\,\gamma^{\text{AdS}}_{\Delta^\pm_1,\,\Delta^\pm_2},
\end{equation}
where we must divide by $c^{\text{dS-AdS}}_{\Delta}$ to account for the change in tree-level two-point coefficient from AdS to dS, as per the definition \eqref{anomdimdef} of the anomalous dimension.

In \cite{Giombi:2017hpr} two-point bubble diagrams in EAdS were studied using the spectral representation \eqref{dricheads} of bulk-to-bulk propagators. Upon evaluating the spectral integrals the following explicit expression was obtained for $d=2$:
\begin{align}\nonumber
    \gamma^{\text{AdS}}_{\Delta_1,\,\Delta_2}&=\frac{1}{16\pi \left(\Delta-1\right)^2}\left[\psi ^{(0)}\left(\tfrac{\Delta+\Delta_1+\Delta_2}{2}-1\right)+\psi ^{(0)}\left(2-\tfrac{\Delta+\Delta_1+\Delta_2}{2}\right)-\psi ^{(0)}\left(\tfrac{\Delta_1+\Delta_2-\Delta}{2}\right)\right.\\&\left.\hspace*{1cm}-\psi ^{(0)}\left(\tfrac{\Delta-\Delta_1-\Delta_2}{2}+1\right)+\frac{2\pi \sin \left(\pi  \left(\Delta-1\right) \right)}{\cos \left(\pi  \left(\Delta-1\right) \right)+\cos \left(\pi  \left(\Delta_1+\Delta_2-2\right)\right)}\right]\,,
\end{align}
where $\psi ^{(0)}$ is the digamma function. Plugging this expression into \eqref{gammadsasads} then gives an explicit expression for the corresponding anomalous dimension in dS.

\subsection{Example: Products of propagators}
\label{subsec::propprods}

The examples of loop diagrams we have considered so far have all been at one-loop. An interesting class of loop diagrams are given by those formed from products of bulk-to-bulk propagators connecting the same two bulk points. The bubble considered in the previous section is an example of such a loop diagram, where one loop formed from the product of two bulk-to-bulk propagators. To extend this to products of more than two propagators, and hence beyond one loop, it is useful to use the following identity satisfied by bulk-to-bulk propagators in EAdS \cite{Fitzpatrick:2010zm,Fitzpatrick:2011hu}:
\begin{align}\label{PropagatorSquared}
\Pi^{\text{AdS}}_{\Delta_1,0}\left(x_1;x_2\right)\Pi^{\text{AdS}}_{\Delta_2,0}\left(x_1;x_2\right)=\sum_n a_{\Delta_1\,\Delta_2}(n)\Pi^{\text{AdS}}_{\Delta_1+\Delta_2+2n,0}(x_1;x_2)\,,
\end{align}
where a product of two bulk-to-bulk propagators is expressed as a sum of bulk-to-bulk propagators with scaling dimensions $\Delta_1+\Delta_2+2n$ weighted by:
\begin{align}
    a_{\Delta_1\,\Delta_2}(n)=\frac{ \left(\frac{d}{2}\right)_n (-d+n+\Delta_1+\Delta_2+1)_n (2 n+\Delta_1+\Delta_2)_{1-\frac{d}{2}}}{2 n!\pi ^{\frac{d}{2}} (n+\Delta_1)_{1-\frac{d}{2}} (n+\Delta_2)_{1-\frac{d}{2}} \left(-\frac{d}{2}+n+\Delta_1+\Delta_2\right)_n}\,.
\end{align}
This identity can be trivially re-written as a spectral integral:
\begin{subequations}\label{prodpropspecrep}
\begin{align}
    \Pi^{\text{AdS}}_{\Delta_1,0}\left(x_1,x_2\right)\Pi^{\text{AdS}}_{\Delta_2,0}\left(x_1,x_2\right)&=\int_{-\infty}^{+\infty}d\nu\,b^{\text{AdS}}_{\Delta_1\,\Delta_2}(\nu)\,\Omega^{\text{AdS}}_{\nu,0}(x_1;x_2),\\
    &=\int_{-\infty}^{+\infty}d\nu\,b^{\text{AdS}}_{\Delta_1\,\Delta_2}(\nu)\,\frac{i\nu}{\pi}\Pi^{\text{AdS}}_{\frac{d}{2}+i\nu,0}(x_1;x_2)\,,
\end{align}
\end{subequations}
with
\begin{align}
    b^{\text{AdS}}_{\Delta_1\,\Delta_2}\left(\nu\right)=\frac12\sum_{n}\frac{1}{ (\Delta_1+\Delta_2-\tfrac{d}{2}+2 n)}\frac{a_{\Delta_1\,\Delta_2}(n)}{\Delta_1+\Delta_2-\tfrac{d}{2} +2 n-i \nu}+(\nu\to-\nu)\,,
\end{align}
and the second line of \eqref{prodpropspecrep} we used the identity \eqref{idsubharm}. The re-summation can be straightforwardly performed using Mathematica in terms of a ${}_7F_6$ hypergeometric function:\footnote{A similar expression was presented very recently in \cite{DiPietro:2021sjt}. Our result disagrees with their expression, which we believe is because of a small typo in their formula.} 
\begin{align}
    b^{\text{AdS}}_{\Delta_1\,\Delta_2}(\nu)&=\frac{\Gamma \left(\tfrac{d}{2}+i \mu_1\right) \Gamma \left(\tfrac{d}{2}+i \mu_2\right) \Gamma (1+i\mu_1+i\mu_2) \Gamma \left(\tfrac{d}{2}+ i (\mu_1+\mu_2)\right) \Gamma \left(\tfrac{d}{4} - \tfrac{i}2 (\nu -\mu_1-\mu_2)\right)}{\pi ^{\frac{d-1}{2}} 2^{d+i (\mu_1+\mu_2-2 i)}}\nonumber\\&{}_7\widetilde{F}_6\left(\begin{matrix}
    \frac{d}{2}&\tfrac{d}{2} + i \mu_1&\tfrac{d}2+i \mu_2&\tfrac{1+i \mu_1+i\mu_2}{2} &\tfrac{ 2+i\mu_1+i\mu_2}{2}  & \tfrac{d+2 i (\mu_1+\mu_2)}{2}&\tfrac{d-2 i (\nu -\mu_1-\mu_2)}{4}\\
    &1+i \mu_1&1+i \mu_2&1+i (\mu_1+\mu_2)&\tfrac{d+i (\mu_1+\mu_2)}2&\tfrac{d+1+i (\mu_1+\mu_2)}{2}&\frac{d+4-2 i (\nu - \mu_1- \mu_2)}{4}
    \end{matrix};1\right)\nonumber\\
    &+(\nu\to-\nu)\,,
\end{align}
with ${}_7\tilde{F}_6$ the regularised hypergeometric function and $\Delta_{1,2}=\frac{d}{2}+i\mu_{1,2}$. 

\begin{figure}[t]
    \centering
    \captionsetup{width=0.95\textwidth}
    \includegraphics[width=0.8\textwidth]{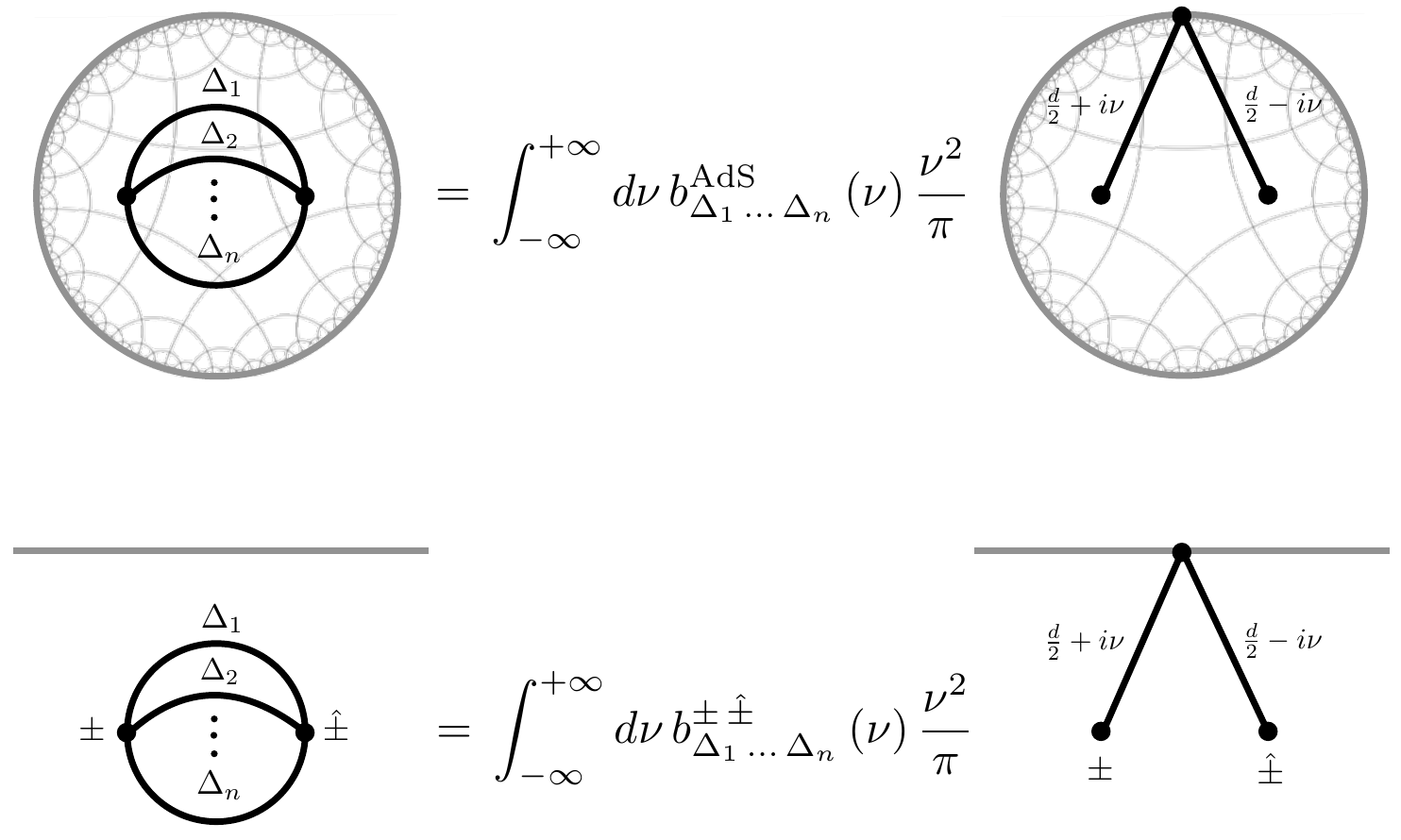}
    \caption{Products of bulk-to-bulk propagators at the same two bulk points admit a spectral decomposition, which is owing to the completeness and orthogonality of harmonic functions $\Omega^{\text{AdS}}_{\nu,J}$. Owing to the relations between propagators in EAdS and dS, the spectral function $b^{\pm {\hat \pm}}_{\Delta_1\ldots \Delta_n}\left(\nu\right)$ in dS can be expressed in terms of the AdS spectral function $b^{\text{AdS}}_{\Delta_1\ldots \Delta_n}\left(\nu\right)$, This relation is given in \eqref{dstoadsbeachballspect}.}
    \label{fig::prodpowersadsds}
\end{figure}

This result can be immediately extended to products of in-in propagators in dS using that they are given by a specific linear combination of analytically continued bulk-to-bulk propagators in EAdS, as reviewed in section \ref{subsec::props(EA)dS}. In particular, we have that
\begin{multline}
    \Pi^{\pm\,{\hat \pm}}_{\mu_1,0}\left(\eta,{\bf x};{\bar \eta},{\bar{\bf x}}\right)\Pi^{\pm\,{\hat \pm}}_{\mu_2,0}\left(\eta,{\bf x};{\bar \eta},{\bar{\bf x}}\right)= c^{\text{dS-AdS}}_{\Delta^+_1}c^{\text{dS-AdS}}_{\Delta^+_2}e^{\mp \tfrac{i\pi }{2}\left(\Delta^+_1+\Delta^+_2\right)}e^{{\hat \mp} \tfrac{i\pi }{2}\left(\Delta^+_1+\Delta^+_2\right)}\\ \hspace*{3.25cm}\times\Pi^{\text{AdS}}_{\Delta^+_1,0}\left(-\eta e^{\pm \frac{i \pi}{2}},{\bf x};-{\bar \eta}e^{{\hat \pm} \frac{i \pi}{2}},{\bar {\bf x}}\right)\Pi^{\text{AdS}}_{\Delta^+_2,0}\left(-\eta e^{\pm \frac{i \pi}{2}},{\bf x};-{\bar \eta}e^{{\hat \pm} \frac{i \pi}{2}},{\bar{\bf x}}\right)\\
   +\left(\Delta^+_1 \to \Delta^-_1\right)+\left(\Delta^+_2 \to \Delta^-_2\right)++\left(\Delta^+_1 \to \Delta^-_1, \Delta^+_2 \to \Delta^-_2\right).
\end{multline}
One then applies the Wick rotation \eqref{wickinin} to the identity \eqref{prodpropspecrep}, giving:
\begin{multline}
    \Pi^{\text{AdS}}_{\Delta_1,0}\left(-\eta e^{\pm \frac{i \pi}{2}},{\bf x};-{\bar \eta}e^{{\hat \pm} \frac{i \pi}{2}},{\bar{\bf x}}\right)\Pi^{\text{AdS}}_{\Delta_2,0}\left(-\eta e^{\pm \frac{i \pi}{2}},{\bf x};-{\bar \eta}e^{{\hat \pm} \frac{i \pi}{2}},{\bar {\bf x}}\right)\\=\frac{1}{4}\int^{+\infty}_{-\infty}d\nu\,b^{\text{AdS}}_{\Delta_1\,\Delta_2}\left(\nu\right)\,\text{csch}\left(\pi \nu\right) e^{\pm \tfrac{i\pi }{2}\left(\tfrac{d}{2}+i\nu\right)}e^{{\hat \pm} \tfrac{i\pi }{2}\left(\tfrac{d}{2}-i\nu\right)}\Omega^{\pm, {\hat \pm}}_{\nu,0}\left(\eta, {\bf x};{\bar \eta},  {\bar {\bf x}}\right). 
\end{multline}
This immediately gives:
\begin{align}
     \Pi^{\pm\,{\hat \pm}}_{\mu_1}(x_1,x_2)\,\Pi^{\pm\,{\hat \pm}}_{\mu_2}(x_1,x_2) &= \int^{+\infty}_{-\infty}d\nu\, b^{\pm\,{\hat \pm}}_{\mu_1\,\mu_2}\left(\nu\right)\,\Omega^{\pm, {\hat \pm}}_{\nu,0}\left(x_1;x_2\right), 
\end{align}
where
\begin{multline}\label{dstoadsbeachballspect}
    b^{\pm\,{\hat \pm}}_{\mu_1\,\mu_2}\left(\nu\right) =\frac{1}{4}\text{csch}\left(\pi \nu\right) \,e^{\pm \tfrac{i\pi }{2}\left(\tfrac{d}{2}+i\nu\right)}e^{{\hat \pm} \tfrac{i\pi }{2}\left(\tfrac{d}{2}-i\nu\right)}\\\hspace*{-2cm}\times\left[ c^{\text{dS-AdS}}_{\Delta^+_1}c^{\text{dS-AdS}}_{\Delta^+_2}e^{\mp \tfrac{i\pi }{2}\left(\Delta^+_1+\Delta^+_2\right)}e^{{\hat \mp} \tfrac{i\pi }{2}\left(\Delta^+_1+\Delta^+_2\right)}b^{\text{AdS}}_{\Delta^+_1\,\Delta^+_2}\left(\nu\right)\right.\\\hspace*{2cm}\left.+\left(\Delta^+_1 \to \Delta^-_1\right)+\left(\Delta^+_2 \to \Delta^-_2\right)++\left(\Delta^+_1 \to \Delta^-_1, \Delta^+_2 \to \Delta^-_2\right)\right].
\end{multline}

Such identities can be extended to any number $n$ of products of bulk-to-bulk propagators in (EA)dS$_{d+1}$ (see figure \ref{fig::prodpowersadsds}):
\begin{subequations}\label{genidpropdsads}
\begin{align}
\Pi^{\text{AdS}}_{\Delta_1,0}\left(x_1,x_2\right)\,\ldots\,\Pi^{\text{AdS}}_{\Delta_n,0}\left(x_1,x_2\right)&=\int_{-\infty}^{+\infty}d\nu\,b^{\text{AdS}}_{\Delta_1\,\ldots\,\Delta_n}(\nu)\,\Omega^{\text{AdS}}_{\nu,0}(x_1;x_2),\\
     \Pi^{\pm\,{\hat \pm}}_{\mu_1}(x_1,x_2)\,\ldots\,\Pi^{\pm\,{\hat \pm}}_{\mu_n}(x_1,x_2) &= \int^{+\infty}_{-\infty}d\nu\, b^{\pm\,{\hat \pm}}_{\mu_1\,\ldots\,\mu_n}\left(\nu\right)\,\Omega^{\pm, {\hat \pm}}_{\nu,0}\left(x_1;x_2\right), 
\end{align}
\end{subequations}
where
\begin{multline}
    b^{\pm\,{\hat \pm}}_{\mu_1\,\ldots\,\mu_n}\left(\nu\right) =\frac{1}{4}\text{csch}\left(\pi \nu\right) \,e^{\pm \tfrac{i\pi }{2}\left(\tfrac{d}{2}+i\nu\right)}e^{{\hat \pm} \tfrac{i\pi }{2}\left(\tfrac{d}{2}-i\nu\right)}
    \\\times \sum\limits_{\Delta^\pm_1 \ldots \Delta^\pm_n}\, c^{\text{dS-AdS}}_{\Delta^+_1}\,\ldots\,c^{\text{dS-AdS}}_{\Delta^+_n}e^{\mp \tfrac{i\pi }{2}\left(\Delta^+_1+\ldots+\Delta^+_n\right)}e^{{\hat \mp} \tfrac{i\pi }{2}\left(\Delta^+_1+\ldots+\Delta^+_n\right)}b^{\text{AdS}}_{\Delta^+_1\,\ldots\,\Delta^+_n}\left(\nu\right).
\end{multline}
One first determines the spectral function $b^{\text{AdS}}_{\Delta_1\,\ldots\,\Delta_n}$ in EAdS, which can be obtained for instance by iterating the identity \eqref{PropagatorSquared} above, from which the spectral function $b^{\pm\,{\hat \pm}}_{\mu_1\,\ldots\,\mu_n}\left(\nu\right)$ in dS follows from the identity above. The identities \eqref{genidpropdsads} simplify the computation of loop diagrams in (EA)dS containing such multiple products of bulk-to-bulk propagators as they are replaced by the spectral integral of a \emph{single} harmonic function which is factorised by virtue of the split representation \eqref{spliteadsds}. 

There are some simplifications in the case $d=2$. For instance, for $n=2$ we have \cite{Carmi:2018qzm}:
\begin{equation}
    b^{\text{AdS}}_{\Delta_1\,\Delta_2}\left(\nu\right) = \frac{i}{8\pi \nu}\left[\psi^{(0)}\left(\frac{-i\nu+i\mu_1+i\mu_2+1}{2}\right)-\psi^{(0)}\left(\frac{i\nu+i\mu_1+i\mu_2+1}{2}\right)\right],
\end{equation}
where $\psi^{(0)}$ is the digamma function, while for $n=3$, by iterating \eqref{PropagatorSquared} we obtain:
\begin{multline}\label{MellonF}
    b^{\text{AdS}}_{\Delta_1\,\Delta_2\,\Delta_3}\left(\nu\right)=\frac{1}{32 \pi ^2 \nu }\left(\mu_1+\mu_2+\mu_3-\nu\right) \Big[\psi^{(0)}\left(\tfrac{i\mu_1+i\mu_2+i\mu_3-i \nu}{2}+1\right)+\gamma\Big]\\+\left(\nu\to-\nu\right)\,,
\end{multline}
which is a new result.\footnote{Note that the sum leading to \eqref{MellonF} diverges logarithmically. However, after including the term $\nu\to -\nu$, \eqref{MellonF} does not depend on the regulator which enters only in the $\nu$-independent terms.}

\section{Analyticity}
\label{sec::analyticity}

This section gives an extended discussion of parts of section 4 and section 5 in \cite{Sleight:2020obc} on the conformal partial wave / conformal block expansion in de Sitter.

The fact that diagrams contributing to the perturbative computation of boundary correlators in dS can be expressed as a linear combination of EAdS Witten diagrams not only implies that we can import a wealth of techniques and results from EAdS to dS. It also implies that dS diagrams inherit properties from their EAdS counterparts. A particularly important property of boundary correlators in AdS is their single valuedness in the Euclidean regime. This in particular implies that that they admit an expansion in terms of the conformal partial waves/harmonic functions \eqref{CPW} \cite{Dobrev:1975ru,Dobrev:1977qv,Mack:2009mi,Costa:2012cb,Caron-Huot:2017vep} which, as reviewed in section \ref{subsec::bootEAdSexch}, provide an orthgonal basis of single-valued Eigenfunctions of the Casimir invariants of the conformal group. As pointed out in \cite{Sleight:2020obc} (section 4 below equation (4.10)), that dS diagrams can be expressed as a linear combination of EAdS Witten diagrams implies that dS boundary correlators are also single-valued (at least in perturbation theory) and hence also admit an expansion in terms of the Conformal Partial Waves \eqref{CPW}.\footnote{This observation, made in the companion paper \cite{Sleight:2020obc}, was very recently reiterated in \cite{DiPietro:2021sjt}.} The Conformal Partial Wave decomposition of a conformal four-point function takes the form
\begin{equation}\label{4ptcpwdecomp}
    \langle {\cal O}_1\,{\cal O}_2\,{\cal O}_3\,{\cal O}_4\rangle = \left(\text{non-normalisable}\right)+\sum\limits_J\int^{+\infty}_{-\infty}d\nu\,a_J\left(\nu\right){\cal F}_{\nu,J},
\end{equation}
where the spectral integral over the parameter $\nu$ captures the normalisable contributions to the four-point function. The spectral function $a_{J}\left(\nu\right)$ is a meromorphic function for boundary correlators in AdS and hence the same is true (at least in perturbation theory) for boundary correlators in dS for the Bunch-Davies vacuum. Note that from the conformal partial wave expansion \eqref{4ptcpwdecomp} one can obtain an expansion of the four-point function into conformal blocks using that the conformal partial wave is given by the linear combination \eqref{CPW} of conformal blocks, deforming the integration contour and evaluating the residues of the corresponding poles in $\nu$. The function $a_{J}\left(\nu\right)$ therefore encodes the operator product expansion (OPE) data (spectrum and OPE coefficients) via the location and residue of the poles in $\nu$, where the exchange of a state with spin-$J$ and (normalisable) scaling dimension $\Delta$ implies that $a_{J}\left(\nu\right)$ has the following pair of poles:
\begin{equation}
    a_J\left(\nu\right)\,\sim\,\frac{\lambda_{12\Delta}\lambda_{\Delta 34}}{C_{\Delta,J}}\frac{1}{\nu^2+\left(\Delta-\frac{d}{2}\right)^2}.
\end{equation}
We will not discuss the procedure of going from the conformal partial wave expansion to the conformal block decomposition in much further detail here, which has been described extensively in the literature e.g. \cite{Costa:2012cb,Caron-Huot:2017vep}.

For example, the Conformal Partial Wave decomposition of AdS exchanges given in \cite{Penedones:2010ue,Costa:2014kfa} and for the normalisable $\left(\Delta^+\right)$ boundary condition on the exchanged particle reads,
\begin{equation}\label{delta+normspect}
    {\cal A}^{\text{EAdS}}_{\Delta^+,J} = \int^{+\infty}_{-\infty} \frac{d\nu}{\nu^2+\left(\Delta^+-\frac{d}{2}\right)^2}\frac{\nu^2}{\pi}\,{\cal F}^{\text{EAdS}}_{\nu,J}\,+\,\text{contact,}
\end{equation}
which follows immediately from the spectral representation of the corresponding bulk-to-bulk propagator \eqref{dricheads} (originally given in \cite{Penedones:2007ns,Penedones:2010ue,Costa:2014kfa}) and the ``+ contact" are lower spin $\left(<J\right)$ contributions which correspond to contact terms in the exchange. The Conformal Partial Wave decomposition for the exchange with $\left(\Delta^-\right)$ boundary condition on the exchanged particle follows from the identity \eqref{idsubharm} states that the difference of bulk-to-bulk propagators with $\Delta^\pm$ boundary conditions is given by a harmonic function. At the level of the exchange Witten diagram the identity reads: 
\begin{equation}\label{delta-nonnormspect}
    {\cal A}^{\text{EAdS}}_{\Delta^-,J} = 2\mu i\,{\cal F}^{\text{EAdS}}_{\mu,J} +{\cal A}^{\text{EAdS}}_{\Delta^+,J}\,,
\end{equation}
where we see that the harmonic function ${\cal F}^{\text{EAdS}}_{\mu,J}$ gives the non-normalisable contribution in the decomposition \eqref{4ptcpwdecomp} while ${\cal A}^{\text{EAdS}}_{\Delta^+,J}$ gives the normalisable contributions given by the spectral integral \eqref{delta+normspect}. Given that the corresponding exchange \eqref{dSasAdSexch} in the Bunch-Davies vacuum of dS$_{d+1}$ is a linear combination of ${\cal A}^{\text{EAdS}}_{\Delta^\pm,J}$, the Conformal Partial Wave expansion of the dS exchange then follows from those \eqref{delta+normspect} and \eqref{delta-nonnormspect} of its EAdS$_{d+1}$ counterpart. It reads:
\begin{multline}\label{CPWdSexch}
  \hspace*{-1.75cm}  {\cal A}^{\text{dS}}_{\mu,J} =   \left(\prod\limits^4_{i=1} c^{\text{dS-AdS}}_{\Delta_i}\right) \left[-8 \mu i\,c^{\text{dS-AdS}}_{\Delta^+} \, \sin\left(\left(\tfrac{-d+\Delta_1+\Delta_2+\Delta^-+J_1+J_2+J}{2}\right)\pi\right)\,\sin\left(\left(\tfrac{-d+\Delta_3+\Delta_4+\Delta^-+J_3+J_4+J}{2}\right)\pi\right)\,{\cal F}^{\text{EAdS}}_{\mu,J}\right.\\\left.+2\sin\left(\left(\tfrac{-d+2J+\Delta_1+\Delta_2+\Delta_3+\Delta_4+J_1+J_2+J_3+J_4}{2}\right)\pi\right)\int^\infty_{-\infty}\frac{d\nu}{\nu^2+\left(\Delta^+-\frac{d}{2}\right)^2}\frac{\nu^2}{\pi}{\cal F}^{\text{EAdS}}_{\nu,J}\right],
\end{multline}
where to arrive to the above from \eqref{dSasAdSexch} we used the identity \eqref{3ptto4pttriid}. This extends the result of \cite{Sleight:2020obc} (equation (5.2)), giving the conformal partial wave expansion of the dS exchange for arbitrary internal and external spinning fields. Note that the conformal partial wave expansion \eqref{CPWdSexch} of the dS exchange contains a discrete contribution (the first line) in addition to the spectral integral on the second line owing to the propagation of the $\Delta^-$ mode, which is non-normalisable. It is also a nice consistency check to see that the contribution from the spectral integral on the second line is given by that \eqref{delta+normspect} and \eqref{delta-nonnormspect} of the corresponding exchanges in EAdS multiplied by the sinusoidal factor \eqref{uniacoupdsads}, which corrects for the change in the coefficient of four-point contact diagrams as we move from EAdS to dS. This is consistent with the fact that only the contribution from the spectral integral can encode contact contributions to the exchange, since the discrete contributions are given by a finite number of conformal partial waves which solve the \emph{homogeneous} conformal invariance equation \eqref{ccasmir}. Note that by using the fact that the conformal partial wave is a linear combination \eqref{CPW} of conformal blocks and deforming the integration contour as described above one recovers the conformal block expansion of the dS exchange given in section \ref{subsec::factunitcbe}.

Similarly, one can obtain the conformal partial wave expansions of loop diagrams in dS from the conformal partial wave expansions of their AdS counterparts. We have already seen such an example in section \ref{subsec::propprods}, where the conformal partial wave expansion of the four-point ``beach ball" diagram is given simply by attaching two pairs of external legs:
\begin{equation}
    \includegraphics[width=0.55\textwidth]{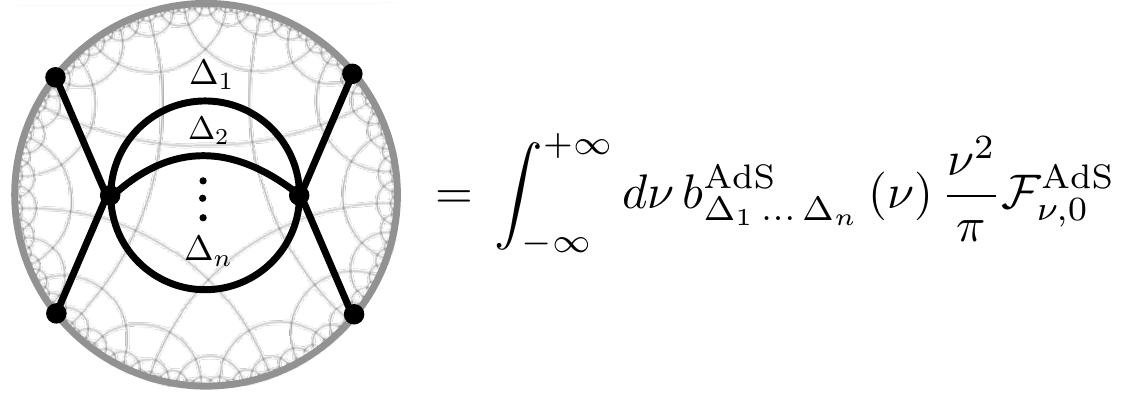}.
\end{equation}
The conformal partial wave expansion of the corresponding diagram in dS then follows immediately from the identity \eqref{dstoadsbeachballspect} relating the dS beach ball to the AdS beach ball. Note that in the conformal partial wave decomposition of such diagrams there is no discrete contribution -- it can be decomposed purely in terms of a spectral integral. This is because such products of bulk-to-bulk propagators can be expressed as a linear combination of \emph{normalisable} bulk-to-bulk propagators as in identity \eqref{PropagatorSquared}.

\paragraph{From dS to EAdS to Lorentzian AdS.} The fact that dS diagrams in the Bunch-Davies vacuum have the same analytic structure as their EAdS Witten diagram counterparts opens up the possibility to apply various well-honed techniques developed in the context of anti-de Sitter space to the perturbative computation of boundary correlators in de Sitter. In particular, the dS to EAdS rules given in section \ref{sec::generalalg}, which immediately provide the precise linear combination of EAdS Witten diagrams that compute a given diagram in dS, make it tempting to take one further step: To go from the EAdS diagrams computing our dS correlators to \emph{Lorentzian} AdS, where numerous powerful techniques have been developed such as the large spin expansion \cite{Fitzpatrick:2012yx,Komargodski:2012ek,Alday:2013cwa,Alday:2015ewa,Alday:2016njk,Simmons-Duffin:2016wlq,Aharony:2016dwx,Kulaxizi:2017ixa,Alday:2017gde,Li:2017lmh} and the Froissart-Gribov inversion formula \cite{Caron-Huot:2017vep}. Let us note that such techniques reconstruct correlators in AdS from their (lightcone) singularities and thus provide alternatives to the Mellin-Barnes cutting rules and dispersion formula \cite{Sleight:2020obc} reviewed in section \ref{subsec::cuttinganddisp} and related works \cite{Goodhew:2020hob,Jazayeri:2021fvk,Melville:2021lst,Goodhew:2021oqg,Baumann:2021fxj,Meltzer:2021zin} at the level of the wavefunction coefficients. In particular, in flat space the discontinuity of the (non-trivial) S-matrix ${\cal M}$ is related to the imaginary part via the optical theorem
\begin{equation}
    2 \text{Im}\left({\cal M}\right) = {\cal M}^\dagger {\cal M},
\end{equation}
which is a consequence of unitarity: ${\cal S}^\dagger {\cal S}=1$, where ${\cal S}=1+{\cal M}$. In Lorentzian CFTs, dual to physics in AdS, the analogue of $\text{Im}\left({\cal M}\right)$ is the double-discontinuity $\text{dDisc}$ of the CFT correlator: The inversion formula \cite{Caron-Huot:2017vep} reconstructs the CFT data from the double discontinuity. This perspective was taken in \cite{Meltzer:2019nbs}, which related the double discontinuity of AdS Witten diagrams to AdS Witten diagrams with internal lines placed on-shell. From the latter, the inversion formula reconstructs the full diagram. For example, the double-discontinuity of a conformal block is (see e.g. \cite{Liu:2018jhs,Meltzer:2019nbs}):
\begin{align}\label{ddcb}
   \text{dDisc}\left[W_{\Delta^\pm,J}\right]=2\sin\left(\left(\tfrac{-d+\Delta_1+\Delta_2+\Delta^ \mp+J}{2}\right)\pi\right)\sin\left(\left(\tfrac{-d+\Delta_3+\Delta_4+\Delta^ \mp+J}{2}\right)\pi\right)\,W_{\Delta^ \pm,J}\,,
\end{align}
where for simplicity we took the external fields to be scalars. It is interesting to note that the double discontinuity of the $\Delta^ \pm$ block gives precisely the sine factors \eqref{spinning3pt} for the three-point contact diagram of the $\Delta^\mp$ mode. The double-discontinuity of the $\Delta^\pm$ AdS exchanges is therefore 
\begin{multline}\label{ddadsexchcpw}
 \hspace*{-0.75cm}  \text{dDisc}\left[{\cal A}^{\text{EAdS}}_{\Delta^\pm,\,J}\right] = 2\sin\left(\left(\tfrac{-d+\Delta_1+\Delta_2+\Delta^ \mp+J}{2}\right)\pi\right)\sin\left(\left(\tfrac{-d+\Delta_3+\Delta_4+\Delta^\mp+J}{2}\right)\pi\right){\cal A}^{\text{EAdS}}_{\Delta^\pm,\,J}\Bigg|_{\text{on-shell}},
\end{multline}
which gives the cut \eqref{osexchcb} multiplied the sinusoidal factors \eqref{spinning3pt} for the $\Delta^ \mp$ mode. Note that the contributions from double-trace operators (which encode the contact terms in the exchange) are projected out by the double-discontinuity owing to the zeros of the sine factors in \eqref{ddcb} for double-trace values of the exchanged scaling dimension. Using that the dS exchange in the Bunch-Davies vacuum is the linear combination \eqref{dSasAdSexch} of $\Delta^ \pm$ AdS exchanges, its double discontinuity is therefore: 
\begin{multline}
   \text{dDisc}\left[ {\cal A}^{\text{dS}}_{\mu,J}\right]=4\sin\left(\left(\tfrac{-d+\Delta_1+\Delta_2+\Delta^++J}{2}\right)\pi\right)\sin\left(\left(\tfrac{-d+\Delta_3+\Delta_4+\Delta^++J}{2}\right)\pi\right)\\
    \times 4 \sin\left(\left(\tfrac{-d+\Delta_1+\Delta_2+\Delta^-+J}{2}\right)\pi\right)\sin\left(\left(\tfrac{-d+\Delta_3+\Delta_4+\Delta^-+J}{2}\right)\pi\right)\\\times \left(\prod\limits^4_{i=1}c^{\text{dS-AdS}}_{\Delta_i}\right)\frac{1}{2}c^{\text{dS-AdS}}_{\Delta^+}\left[{\cal A}^{\text{EAdS}}_{\Delta^+,J}-{\cal A}^{\text{EAdS}}_{\Delta^-,J}\right]\Bigg|_{\text{on-shell}}\\
    =4\sin\left(\left(\tfrac{-d+\Delta_1+\Delta_2+\Delta^++J}{2}\right)\pi\right)\sin\left(\left(\tfrac{-d+\Delta_3+\Delta_4+\Delta^++J}{2}\right)\pi\right)\\
    \times 4 \sin\left(\left(\tfrac{-d+\Delta_1+\Delta_2+\Delta^-+J}{2}\right)\pi\right)\sin\left(\left(\tfrac{-d+\Delta_3+\Delta_4+\Delta^-+J}{2}\right)\pi\right)\\\times \left(\prod\limits^4_{i=1}c^{\text{dS-AdS}}_{\Delta_i}\right)\frac{\mu^2 \Gamma\left(i\mu\right)\Gamma\left(-i\mu\right)}{2\pi}{\cal F}^{\text{AdS}}_{\mu,J},
\end{multline}
where in the second equality we used the identity \eqref{delta-nonnormspect}. It is interesting to note that the double discontinuity of the dS exchange projects onto the corresponding conformal partial wave. To tidy up the expression let us re-write this in terms of the dS normalised conformal partial wave, 
\begin{equation}
    {\cal F}^{\text{dS}}_{\mu,J}=\frac{\lambda^{\text{dS}}_{\Delta_1\,\Delta_2\,\Delta^+}\lambda^{\text{dS}}_{\Delta^-\,\Delta_3\,\Delta_4}}{\lambda^{\text{AdS}}_{\Delta_1\,\Delta_2\,\Delta^+}\lambda^{\text{AdS}}_{\Delta^-\,\Delta_3\,\Delta_4}}{\cal F}^{\text{AdS}}_{\mu,J},
\end{equation}
so that the double-discontinuity of the dS exchange can be written as:
\begin{multline}\label{dddsexchcpw}
   \text{dDisc}\left[ {\cal A}^{\text{dS}}_{\mu,J}\right]
    =4\sin\left(\left(\tfrac{-d+\Delta_3+\Delta_4+\Delta^++J}{2}\right)\pi\right) \sin\left(\left(\tfrac{-d+\Delta_1+\Delta_2+\Delta^-+J}{2}\right)\pi\right)\\\times  \frac{2\pi}{\Gamma\left(+i\mu\right)\Gamma\left(-i\mu\right)}{\cal F}^{\text{dS}}_{\mu,J}.
\end{multline}
It would be interesting to make contact between the above and the Cosmological Optical Theorem presented in \cite{Goodhew:2020hob}. The above relations also formally extend to loop level, taking the discontinuity of each internal leg. 

It is interesting to note that while the double discontinuity \eqref{ddadsexchcpw} of $\Delta^\pm$ AdS exchanges is not single-valued since it is proportional to individual conformal blocks \eqref{osexchcb}, the double discontinuity \eqref{dddsexchcpw} of exchanges in the Bunch-Davies vacuum of dS is single-valued since it is proportional to a conformal partial wave.

\section*{Acknowledgments}

C.S. is grateful to the Institute for Advanced Study for hospitality and support during 2018-2020, where the main part of this work was carried out. The research of C.S. was partially supported by the l'Universit\'e libre de Bruxelles, the European Union's Horizon 2020 research and innovation programme under the Marie Sk\l odowska-Curie grant agreement No 793661 and the STFC grant ST/T000708/1. The research of M.T. was partially supported by the program  “Rita  Levi  Montalcini”  of the MIUR (Minister for Instruction, University and Research) and the INFN initiative STEFI.

\appendix

\section{(EA)dS bulk-to-bulk propagators}
\label{app}

\subsection{Mellin-Barnes representation of EAdS bulk-to-bulk propagators}
\label{app::bubuEAdS}

In this appendix we determine the Mellin-Barnes representation \eqref{MBrepbubuads} of EAdS bulk-to-bulk propagators. The starting point is the spectral representation \eqref{dricheads} of the propagators with $\Delta^+$ boundary condition, which for scalar fields $\left(J=0\right)$ reads:
\begin{equation}
    \Pi^{\text{AdS}}_{\Delta^+,0}\left(x;{\bar x}\right)=\int^{+\infty}_{-\infty}\frac{d\nu}{\nu^2+\left(\Delta_+-\frac{d}{2}\right)^2}\,\Omega^{\text{AdS}}_{\nu,0}\left(x;{\bar x}\right).
\end{equation}
To determine the Mellin-Barnes representation we go to Fourier space and take the Mellin transform:
\begin{align}
 \hspace*{-0.5cm}    \Pi^{\text{AdS}}_{\Delta^+,0}\left(u, {\bf p};{\bar u},-{\bf p}\right)&=\int^{+\infty}_{-\infty}\frac{d\nu}{\nu^2+\left(\Delta_+-\frac{d}{2}\right)^2}\,\Omega^{\text{AdS}}_{\nu,0}\left(u, {\bf p};{\bar u},-{\bf p}\right)\\
     &=\int^{+\infty}_{-\infty}\frac{d\nu}{\nu^2+\left(\Delta_+-\frac{d}{2}\right)^2} \frac{\Gamma\left(u+\tfrac{i\nu}{2}\right)\Gamma\left(u-\tfrac{i\nu}{2}\right)\Gamma\left({\bar u}+\tfrac{i\nu}{2}\right)\Gamma\left({\bar u}-\tfrac{i\nu}{2}\right)}{4\pi\Gamma\left(+i\nu\right)\Gamma\left(-i\nu\right)}\left(\frac{p}{2}\right)^{-2\left(u+{\bar u}\right)},
\end{align}
where in the second equality we plugged in the Mellin-Barnes representation \eqref{spliteadsds} of the harmonic function, which is inherited from that \eqref{MBeadssc} of its constituent bulk-to-boundary propagators.

As noted in section 4.7 of \cite{Sleight:2019hfp}, the spectral integral in $\nu$ can be lifted using the identity:
\begin{multline}\label{specident}
    \int_{-\infty}^\infty d\nu \frac{\Gamma \left(a_1+\frac{i \nu }{2}\right)\Gamma \left(a_1-\frac{i \nu }{2}\right)\Gamma \left(a_2+\frac{i \nu }{2}\right)\Gamma \left(a_2-\frac{i \nu }{2}\right)\Gamma \left(a_3+\frac{i \nu }{2}\right)\Gamma \left(a_3-\frac{i \nu }{2}\right) }{\Gamma (-i \nu ) \Gamma (i \nu ) \Gamma \left(a_4+\frac{i \nu }{2}+1\right) \Gamma \left(a_4-\frac{i \nu }{2}+1\right)}\\=\frac{8 \pi  \Gamma (a_1+a_2) \Gamma (a_1+a_3) \Gamma (a_2+a_3) \Gamma (-a_1-a_2-a_3+a_4+1)}{\Gamma (1-a_1+a_4) \Gamma (1-a_2+a_4) \Gamma (1-a_3+a_4)}.
\end{multline}
A proof can be found in \cite{Sleight:2018ryu}. This gives:
\begin{multline}
    \int^{+\infty}_{-\infty}\frac{d\nu}{\nu^2+\left(\Delta_+-\frac{d}{2}\right)^2}\,\frac{\Gamma(u+\tfrac{i\mu}2)\Gamma(u-\tfrac{i\mu}2)\Gamma({\bar u}+\tfrac{i\mu}2)\Gamma({\bar u}-\tfrac{i\mu}2)}{\Gamma (i \mu ) \Gamma (-i \mu )}\\
    =\frac{2 \pi ^2 \csc (\pi  (u+\bar{u}))  \Gamma \left(u+\frac{i \mu }{2}\right)\Gamma \left(\bar{u}+\frac{i \mu }{2}\right)}{\Gamma \left(1-u+\frac{i \mu }{2}\right) \Gamma \left(1-\bar{u}+\frac{i \mu }{2}\right)}\,.
\end{multline}
Using that
\begin{equation}
    \frac{1}{\Gamma\left(1-u+\tfrac{i\mu}{2}\right)\Gamma\left(u-\tfrac{i\mu}{2}\right)} = \frac{1}{\pi} \sin\left(\pi\left(u-\tfrac{i\mu}{2}\right)\right),
\end{equation}
we obtain
\begin{equation}
\Pi^{\text{AdS}}_{\Delta^+,0}\left(u, {\bf p};{\bar u},-{\bf p}\right)=\csc\left(\pi\left(u+{\bar u}\right)\right)\omega_{\Delta^{+}}\left(u,{\bar u}\right) \Gamma\left(i\mu\right)\Gamma\left(-i\mu\right) \Omega^{\text{AdS}}_{\mu,0}\left(u,{\bf p};{\bar u},-{\bf p}\right),    
\end{equation}
where
\begin{equation}
    \omega_{\Delta^+}\left(u,{\bar u}\right) = 2 \sin\left(\pi\left(u - \tfrac{i\mu}{2}\right)\right)\sin\left(\pi\left({\bar u} - \tfrac{i\mu}{2}\right)\right).
\end{equation}
Now that we have lifted the spectral integral, which is only valid for the normalisable $\Delta^+$ boundary condition, we can replace $\Delta^+\to \Delta^-$ to obtain the $\Delta^-$ propagator too:
\begin{equation}\label{bubupropAdSapp}
\Pi^{\text{AdS}}_{\Delta^\pm,0}\left(u, {\bf p};{\bar u},-{\bf p}\right)=\csc\left(\pi\left(u+{\bar u}\right)\right)\omega_{\Delta^{\pm}}\left(u,{\bar u}\right) \Gamma\left(i\mu\right)\Gamma\left(-i\mu\right) \Omega^{\text{AdS}}_{\mu,0}\left(u,{\bf p};{\bar u},-{\bf p}\right),    
\end{equation}
with
\begin{equation}
    \omega_{\Delta^\pm}\left(u,{\bar u}\right) = 2 \sin\left(\pi\left(u \mp \tfrac{i\mu}{2}\right)\right)\sin\left(\pi\left({\bar u} \mp \tfrac{i\mu}{2}\right)\right).
\end{equation}
That this step is valid is confirmed by the identity \eqref{idsubharm}.

That the bulk-to-bulk propagators for spin-$J$ take the same form \eqref{bubupropAdSapp} (with the harmonic function for $J=0$ replaced by the harmonic function for general spin-$J$), up to contact terms, can be seen by comparing with the Mellin-Barnes representation of the spin-$J$ exchange in EAdS computed as described in \cite{Sleight:2019hfp}.

\subsection{dS propagators as Wick rotated AdS propagators}
\label{app::bubudS}

In this appendix we explain how in-in bulk-to-bulk propagators in dS can be written as linear combinations of analytically continued AdS ones, giving the details behind the result presented in \cite{Sleight:2020obc}. The starting point is the formulas presented in \cite{Sleight:2019mgd,Sleight:2019hfp} for the in-in bulk-to-bulk propagators in terms of analytically continued EAdS harmonic functions (which were identified in \cite{Sleight:2019mgd,Sleight:2019hfp} as the dS Wightman functions):
\begin{subequations}
\begin{align}
    \Pi^{\pm \mp}_{\mu,J}\left(\eta,{\bf p};{\bar \eta},-{\bf p}\right)&= \Gamma\left(+i\mu\right)\Gamma\left(-i\mu\right)\Omega^{\text{AdS}}_{\mu,J}(-\eta e^{\pm \frac{i\pi}{2}},{\bf p};-{\bar \eta} e^{\mp \frac{i\pi}{2}},-{\bf p}),\\
\Pi^{\pm \pm}_{\mu,J}\left(\eta,{\bf p};{\bar \eta},-{\bf p}\right) &= \Gamma\left(+i\mu\right)\Gamma\left(-i\mu\right)\left[\theta\left({\bar\eta}- \eta\right)\Omega^{\text{AdS}}_{\mu,J}(-\eta e^{\pm \frac{i\pi}{2}},{\bf p};-{\bar \eta} e^{\mp \frac{i\pi}{2}},-{\bf p})\right.\\&\left.\hspace*{5cm}+\theta\left(\eta-{\bar \eta}\right)\Omega^{\text{AdS}}_{\mu,J}(-\eta e^{\mp \frac{i\pi}{2}},{\bf p};-{\bar \eta} e^{\pm \frac{i\pi}{2}},-{\bf p})\right]. \nonumber
\end{align}
\end{subequations}
For the $\pm \mp$ propagators, using the identity \eqref{idsubharm} we can immediately write
\begin{multline}
    \Pi^{\pm \mp}_{\mu,J}\left(\eta,{\bf p};{\bar \eta},-{\bf p}\right)= c^{\text{dS-AdS}}_{\frac{d}{2}+i\mu}\,\Pi^{\text{AdS}}_{\frac{d}{2}+i\mu,J}(-\eta e^{\pm \frac{i\pi}{2}},{\bf p};-{\bar \eta} e^{\mp \frac{i\pi}{2}},-{\bf p})\,\\+\,c^{\text{dS-AdS}}_{\frac{d}{2}-i\mu}\,\Pi^{\text{AdS}}_{\frac{d}{2}-i\mu,J}(-\eta e^{\pm \frac{i\pi}{2}},{\bf p};-{\bar \eta} e^{\mp \frac{i\pi}{2}},-{\bf p}).
\end{multline}
This implies that in the ansatz \eqref{dslcadsbubu} we have, 
\begin{align}
    \alpha^{\pm \mp} =\frac{1}{c^{\text{dS-AdS}}_{\frac{d}{2}-i\mu}}e^{\mp \pi \mu}, \qquad \beta^{ \pm \mp}=\frac{1}{c^{\text{dS-AdS}}_{\frac{d}{2}+i\mu}}e^{\mp \pi \mu}.
\end{align}

To determine these coefficients for the $\pm\,\pm$ propagators we take the Mellin transform, which resolves the $\theta$-function singularities: 
\begin{align}
\hspace*{-1cm} \Pi^{\pm\, {\hat \pm}}_{\mu,J}\left(u,{\bf p};{\bar u},-{\bf p}\right) &= \int^0_{-\infty} d\eta\,d{\bar \eta}\left(-\eta\right)^{u-1}\left(-{\bar \eta}\right)^{{\bar u}-1} \Pi^{\pm\, {\hat \pm}}_{\mu,J}\left(\eta,{\bf p};{\bar \eta},-{\bf p}\right)  \\
 &=\Gamma\left(+i\mu\right)\Gamma\left(-i\mu\right)\int^0_{-\infty} d\eta\,d{\bar \eta}\left(-\eta\right)^{u-1}\left(-{\bar \eta}\right)^{{\bar u}-1}\nonumber\\&\hspace*{3.5cm}\times \left[\theta\left({\bar \eta}-\eta\right)\Omega^{\text{AdS}}_{\mu,J}(-\eta e^{\pm \frac{i\pi}{2}},{\bf k};-{\bar \eta} e^{\mp \frac{i\pi}{2}},-{\bf k})+\left(\eta \leftrightarrow {\bar \eta}\right)\right]\,.\nonumber
\end{align}
To evaluate the integrals in $\eta$ and ${\bar \eta}$ it is convenient to plug in the Mellin representation of the harmonic function, which for spin $J=0$ reads:
\begin{multline}
    \Gamma\left(+i\mu\right)\Gamma\left(-i\mu\right)\Omega^{\text{AdS}}_{\mu,J}(-\eta e^{\pm \frac{i\pi}{2}},{\bf k};-{\bar \eta} e^{\mp \frac{i\pi}{2}},-{\bf k})\\
    =\Gamma\left(i\mu\right)\Gamma\left(-i\mu\right)\int^{+i\infty}_{-i\infty}\frac{ds_1ds_2}{(2\pi i)^2}\,(-\eta )^{\frac{d}{2}-2 s_1} (-\bar{\eta})^{\frac{d}{2}-2 s_2}e^{\pm i \pi  (s_1-s_2)}\\ \times \underbrace{\frac{1}{4\pi\,\Gamma\left(i\mu\right)\Gamma\left(-i\mu\right)}\Gamma \left(s_1-\tfrac{i \mu }{2}\right) \Gamma \left(s_1+\tfrac{i \mu }{2}\right) \Gamma \left(s_2-\tfrac{i \mu }{2}\right) \Gamma \left(s_2+\tfrac{i \mu }{2}\right)\left(\frac{p}{2}\right)^{-2(s_1+s_2)}}_{\Omega^{\text{AdS}}_{\mu,0}(s_1,{\bf p};s_2,-{\bf{p})}}\,.\nonumber
\end{multline}
The $\eta$ and $\bar{\eta}$ integrals can be regulated to give:
\begin{subequations}
\begin{align}
    &\int^0_{-\infty} d\eta\,d{\bar \eta}\,\theta (\eta -\bar{\eta}) (-\eta )^{\frac{d}{2}-2 s_1+u-1} (-\bar{\eta})^{\frac{d}{2}-2 s_2+\bar{u}-1}\label{Ig}\\&\hspace*{6cm}=\lim_{\eta_0\to0}\frac{2 (-\eta_0)^{d-2 s_1-2 s_2+u_1+\bar{u}}}{(d-4 s_2+2 \bar{u}) (d-2 s_1-2 s_2+u+\bar{u})}\,,\nonumber\\
    &\int^0_{-\infty} d\eta\,d{\bar \eta}\,\theta (\bar{\eta}-\eta) (-\eta )^{\frac{d}{2}-2 s_1+u-1} (-\bar{\eta})^{\frac{d}{2}-2 s_2+\bar{u}-1}\label{Il}\\&\hspace*{6cm}=\lim_{\eta_0\to0}\frac{2 (-\eta_0)^{d-2 s_1-2 s_2+u+\bar{u}}}{(d-4 s_1+2 u) (d-2 s_1-2 s_2+u+\bar{u})}\,,\nonumber
\end{align}
\end{subequations}
where the following conditions specify how to treat the integration contour:
\begin{align}
    \Re\left(\frac{d}{2}-2 s_1+u\right)&<0\,,& \Re\left(\frac{d}{2}-2 s_2+\bar{u}\right)&<0\,,& \Re(d-2 s_1-2 s_2+u+\bar{u})&<0\,.
\end{align}
Splitting now the integral of the sum of two $\theta$ function terms into the sum of two separate Mellin-Barnes integrals, we can evaluate the $s_1$ integral for the term proportional to \eqref{Ig} and the $s_2$ integral for the term proportional to $\eqref{Il}$. After a change of variables the leftover Mellin-Barnes integral simplifies to:
\begin{align}
   & \Pi^{\pm\, {\hat \pm}}_{\mu,0}\left(u,{\bf p};{\bar u},-{\bf p}\right)=\frac1{4\pi}\int^{+i\infty}_{-i\infty}\frac{ds}{2\pi i}\left[\frac{e^{\pm i \pi  (2 s+u-\bar{u})}}{s-\epsilon }-\frac{e^{\mp i \pi  (2 s+u-\bar{u})}}{s+\epsilon }\right]\\&\hspace*{2cm}\times \Gamma \left(s+u-\tfrac{i \mu }{2}\right) \Gamma \left(s+u+\tfrac{i \mu }{2}\right) \Gamma \left(-s+\bar{u}-\tfrac{i \mu }{2}\right) \Gamma \left(-s+\bar{u}+\tfrac{i \mu }{2}\right)\left(\frac{p}{2}\right)^{-2 (u+\bar{u})}\,.\nonumber
\end{align}
This integral has been studied in detail in \cite{Sleight:2019mgd,Sleight:2019hfp} and can be evaluated explicitly in the form:
\begin{multline}\label{bubudsapp1}
    \Pi^{\pm\, {\hat \pm}}_{\mu,0}\left(u,{\bf p};{\bar u},-{\bf p}\right)=\omega^{\pm\pm}_\mu(u,\bar{u})
    \,\csc (\pi  (u+\bar{u}))\Gamma(i\mu)\Gamma(-i\mu)\\ \times \underbrace{c^{\text{dS-AdS}}_{\frac{d}{2}+i\mu}c^{\text{dS-AdS}}_{\frac{d}{2}-i\mu}\,e^{\mp \left(u+{\bar u}\right)\pi i}\Omega^{\text{AdS}}_{\mu,0}(u,{\bf p};{\bar u},-{\bf{p})}}_{\Omega^{\pm\,\pm}_{\mu,J}\left(u, {\bf p};{\bar u},-{\bf p}\right)}\,,\nonumber
\end{multline}
where
\begin{align}
    \omega^{\pm\pm}_\mu(u,\bar{u})=\mp \frac{i}{2}\Big(e^{\mp 2 i \pi  u}+e^{\mp 2 i \pi  \bar{u}}&-e^{-\pi  \mu }-e^{\pi  \mu }\Big)\,\frac{e^{\pm \left(u+{\bar u}\right)\pi i}}{c^{\text{dS-AdS}}_{\frac{d}{2}+i\mu}c^{\text{dS-AdS}}_{\frac{d}{2}-i\mu}}.
\end{align}
Note that we defined $\omega^{\pm\pm}_\mu(u,\bar{u})$ so that the phases and factors of $c^{\text{dS-AdS}}_{\frac{d}{2}\pm i\mu}$ appearing in \eqref{bubudsapp1} implement the analytic continuation of the EAdS harmonic function \eqref{dsharmphases} according to the Wick rotations \eqref{wickinin}. To determine the precise linear combination of analytically continued AdS propagators we compare this expression to the corresponding one for the bulk-to-bulk propagators in EAdS derived in the previous section:
\begin{equation}
\Pi^{\text{AdS}}_{\Delta^\pm,0}\left(u, {\bf p};{\bar u},-{\bf p}\right)=\csc\left(\pi\left(u+{\bar u}\right)\right)\omega_{\Delta^{\pm}}\left(u,{\bar u}\right) \Gamma\left(i\mu\right)\Gamma\left(-i\mu\right) \Omega^{\text{AdS}}_{\mu,0}\left(u,{\bf p};{\bar u},-{\bf p}\right). 
\end{equation}
In particular, the problem is to solve the following equations for $\alpha^{\pm\,\pm}$ and $\beta^{\pm\,\pm}$:
\begin{align}
    \omega^{\pm\,\pm}_\mu(u,\bar{u})&=\alpha^{\pm\,\pm} \omega^{\text{AdS}}_{\Delta^+}(u,\bar{u})+\beta^{\pm\,\pm} \omega^{\text{AdS}}_{\Delta^-}(u,\bar{u}).
\end{align}
This is straightforward to implement e.g. in Mathematica, giving:
\begin{align}
    \alpha^{\pm\, \pm} =\frac{1}{c^{\text{dS-AdS}}_{\frac{d}{2}-i\mu}}e^{\pm \pi \mu}, \qquad \beta^{\pm \,\pm}=\frac{1}{c^{\text{dS-AdS}}_{\frac{d}{2}+i\mu}}e^{\mp \pi \mu}.
\end{align}

That the bulk-to-bulk propagators for spin-$J$ take the same form (with the harmonic function for $J=0$ replaced by the harmonic function for general spin-$J$), up to contact terms, can be seen by comparing with the Mellin-Barnes representation of the spin-$J$ exchange in dS computed in \cite{Sleight:2019hfp}.

\newpage

%%%%%% References %%%%%%%

\providecommand{\href}[2]{#2}\begingroup\raggedright\endgroup

\end{document}